\documentclass[a4paper,11pt]{article}
\pdfoutput=1
\usepackage{jheppub}
\usepackage[T1]{fontenc}
\usepackage{mathtools}

\title{Inflation in Extra-Dimensions with one or two branes}

\author[a]{Nicolás Bernal,$^1$\note{ORCID: \href{https://orcid.org/0000-0003-1069-490X}{0000-0003-1069-490X}}}
\author[b,c]{Catarina Cosme,$^2$\note{ORCID: \href{https://orcid.org/0000-0001-5148-8868}{0000-0001-5148-8868}}}
\author[b]{Andrea Donini,$^3$\note{ORCID: \href{https://orcid.org/0000-0001-6668-5477}{0000-0001-6668-5477}}}
\author[b]{and Nuria Rius.$^4$\note{ORCID: \href{http://orcid.org/0000-0002-0606-4297}{0000-0002-0606-4297}}}

\affiliation[a]{New York University Abu Dhabi\\
PO Box 129188, Saadiyat Island, Abu Dhabi, United Arab Emirates}
\affiliation[b]{Instituto de Física Corpuscular, Universidad de Valencia and CSIC\\
Edificio Institutos Investigación, Catedrático Jose Beltrán 2, Paterna, 46980 Spain}
\affiliation[c]{Univ Coimbra, Faculdade de Ci\^{e}ncias e Tecnologia da Universidade de Coimbra and CFisUC,
Rua Larga, 3004-516 Coimbra, Portugal}

\emailAdd{nicolas.bernal@nyu.edu}
\emailAdd{ccosme@uc.pt}
\emailAdd{\\donini@ific.uv.es}
\emailAdd{nuria.rius@ific.uv.es}

\abstract{In this paper, we study two inflationary models, namely, {\em monomial inflation} and the simplest {\em $\alpha$-attractor inflation}, within extra-dimensional frameworks. We consider three extra-dimensional setups: {\em Dark Dimension}, which embeds one flat extra-dimension to explain the observed smallness of the 4D cosmological constant $\Lambda_4$; and the two {\em Randall-Sundrum} scenarios with one warped extra-dimension, namely {\em RS1} with two branes  and {\em RS2} with one brane. We derive the corresponding Friedmann equations, compute the slow-roll parameters in each case, and we fit the experimental data for ($n_s - 1$, $\alpha$, $\Delta_s^2$, $r$), using Planck, BICEP and ACT data. We find that monomial inflation is strongly disfavored in all scenarios, while $\alpha$-attractor inflation provides an excellent fit to current observations, with extra-dimensional setups offering additional flexibility compared to the standard 4D case.}

\begin{document}

\begin{flushright}
\end{flushright}

\maketitle

\section{Introduction}
Despite the remarkable success of the Standard Model (SM) in explaining the observed elementary particles and their interactions, there are hints both from the experimental and theoretical sides that it is not a complete theory. We have evidence of the existence of Dark Matter (DM), neutrino masses, and the baryon asymmetry of the Universe, which cannot be accommodated within the SM. From a theoretical perspective, we would like to understand the origin of the many free parameters of the SM, in particular, those related to its flavor structure, to find a unified description of the four fundamental interactions,  or to explain the huge hierarchy between the electroweak and Planck scales. 

After the discovery of the Brout-Englert-Higgs boson in 2012, the LHC has not provided any clue about the possible solutions to the above problems, except for suggesting that new physics may be (much) above the TeV scale or feebly interacting, since no signal has been found so far. However, the evidence for physics beyond the SM is quite robust, and it is worth continuing to explore scenarios that may account for some or ideally all the experimental observations and also provide more complete theoretical frameworks. 

Among various possibilities, the existence of extra spatial dimensions has been thoroughly studied over the years. Originally introduced by Nordström~\cite{Nordstrom:1914ejq}, Kaluza~\cite{Kaluza:1921tu} and Klein~\cite{Klein:1926tv} in the early twentieth century in order to unify gravity and electromagnetism, a renewed interest in this idea arose in the 1980s when trying to combine the principles of quantum mechanics and relativity with the development of string theory, which was only consistent in more than 4D. From a more phenomenological perspective, extra dimensions were also proposed as a solution to the hierarchy problem in the late 1990s, by relating the observed 4D Planck scale of order $10^{18}$~GeV to the fundamental gravity scale in D dimensions, $M_D \sim \cal{O}$(TeV) via a volume factor in Large Extra Dimensions (LED)~\cite{Antoniadis:1990ew, Antoniadis:1997zg, Arkani-Hamed:1998jmv, Antoniadis:1998ig, Arkani-Hamed:1998sfv}, or by a warping of spacetime which induces an effective Planck scale in the 4D brane, $\Lambda \ll M_\text{Planck}$ such as in the Randall-Sundrum (RS) scenarios~\cite{Randall:1999ee, Randall:1999vf}. Recently,  a mixture of the two mechanisms was proposed in the Clockwork/Linear Dilaton (CW/LD) model~\cite{Giudice:2016yja, Giudice:2017fmj}. Moreover, applications of the holographic AdS/CFT correspondence seem to indicate a duality between strongly coupled theories in $D$ dimensions and a gravitational dual in $D+1$ dimensions, even beyond supersymmetric or exactly conformal theories~\cite{Aharony:1999ti}. Thus, computations in extra dimensions can also be viewed as a handle to better understand strongly interacting theories; in particular, trying to find the QCD dual has received considerable attention.

As a consequence, the phenomenological signatures of extra dimensions at the TeV scale were scrutinized and searched at LHC. Even though we have not found them yet, given that it is conceivable (and sometimes theoretically required for consistency reasons) that they exist, it is interesting to keep looking for their possible impact on currently accessible observables. Consider, for example, DM. Up to now, the only evidence of its existence has been due to gravitational effects, and it could well be that it does not have any other kind of interaction. In this case, it will be completely undetectable in current and future particle physics experiments. However, this is only true if we live in a 4D space-time: in extra-dimensional scenarios the gravitational interaction may be enhanced, and a DM particle with just such an interaction could become a WIMP~\cite{Arcadi:2024ukq}, that is, a stable or cosmologically long-lived weakly interactive massive particle, with mass typically in the range $\sim 10$~GeV to $\sim 100$~TeV, and whose relic abundance is set via the freeze-out mechanism~\cite{Lee:2013bua, Lee:2014caa, Han:2015cty, Rueter:2017nbk, Rizzo:2018obe, Rizzo:2018joy, Carrillo-Monteverde:2018phy, Folgado:2019sgz, Folgado:2019gie, deGiorgi:2021xvm, Chivukula:2024nzt, Donini:2025cpl, Donini:2025qrf}, or a FIMP~\cite{McDonald:2001vt, Choi:2005vq, Kusenko:2006rh, Petraki:2007gq, Hall:2009bx, Bernal:2017kxu}, feebly interacting massive particle whose relic abundance is determined by freeze-in~\cite{Brax:2019koq, Bernal:2020fvw, Bernal:2020yqg, deGiorgi:2022yha}. In the latter case, the final abundance of DM depends on the reheating temperature and, therefore, on the complete theoretical model including inflation. 

On the other hand, the existence of extra dimensions itself also affects the inflationary paradigm in several aspects: namely, in LED, to reproduce the observed scalar perturbations, the fundamental Planck scale must be much larger than ${\cal O}$(TeV), unless the inflaton field propagates in the bulk~\cite{Mohapatra:2000cm}. This possibility has been recently explored~\cite{Antoniadis:2023sya} in the framework of the {\em Dark Dimension} paradigm, {\em i.e.} an attempt of using one flat spatial dimension to explain the observed value of the 4D cosmological constant $\Lambda_4$. Moreover, it has been shown that during inflation, living on a 4D brane of a 5D bulk can have observable effects, leading to an additional contribution to the Hubble parameter related to the brane tension~\cite{Langlois:2002bb}.

In this work, our aim is to analyze the viability of several extra-dimensional inflationary scenarios by comparing their predictions with the current available cosmological data from the Planck~\cite{Planck:2018jri}, BICEP~\cite{BICEP:2021xfz} and ACT~\cite{AtacamaCosmologyTelescope:2025blo, AtacamaCosmologyTelescope:2025nti} collaborations. We are agnostic about the hierarchy problem, so we allow varying the Planck scale in extra-dimensional models (either fundamental or effective) in a wide range from a purely phenomenological perspective, drawing conclusions about the values which are consistent with present data. In this paper, we focus on both the two-brane and one-brane RS models proposed in Refs.~\cite{Randall:1999ee} and~\cite{Randall:1999vf}, respectively. The first one (known as RS1, even though it considers two branes) is traditionally explored for its low-energy phenomenology, whereas the second one (known commonly as RS2, even though it only considers one brane) is more popular in the literature concerning cosmology. The latter model, in fact, cannot solve the hierarchy problem in its simplest form, although it contains a non-compact extra-dimension.

The outline of the paper is as follows: in Section~\ref{sec:extraD}, we briefly review the main features of the two extra-dimensional frameworks considered, the Dark Dimension (DD) and the two versions of the Randall-Sundrum model, RS1 and RS2. Section~\ref{sec:Friedmann} is devoted to the derivation of Friedmann's equation and the Hubble parameter on the IR-brane starting from the 5D Einstein equations. In Section~\ref{sec:inflationBrane}, we obtain analytic approximated expressions for the slow-roll (SR) conditions as well as for the scalar and tensor perturbations, taking into account the non-standard Hubble parameter due to the extra-dimensional framework, when needed. In Section~\ref{sect:inflapot}, we test the monomial and $\alpha$-attractor inflationary potentials against Planck, BICEP and ACT data, under the assumption that the inflaton field is confined to the same brane as the SM particles, identifying the allowed regions of the free parameters consistent with current observations. We finally conclude in Section~\ref{sec:conclusions}. Some technical details are collected in the appendices: Appendix~\ref{sec:EEqand5D} is devoted to reconcile different versions of the action for extra-dimensional models that appear in the literature; in Appendix~\ref{sec:AppHdot}, we present the time-derivatives of the Hubble parameter in the  4D, DD, RS1 and RS2 scenarios; in Appendix~\ref{sec:AppSRDD} we present the SR parameters when a constant is added to the Friedmann equation (such as in the case of the DD scenario), while in Appendix~\ref{sec:FOcorrectionsRS2small} we give the expressions for the SR parameters and the physical observables at first order in $V/\sigma_0$ in the RS2 scenario.

\section{Brief review on extra-dimensional frameworks} 
\label{sec:extraD}
Even if the original motivation for extra-dimensional extensions of gravity was the attempt to unify gravity with electromagnetism, the most recent proposals were, in fact, due to the less ambitious goal to solve the so-called {\em hierarchy problem} of the SM, that is, the huge difference between the typical scale of SM processes (the {\em electroweak scale}), and the fundamental scale of gravity. In this paper, we will focus on two popular extra-dimensional scenarios that have been proposed twenty years ago to address the hierarchy problem and some other shortcomings of the SM, the {\em Large Extra-Dimensions} (LED) model~\cite{Antoniadis:1997zg, Arkani-Hamed:1998jmv} and the Randall-Sundrum setups RS1~\cite{Randall:1999ee} and RS2~\cite{Randall:1999vf}, also called {\em warped extra-dimensions}. In both models, a factorizable 5-dimensional space-time ${\cal M} = {\cal M}_4 \oplus {\cal M}_1$ is considered. In the former approach, the 5$^\text{th}$ dimension is flat, whereas in the latter two models, a negative 5D cosmological constant $\Lambda_5$ is considered and, as a consequence, the background metric is an anti-de Sitter space-time.

They will be briefly reviewed in Sections~\ref{sec:darkdimpheno}, \ref{sec:RS1pheno}, and~\ref{sec:RS2pheno}, respectively. 

\subsection{Dark Dimension: a 5-dimensional Large Extra-Dimensions model} 
\label{sec:darkdimpheno}
The LED model was proposed in Refs.~\cite{Antoniadis:1997zg, Arkani-Hamed:1998jmv} (see also Ref.~\cite{Antoniadis:1990ew}) as a way to solve the hierarchy problem via the compactification of $n$ extra-spatial dimensions. If all extra-dimensions are compactified on circles of common radius $r_1 = \dots = r_n = r_c$, the fundamental scale of gravity in $D = 4 + n$ dimensions, $M_{\rm D}$, and the Planck mass, $M_{\rm P}$, are related as: 
\begin{equation} \label{eq:MplanckLED}
    M_{\rm P}^2 = V_n \, M_{\rm D}^{2+n} = (2 \pi\, r_c)^n \, M_{\rm D}^{2 + n} \, ,
\end{equation}
where $M_{\rm P}$ is the reduced Planck mass, $M_{\rm P} = M_{\rm Planck}/\sqrt{8 \pi}$. To solve the hierarchy problem, the compactification radius $r_c$ has to be large enough (thus the name) to bring $M_{\rm D}$ down towards the electroweak scale. If we ask for $M_{\rm D}$ to be as low as $M_{\rm D} \sim 1$~TeV, it turns out that $n=1$ is forbidden, as the corresponding radius should be of astronomical size (and thus excluded by observations). On the other hand, for $n \geq 2$, a radius in the range of some tenths of a millimeter or less suffices. Bounds on the size of the compactification radius can be obtained in experiments looking for deviations from the $1/r^2$ Newton's law (see, {\em e.g.}, Refs.~\cite{Kapner:2006si, Adelberger:2009zz, Lee:2020zjt} and references therein). The current bounds give $r_c \leq 40$ $\mu$m for $n \in [1, 6]$. On the other hand, for $n \geq 2$, stringent bounds can be obtained in detailed studies of supernova explosions~\cite{Hannestad:2003yd}. For $n=2$, the limits range from $r_c < 0.96~\mu$m to $r_c < 1.6 \times 10^{-4}~\mu$m, depending on the specific data set and the models used. In a more recent work~\cite{Fermi-LAT:2012zxd}, the ultimate bound combining different data sets is somewhat less stringent, $r_c < 8.7 \times 10^{-3}~\mu$m, whilst still well below the present sensitivity of tests of Newton's law.

Bounds on $r_c$ are, in any case, strong enough to push the fundamental scale of gravity $M_{\rm D}$ beyond the reach of the LHC. This is one of the reasons the LED model has lost part of its appeal as a solution to the hierarchy problem. However, recent revival has been motivated within the framework of the so-called ``swampland conjectures''~\cite{Montero:2022prj}. Within this approach, it was suggested that another scale could be explained by the existence of one ``large'' extra-dimension, namely the 4D cosmological constant $\Lambda_4$.

\begin{figure}
	\centering
    \includegraphics[scale=0.49]{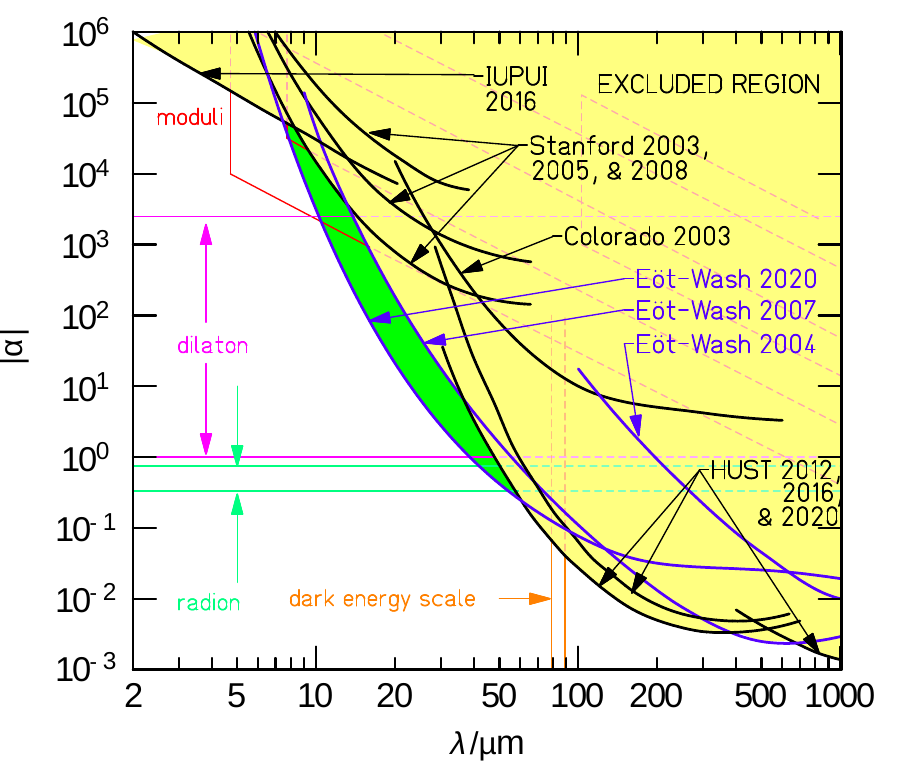}
    \caption{Bounds on the coupling $\alpha$ and distance $\lambda$ at which deviations from the $1/r^2$ Newton's law should be observed, once new physics is cast in terms of a Yukawa potential $V = - G_{\rm N} \, m/r \left [ 1 + \alpha \, \exp \left ( - r/\lambda \right )\right ] $, from Ref.~\cite{Lee:2020zjt}.}
    \label{fig:DDbounds}
\end{figure}

Explaining the theoretical motivation behind the possible existence of a fourth spatial dimension within the swampland paradigm is certainly beyond the scope of this paper. It suffices here to recall that, according to this proposal, the numerical value of the 4D cosmological constant $\Lambda_4$ can be related to the mass of a tower of unspecified Kaluza-Klein states, $m_{\rm KK} = {\cal O}(1/r_c)$, as follows: 
\begin{equation} \label{eq:lambda4_DD}
    \Lambda^2_4 = (g_{\rm DD} \, m_{\rm KK} )^4 = 10^{-122} \, 
    M_{\rm P}^4 \, .
\end{equation}
Note that for consistency with the rest of the paper we have defined the cosmological constant as a dimension-2 operator, $[\Lambda_4] = 2$. The coupling $g_{\rm DD}$ should be computed for a specific model using an effective field theory approach. The authors of Ref.~\cite{Montero:2022prj} consider as an example a 5D LED model in which gravity is the only field that propagates into the extra-dimension. Summing the graviton KK tower up to the fundamental scale of gravity $M_5$ in the computation of the vacuum energy should then give a calculable contribution to the Casimir energy~\cite{Appelquist:1982zs, Appelquist:1983vs}, such that the corresponding $g_{\rm DD} = {\cal O}(0.1)$. Combining this value for $g_{\rm DD}$ with the experimental value for $\Lambda_4$ gives that the required compactification radius is $r_c = 7.42~\mu$m, just below the experimental bound. In Fig.~\ref{fig:DDbounds}, we show the results of a relatively recent review~\cite{Lee:2020zjt} of bounds on the length $\lambda$ at which deviations from Newton's law should be observed and on the coupling $\alpha$ of a possible correction to Newtonian gravity, cast in terms of the Yukawa potential:
\begin{equation}
    V = - \frac{G_{\rm N} \, m}{r} \left [ 1 + \alpha \, e^{- r/\lambda }\right ] \ .
\end{equation}

More stringent bounds on deviations from the Newton's law are clearly needed to test this hypothesis. However, the present experimental techniques are strongly limited by electric backgrounds. Although it is possible that some improvement of the standard techniques may reach the microns regions, some new idea has been advanced to improve these bounds by eliminating most electric backgrounds (see Refs.~\cite{Donini:2016kgu, Baeza-Ballesteros:2021tha, Baeza-Ballesteros:2023par}). It is certainly intriguing that such an important quantity as $\Lambda_4$ could be explained by an extra-dimensional model ``within reach'' of experiments. However, it should be stressed that the computation of $g_{\rm DD}$ is model-dependent and, as has been very recently pointed out (see Refs.~\cite{Branchina:2023ogv, Branchina:2024ljd}), it is not free of ambiguities related to the correct definition of the cut-off scale, $M_5$. 

We will study here what might be the impact of the existence of one large (flat) extra-dimension with compactification radius in the micron range on cosmology. Since we are interested  in the case of $n=1$, {\em i.e.} extending the Minkowski spacetime with a single, flat, spatial extra-dimension, it is clear that we cannot solve the hierarchy problem. If we ask for $r_c$ to saturate the present bound on deviations from Newton's law, the corresponding fundamental scale of gravity will be $M_5 = {\cal O} (10^9)$~TeV (for such a large value of $M_5$ bounds from supernovae do not apply). The action of the model is:
\begin{equation}
    S_\text{DD}= - M_5^3\,\int d^{4}x \int_0^{2 \pi\, r_c} dy\, \sqrt{g^{(5)}} \, R^{\left(5\right)} +
    S_{\rm brane}
    \label{eq:DDaction}
\end{equation}
where $R^{\left(5\right)}$ is the 5D Ricci scalar, $g ^{(5)}$ is the determinant of the 5D metric $g^{(5)}_{MN}$, with signature $(+,-,-,-,-)$.\footnote{Note that here we are using the signature widely used in the realm of high-energy physics. On the other hand, the signature $(-,+,+,+,+)$ is more common in the literature on gravitation and cosmology.} The brane action is given by: 
\begin{equation}
    \label{eq:DDbraneterms}
    S_{\rm brane} = \int d^4 x \int_0^{2 \pi\, r_c} dy \sqrt{-g_i^{(4)}} \delta (y) \left \{ - \sigma_0 + \dots \right \} \, , 
\end{equation}
where $\dots$ refers to fields that can be localized in the brane (such as, for example, SM fields). The determinant of the induced metric is $-g_i^{(4)} = g^{(5)}/g_{55} = 1$, since $g_{55} = -1$. Notice that in this model, the tension of the brane is a free parameter, since it is not fixed by periodic boundary conditions $g^{(5)}_{MN}(x,y) = g^{(5)}_{MN}(x,y + 2 \pi R)$. This is an important difference with respect to warped models that are compactified in an orbifold and, in addition to $(2 \pi\, r_c)$-periodicity, assume reflectivity with respect to the singular points of the orbifold $y = 0$ and $y = \pi\, r_c$.  

Notice that a different action has been considered in Refs.~\cite{Anchordoqui:2022svl, Antoniadis:2023sya}, where a positive 5D cosmological constant was added to the action, and a bulk inflaton field is taken into account. In the case of a positive 5D cosmological constant $\Lambda_5$, the background metric would be de Sitter. In this case, however, we restrict ourselves to the case of a flat extra-dimension with no field propagating into the bulk, except for the graviton. 

\subsection{Randall-Sundrum two-branes models (aka RS1)}
\label{sec:RS1pheno}
Using conventions inspired by (but not equal to) Refs.~\cite{Randall:1999ee, Csaki:2004ay}, the bulk action of the RS1 model is given by:
\begin{equation} \label{eq:RSaction}
    S_\text{RS1}= - M_5^3 \,\int d^{4}x \int_0^{\pi\, r_c} dy\, \sqrt{g^{(5)}} \, \left[ R^{\left(5\right)} + 2 \, \Lambda_5\right], 
\end{equation}
where $R^{\left(5\right)}$ is the 5D Ricci scalar, $g ^{(5)}$ is the determinant of the 5D metric $g^{(5)}_{MN}$, with signature $(+,-,-,-,-)$, and $\Lambda_5$ is the 5D cosmological constant. The scale $M_5$ is the fundamental scale of gravity in 5D. The extra-dimension is assumed to be compactified in an orbifold of radius $r_c$, with fixed points at $y = 0$ and $y = \pi\, r_c$, {\em i.e.} periodic boundary conditions are implied for $y = 0$ and $y = 2 \pi\, r_c$, together with reflectivity of the metric in the intervals $]- \pi\, r_c, 0[$ and $]0, \pi\, r_c[$. Note that both the dimension of the Ricci scalar and of the cosmological constant is $[R^{(5)}] = [\Lambda_5] = 2$, according to this assumption for the action. Solving the Einstein equations with the ansatz: 
\begin{equation} \label{eq:metricRS}
    ds^2 = e^{- 2 k y} \, \eta_{\mu \nu} dx^\mu dx^\nu - dy^2 \, ,
\end{equation}
we obtain for the curvature $k$ along the 5$^\text{th}$-dimension the following relation with the 5D cosmological constant $\Lambda_5$
\begin{equation} \label{k curvature RS1}
    k=\sqrt{-\frac{\Lambda_5}{6}} \, ,
\end{equation}
from which we see that this scenario has the bulk geometry of an anti-de Sitter 5D spacetime, with negative cosmological constant $\Lambda_5$. The relation between the Planck mass in 4D and the fundamental scale in 5D, $M_5$, is:
\begin{equation} \label{eq:MplanckRS}
    M_{\rm P}^2 = \frac{M_5^3}{k} \, \left (1 -e^{-2 \pi k r_c} \right ).
\end{equation}

The phenomenological assumption of the RS1 model is that {\it all} energy scales are ${\cal O} (M_{\rm P})$, {\em i.e.} $k \sim M_5 \sim M_{\rm P}$.\footnote{This is strictly true only in the so-called ``$\pi$-frame'', see Appendix~C of Ref.~\cite{Giudice:2016yja}.} On the other hand, the scale of the gravitational interactions between massive Kaluza-Klein (KK) modes and standard matter localized at $y= \pi\, r_c$ is: 
\begin{equation} \label{eq:lambdacoupling}
    \Lambda = M_{\rm P} \, e^{- \pi\, k\, r_c} = {\cal O} (1 \; \rm TeV) \, .
\end{equation}
Therefore, to solve the hierarchy problem, we need $k\, r_c \sim {\cal O} (10)$. Within this hypothesis, we would get the lightest (massive) KK graviton with mass
\begin{equation} \label{eq:kkgrav1}
    m_1 = k\, x_1 e^{- \pi\, k\, r_c} = {\cal O} (1 \; \rm TeV)
\end{equation}
(where $x_1$ is the first zero of the Bessel function $J_1$). Notice that the present LHC bounds exclude the possibility that both $\Lambda$ and $m_1$ are as low as 1~TeV. Typical limits, using searches for resonant KK-graviton production with decay into photons or leptons at the LHC Run~II~\cite{ATLAS:2017ayi, ATLAS:2017fih, CMS:2018dqv, ATLAS:2019erb}, state that $m_1 \ge 4$~TeV for $\Lambda \sim 10$~TeV and $\Lambda \ge 50$~TeV for $m_1 \sim 1$~TeV, approximately. 

\begin{figure}
    \centering
    \includegraphics[scale=0.49]{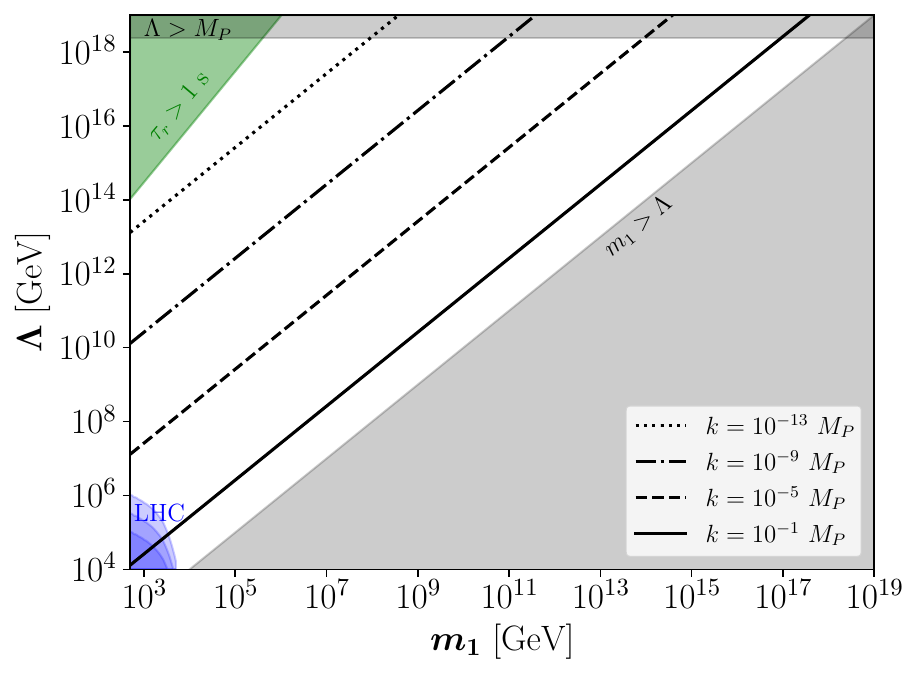}
    \includegraphics[scale=0.49]{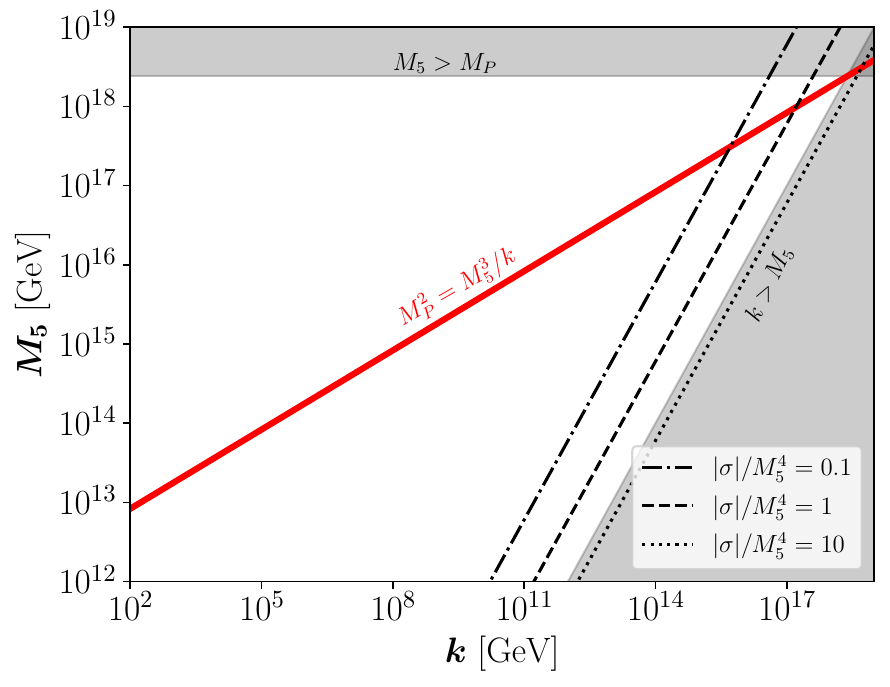}
    \caption{Left panel: The allowed parameter space in the RS1 model. A ratio $m_1/m_r = 10$ was assumed. The blue region is the current exclusion bound at the LHC Run II. The green-shaded region in the upper left corner is disfavored because the radion lifetime is too long. The gray-shaded region in the lower right corner is excluded, as the effective theory is not reliable for particles whose mass is larger than the effective scale of the model.  
    Right panel: The allowed parameter space in the RS2 model. The two gray regions reflect our theoretical prejudice that the hierarchy
    $k \leq M_5 \leq M_{\rm P}$ should be satisfied. In the plot, 
    the red solid line shows the values of $(k, M_5)$, which lead to the observed 4D $M_\text{P}$. We also show three lines for 
    the $\sigma_0/M_5^4$ ratio as a function of $k$. The dot-dashed, 
    dashed and dotted lines co correspond to $\sigma_0/M_5^4 = 0.1, 1$ and $10$, respectively. These values will be useful in Section~\ref{sect:inflapot}, as in the region to the left of the line $\sigma_0/M_5^4 = 0.1$ the  effect from extra-dimensional physics is dominant, according to Eq.~\eqref{Hubble final}. On the other hand, to the right of the line $\sigma_0/M_5^4 = 10$ the extra-dimensional corrections to the
    4D scenario are negligible. 
    }
    \label{fig:sigma}
\end{figure}
To obtain a stable anti-de Sitter background metric on the segment $y \in [0,\pi\, r_c]$ it is mandatory to introduce additional terms in the action localized at the fixed points of the orbifold, $y = 0$ and $y = \pi\, r_c$:
\begin{equation} \label{eq:RSbraneterms RS1}
    S_{\rm brane} = \sum_{i = \rm UV, \rm IR} \int d^4 x \int_0^{\pi\, r_c} dy \sqrt{-g_i^{(4)}} \delta (y - y_i) \left \{ - \sigma_i + \dots \right \} \, , 
\end{equation}
where $\dots$ refers to fields that can be localized in either the UV or IR-brane and that, in the RS1 model, are localized in $y_i = 0, \pi\, r_c$, respectively. The determinant of the induced metric, $-g_i^{(4)} = g^{(5)}/g_{55}$, in both RS scenarios is only $g^{(5)}(x,y)$ computed at the brane locations, since $g_{55} = -1$. The values of the brane tensions $\sigma_i$ must be fine-tuned to properly glue the metric in the intervals $y \in ]-\pi\, r_c, 0[$ and $]0, \pi\, r_c[$: 
\begin{equation} \label{eq:branetensionfinetuningRS}
    \sigma_{\rm IR}= - \sigma_{\rm UV} = 6 \,M_5^3\,k \, .
\end{equation}
Notice that, since {\it all} scales in the RS1 model ${\cal O} (M_{\rm P})$ are assumed, the brane tensions are also of this order: $\sigma_i = {\cal O} (M_{\rm P}^4)$. The parameter space allowed for the RS1 model is shown in Fig.~\ref{fig:sigma} (left panel). The tiny regions in the lower left corner of the picture depict present- and future experimental bounds coming from resonance searches at the LHC. The proton-proton collision can generate resonant KK-gravitons that later decay into SM particles. ATLAS and CMS put bounds on these processes in the $\gamma\gamma$ and lepton-lepton channels as a function of the mass of the resonance (the lightest KK-graviton). These limits can be translated into limits on $\Lambda$ as a function of the mass of the first graviton $m_1$. The present bounds (dark blue) come from the resonant searches at LHC with 36~fb$^{-1}$~\cite{ATLAS:2017fih, ATLAS:2017ayi}, whereas future bounds are estimated assuming 300~fb$^{-1}$ (medium blue) and 3000~fb$^{-1}$ (light blue) for LHC Run-III and High-Luminosity LHC, respectively. Notice that the relevant regions for cosmology cover much larger values of $\Lambda$ and $m_1$ than those typically required to solve the hierarchy problem. The green-shaded region in the upper left corner is disfavored because the radion lifetime is too long, while the gray-shaded region in the lower right corner is where the effective theory is not reliable, as the KK-gravitons are heavier than the effective scale of the theory $\Lambda$, and should have been integrated out.

\subsection{Randall-Sundrum one-brane model (aka RS2)} 
\label{sec:RS2pheno}
The second RS model, conventionally called RS2 although it refers to one single brane at $y = 0$, has the same action as the RS1 (see Refs.~\cite{Randall:1999ee, Csaki:2004ay}) with $r_c \to \infty$:
\begin{equation} \label{eq:RSactionRS2}
    S_\text{RS2}= - M_5^3\,\int d^{4}x \int_0^\infty dy\, \sqrt{g^{(5)}} \, \left[ R^{\left(5\right)} + 2 \, \Lambda_5\right] \, .
\end{equation}
The curvature $k$ along the 5$^\text{th}$-dimension is related to the 5D cosmological constant as in the RS1 setup
\begin{equation} \label{k curvature RS2}
    k=\sqrt{-\frac{\Lambda_5}{6}}\,;
\end{equation}
also in this case the bulk geometry is an anti-de Sitter 5D spacetime. Notice that the one-brane setup is the one most widely adopted when studying the cosmology of braneworld scenarios. In this case, the relation between the Planck mass in 4D and the fundamental scale in 5D, $M_5$, is just $ M_{\rm P}^2 = M_5^3/k$. The KK-tower present in the RS1 model, whose discretization arises due to the boundary at $y = \pi\, r_c$, becomes a continuum spectrum of extra-dimensional modes, and it is not possible to derive individual KK-graviton modes coupling with SM fields. In this model, the hierarchy problem is not addressed, even though an effective 4D gravity can be felt nearby the brane location. In a sense, the curvature in the extra-dimension is so large that the gravitational field is confined on the 4D brane at energies low enough. Since the SM fields are also located at the brane in $y = 0$, no rescaling of the fields is needed, and all fundamental scales are ${\cal O} (M_{\rm P})$.

In order to obtain a stable anti-de Sitter background metric on the segment $y \in [0, \infty [$, it is mandatory to introduce a brane-localized term at $y = 0$, though:
\begin{equation} \label{eq:RSbraneterms RS2}
    S_{\rm brane} = \int d^4 x \int_0^\infty dy \sqrt{-g_i^{(4)}} \delta (y) \left \{ - \sigma_0 + \dots \right \},
\end{equation}
where $\dots$ refers to fields that can be localized in the brane ({\em e.g.}, the SM fields). The value of the brane tension $\sigma_0$ must again be fine-tuned to properly glue the metric to the left and to the right of $y = 0$:
\begin{equation} \label{eq:branetensionfinetuningRS2}
    \sigma_0 = 6 \,M_5^3\,k \, .
\end{equation}
Notice that in the RS2 model, the brane at $y=0$ has positive tension (differently from the case of RS1). We will see in Section~\ref{sect:inflapot} that this is crucial in order to have viable inflationary models in RS2. 

In this model, we cannot constrain the parameter space using low-energy experimental data. However, we can still show the region of the parameter space that is of interest in the RS2 model. This is depicted in Fig.~\ref{fig:sigma} (right panel), where we show two theoretical constraints: the upper gray band shows the region that is excluded if we ask $M_5$ to be smaller than $M_{\rm P}$; the gray triangle on the bottom right corner is the region excluded if we ask $k$ to be smaller than $M_5$. We also show the relation that $k$ and $M_5$ must satisfy in order to obtain the observed value of $M_{\rm P}$. This is depicted as the red solid line. All physical points are on this line. Eventually, it is useful to plot the typical range of values assumed by the tension of the brane $\sigma_0$ (something that will be handy in Section~\ref{sect:inflapot}). We therefore draw the values of $\sigma_0$ in the RS2 model in the $(k, M_5)$ plane, in units of $M_5^4$. The black lines correspond to $\sigma_0/M_5^4 = 0.1$ (dashed dotted), $1$ (dashed), and $10$ (dotted). The region to the left of the dot-dashed line corresponds, roughly speaking, to the region where non-standard cosmology dominates (as $V = {\cal O} (M_5^4) \gg \sigma_0$), whereas the region to the right of the dotted line is ruled by standard cosmology (as $V \ll \sigma_0$).

\section{Friedmann's equation and Hubble parameter from 5D Einstein's equations} 
\label{sec:Friedmann}
To study the phenomenology of brane inflation, we need to derive the Friedmann equations for the model at hand. After a short introduction to the Einstein equations in 5-dimensions, we particularize to the three setups we want to study: the flat Dark Dimension (DD) scenario in Section~\ref{sec:DarkDimcosmo}, the one-brane Randall-Sundrum model (conventionally called RS2) in Section~\ref{sec:RS2cosmo}, and the two-branes Randall-Sundrum model (conventionally called RS1) in Section~\ref{sec:RS1cosmo}.

Let us start by considering the following factorizable 5D metric (in the sense that ${\cal M}_5 = {\cal M}_4 \times {\cal M}_1$):
\begin{equation} \label{eq:metric1}
    ds^2 = g_{\mu\nu}(x, y)\, dx^\mu\, dx^\nu - dy^2 \, ,
\end{equation}
where $g_{\mu\nu}$ is the 4D metric, with signature $(+,-,-,-,-)$. Focusing on the cosmology of the infrared brane, where the SM fields are confined, we can follow Ref.~\cite{Langlois:2002bb} and choose a local coordinate system such that homogeneity and isotropy in the three standard spatial dimensions are understood:
\begin{equation} \label{eq:metric2}
    ds^2 = g_{ab}(z)\, dz^a\, dz^b - a^2(z)\, \delta_{ij}\, dx^i\, dx^j\, ,
\end{equation}
with $z = (t,y)$. The origin of this coordinate system is fixed at the position of the would-be UV-brane of the RS1 model, $y_{\rm UV} = 0$. Notice that in RS2 and DD there is only one brane, and in both models we still consider the origin of the coordinate system at the location of the brane. For the interested reader, in Ref.~\cite{Langlois:2000ns} a different coordinate system was used to describe the dynamics of a two-branes model from the point of view of a bulk observer. 

We can then diagonalize the two-dimensional metric so that $g_{ab}(z) = {\rm diag} \left (n^2(t,y),-1\right)$. Within this ansatz, we have: 
\begin{equation} \label{eq:metricRS2}
    ds^2 = n^2(t,y)\, dt^2 - a^2(t,y)\, \delta_{ij}\, dx^i\, dx^j - dy^2,
\end{equation}
where $t$ is the proper time on the brane, and the coordinate $y$ is the proper distance along the space-like geodesic (locally) normal to the brane.\footnote{Notice that in the literature a different normalization of the 5$^\text{th}$-coordinate $y$ can be found: in Ref.~\cite{Lesgourgues:2000tj} the size of the extra-dimension is normalized to $b_0$ and $y \in [0,1]$.} The time coordinate defined only on the brane can be propagated off the brane along these normal geodesics. 

The action used to derive the cosmological evolution of the extra-dimensional space-time is (see Appendix~\ref{sec:EEqand5D}): 
\begin{equation} \label{eq:RSaction2}
    S_\text{5D}=  - \frac{1}{\kappa_5^2} \, \int d^{4}x\int_0^{\pi\, r_c} dy\, \sqrt{g^{(5)}} \, \left[R^{\left(5\right)} + 2 \, \left ( \frac{D}{2} - 1 \right ) \,  \Lambda_{5}\right],
\end{equation}
where $\kappa^2_5 = 8 \pi G^{(5)}_{\rm N}/c^4 = \left ( D/2 -1\right )/\left(M_5^{\rm RS}\right)^3$, with $G_N^{(5)}$ the 5D Newton constant. Notice that this action can be used for the three models at hand, with the only difference that $\Lambda_5 = 0$ in the DD case, whereas is non-zero (and negative) in the one-brane and two-branes RS cases.

From this action, we get the following Einstein equations in 5D:
\begin{equation} \label{eq:Einstein}
    G_{AB} + \Lambda_{\rm RS} \, g_{AB} = \kappa_5^2\, T_{AB} \, ,
\end{equation}
where $G_{AB}$ is the Einstein tensor, $G_{AB} = R_{AB} - \frac{1}{2} g^{(5)}_{AB} R$. The energy-momentum tensors $T_{AB}$ on the branes are given by: 
\begin{equation} \label{eq:Tbrane0}
    [T^{A}_{\; \; B}]_0 = S^{A}_{\; \; B} \, \delta (y) = \delta (y ) \, {\rm diag} \left ( \sigma_0 + \rho_0, - p_0, - p_0, - p_0, 0 \right ), 
\end{equation}
and 
\begin{equation} \label{eq:Tbranepi}
    [T^{A}_{\; \; B}]_\pi = S^{A}_{\; \; B} \, \delta (y - \pi\, r_c) = \delta (y - \pi\, r_c) \, {\rm diag} \left ( \sigma_\pi + \rho_\pi, - p_\pi, - p_\pi, - p_\pi, 0 \right ), 
\end{equation}
where $\sigma_0, \sigma_\pi$ are the brane tensions for the branes located at $y = 0$ (the only brane in the DD and RS2 models, the UV-brane in the RS1 model) and $y = \pi\, r_c$ (the IR-brane in the RS1 model), respectively. The two brane tensions $\sigma_0$ and $\sigma_\pi$ are fine-tuned to the specific values defined in Sections~\ref{sec:RS1pheno} and~\ref{sec:RS2pheno}. On the other hand, in the DD model, the brane tension is a ``free'' parameter of the theory (we will come back on this statement further in this Section). $\rho_0, \rho_\pi$ and $p_0, p_\pi$ are the matter densities and pressures on the two branes, respectively. Notice that we are thus assuming an empty bulk and that no energy flows out of the brane ({\em i.e.}, $T^0_y = T^i_y = 0$). 

The Einstein tensor components for the action above are~\cite{Langlois:2002bb}: 
\begin{align}
    G_{00}/3 & =\frac{\dot{a}^2}{a^2}-n^2\left(\frac{a^{\prime\prime}}{a}+\frac{a^{\prime2}}{a^2}\right)+k_\text{curv}\frac{n^2}{a^2}\,,\label{G00 RS}\\
    G_{ii} & =\delta_{ii}\,\left[ a^2\left(2\frac{a^{\prime\prime}}{a}+\frac{n^{\prime\prime}}{n}+\frac{a^{\prime2}}{a^2}+2\frac{a^{\prime}}{a}\frac{n^{\prime}}{n}\right)+\frac{a^2}{n^2}\left(-2\frac{\ddot{a}}{a}-\frac{\dot{a}^2}{a^2}+2\frac{\dot{a}}{a}\frac{\dot{n}}{n}\right)-k_\text{curv}\,\right],\label{Gii RS}\\
    G_{0y}/3 & =\frac{n^{\prime}}{n}\frac{\dot{a}}{a}-\frac{\dot{a}^{\prime}}{a}\,,\label{G0y RS}\\
    G_{yy}/3 & =\left(\frac{a^{\prime2}}{a^2}+\frac{a^{\prime}}{a}\frac{n^{\prime}}{n}\right)-\frac{1}{n^2}\left(\frac{\ddot{a}}{a}+\frac{\dot{a}^2}{a^2}-\frac{\dot{a}}{a}\frac{\dot{n}}{n}\right)-\frac{k_\text{curv}}{a^2}\,,\label{Gyy RS}
\end{align}
where the dot ($\dot{}$) represents time derivatives and the prime ($'$) represents derivatives along the extra-dimension. $k_\text{curv}$ is the three-dimensional spatial curvature, with $k_\text{curv} = (-1,0,+1)$.

\subsection{Dark dimension cosmology} 
\label{sec:DarkDimcosmo}
Using Eq.~\eqref{eq:Einstein} and Eqs.~\eqref{G00 RS} to~\eqref{Gyy RS}, the Einstein equations in the absence of a (tree-level) 5D cosmological constant are:
\begin{align}
    &\frac{\dot{a}^2}{a^2}-n^2\left(\frac{a^{\prime\prime}}{a}+\frac{a^{\prime2}}{a^2}\right)+k_\text{curv}\frac{n^2}{a^2}= -\frac{\kappa_{5}^2}{3}\,\rho_{b}\,
    \delta\left(y\right) \, , \label{G00 Einstein DD}\\
    &a^2\left(2\frac{a^{\prime\prime}}{a}+\frac{n^{\prime\prime}}{n}+\frac{a^{\prime2}}{a^2}+2\frac{a^{\prime}}{a}\frac{n^{\prime}}{n}\right)+\frac{a^2}{n^2}\left(2\frac{\dot{a}}{a}\frac{\dot{n}}{n}-2\frac{\ddot{a}}{a}-\frac{\dot{a}^2}{a^2}\right)-k_\text{curv} 
    =\frac{\kappa_{5}^2}{3}\,p_{b}\,
    \delta(y), \label{Gii Einstein DD}\\
    &\frac{n^{\prime}}{n}\frac{\dot{a}}{a}-\frac{\dot{a}^{\prime}}{a}=0\,, \label{G0y Einstein DD}\\
    &\left(\frac{a^{\prime2}}{a^2}+\frac{a^{\prime}}{a}\frac{n^{\prime}}{n}\right)-\frac{1}{n^2}\left(\frac{\ddot{a}}{a}+\frac{\dot{a}^2}{a^2}-\frac{\dot{a}}{a}\frac{\dot{n}}{n}\right)-\frac{k_{\text{curv}}}{a^2} = 0 . 
    \label{Gyy Einstein DD}
\end{align}
Notice that the signature change with respect to the standard literature on braneworld cosmology (see Refs.~\cite{Langlois:2000ns, Langlois:2002bb}) does not modify the Einstein equations, as it should be. 

Since Eq.~\eqref{eq:Tbrane0} implies that there is no flow of energy into the bulk, Eq.~\eqref{G0y RS} vanishes and we get that:
\begin{equation}
    \frac{d}{dy}\left(\frac{\dot{a}}{n}\right)=0 \, .
\end{equation}
Therefore, we can write $\dot a (t,y)$ as the following product of functions:
\begin{equation} \label{eq:ansatz}
    \begin{dcases}
        \dot a (t,y) &= \nu(t) \, n(t,y) \, ,\\
        \ddot a(t,y) &= \dot \nu(t) \, n(t,y) + \nu(t) \, \dot n(t,y) \,.
    \end{dcases}
\end{equation}
Inserting the above ansatz into Eq.~\eqref{G00 Einstein DD}, we immediately get: 
\begin{equation}
    \frac{a^{\prime \prime}}{a} + \left ( \frac{a^\prime}{a} \right )^2
    = \frac{\nu^2 + k_{\text{curv}}}{a^2} \, .
\end{equation}
In this way, we get the following expression: 
\begin{equation}
    a^2(t,y) = \left [\nu^2(t) + k_{\text{curv}} \right ] \, y^2 + C_1(t) \, y + C_2(t) \, .
\end{equation}
Solving the Einstein equations in the bulk is not sufficient to completely determine the geometry of the space-time, as the boundary conditions at the locations of branes must be considered. This makes a difference between the three solutions we are interested in, as we will see  in the following subsections.

We first introduce the so-called {\em junction conditions}, which allow one to glue together piecewise solutions for $y < 0$ and $y > 0$. In practice, junction conditions are obtained by integrating $a^{\prime\prime}$ over $y$ in Eqs.~\eqref{G00 Einstein DD} and~\eqref{Gii Einstein DD}, which gives the required discontinuities of the first derivative of $a$ and $n$ at the location of the IR-brane:
\begin{align} \label{eq:junctionDD}
    \left . \left ( \frac{a^\prime}{a} \right ) \right |_{y \to\, 0^+} & = - \frac{\kappa^2_5}{6} (\sigma_0 +  \rho_0) \,,\\
    \left . \left ( \frac{n^\prime}{n} \right ) \right |_{y \to\, 0^+} & =  \frac{\kappa^2_5}{6} \left [ 3 p_0 + 2 (\sigma_0 + \rho_0) \right ] \, , \\
     \left . \left ( \frac{a^\prime}{a} \right ) \right |_{y \to\, (2 \pi\, r_c)^-} & = + \frac{\kappa^2_5}{6} (\sigma_0 +  \rho_0) \,,\\
    \left . \left ( \frac{n^\prime}{n} \right ) \right |_{y \to\, (2 \pi\, r_c)^-} & = - \frac{\kappa^2_5}{6} \left [ 3 p_0 + 2 (\sigma_0 + \rho_0) \right ] \, ,
\end{align}
where, for simplicity, we treat the discontinuity at $2 \pi\, r_c$ as approaching a second brane from the left. Notice that the matter density on the brane, $\rho_b = \sigma_0 + \rho_0$, is the same at $y = 0$ and $y = 2 \pi\, r_c$. This will differ in the case of the Randall-Sundrum two-branes setup (see Section~\ref{sec:RS1cosmo}). After some algebra, we get:
\begin{equation} \label{eq:a2flatcase}
    a^2(t,y) = \left [ \nu^2(t) + k_{\text{curv}} \right ] \left [ y \, (y - 2 \pi\, r_c ) + \frac{\pi\, r_c}{\alpha_{\rm DD} \eta}\right ] \, , \qquad ({\rm for} \; y \in [0, 2 \pi\, r_c])
\end{equation}
where we have defined
\begin{equation} \label{alphaDD}
    \alpha_{\rm DD}  \equiv  \frac{\kappa_5^2}{6} \, \sigma_0 = \frac{\sigma_0}{6 \, M_5^3}\, , 
\end{equation}
and 
\begin{equation}
    \eta \equiv 1 + \frac{\rho_0}{\sigma_0} \, .
\end{equation}
Notice that $\eta$ could, in principle, depend on time. However, if we forbid leakage into the extra-dimension (as we can see from our choice of the $T^0_y$ and $T^i_y$ components of the energy-momentum tensor), then $\dot \eta = 0$. 

In order to derive $n(t,y)$, we may use the relation: 
\begin{equation}
    n(t,y) = \frac{\dot a(t,y)}{\nu (t)} = \frac{\dot \nu(t)}{\sqrt{\nu^2(t) + k_{\rm curv}}} \, \left [ y \, (y - 2 \pi\, r_c ) + \frac{\pi\, r_c}{\alpha_{\rm DD} \eta}\right ]^{1/2}.
\end{equation}
If we impose the proper time on our brane as $n(t,0) = 1$, we get:
\begin{equation} \label{eq:nufunction}
    n(t,0) = 1 = \sqrt{\frac{\pi\, r_c}{ \alpha_{\rm DD} \, \eta}} \, \frac{\dot \nu (t)}{\sqrt{\nu^2(t) + k_{\rm curv}}} \, ,
\end{equation}
from which we derive:
\begin{equation} \label{nDD}
    n(t,y) = \sqrt{\frac{\alpha_{\rm DD} \, \eta}{\pi\, r_c}} \, 
    \left [ y (y - 2 \pi\, r_c) + \frac{\pi\, r_c}{\alpha_{\rm DD} \, \eta} \right]^{1/2} \, . 
\end{equation}
By solving Eq.~\eqref{eq:nufunction}, we get:
\begin{equation}
    \nu (t) = \frac{1}{2} \left [ \exp \left ( \sqrt{\frac{\alpha_{\rm DD} \, \eta}{\pi \, r_c}} \, t +C_0 \right ) - k_{\rm curv} \exp \left ( -\sqrt{\frac{\alpha_{\rm DD} \, \eta}{\pi \, r_c}} \, t - C_0 \right ) \right ],
\end{equation}
where $\nu(0) = 1/2 \left ( e^{C_0} - k_{\rm curv} e^{- C_0}\right ) $ should be fixed by the initial conditions. The boundary condition for $n(t,y)$ in $y = 0$ and $y = 2 \pi\, r_c$ establishes the equation of state that relates $\rho_0$ and pressure $p_0$, and is irrelevant here. 

Using the solution for $a^2(t,y)$, we can derive the Friedmann equation. Consider Eq.~\eqref{G00 Einstein DD} and compute it in the bulk (that is, for $y \neq 0$):
\begin{equation}
    \frac{\nu^2}{a^2} - \left ( \frac{a^{\prime \prime}}{a} + \frac{a^{\prime 2}}{a^2} \right ) + \frac{k_\text{curv}}{a^2} = 0 \,.
\end{equation}
This can be trivially written as:
\begin{equation}
    \frac{d}{dy} \left ( a \, a^\prime \right )  = \nu^2(t) + k_\text{curv} \, .
\end{equation}
Performing the following change of variables
\begin{equation}
    x(t,y) = a^2 (t,y),
\end{equation}
we obtain
\begin{equation}
    \frac{d}{dy} = \frac{dx}{dy} \, \frac{d}{dx} = 2 a \, a^\prime \frac{d}{d a^2}
\end{equation}
for fixed $t$, meaning that
\begin{equation}
    \frac{d}{da^2} (a \, a^\prime)^2 = (\nu^2 + k_\text{curv}) \, .
\end{equation}
Thus, integrating over $a^2$, we are left with:
\begin{equation} \label{eq:Langloismethodfora2}
    \left ( a \, a^\prime \right )^2 +  {\cal C} = \left ( \nu^2 + k_\text{curv} \right ) a^2  \, ,
\end{equation}
with ${\cal C}$ an integration constant in $y$. Using the junction condition for $y \to 0^+$, we obtain for the Hubble parameter:
\begin{equation} \label{eq:FriedmannFlat}
    \left . \frac{\nu^2 (t)}{a^2(t,y)} \right |_{y = 0^+} = \left ( \frac{\dot a_0 (t)}{a_0 (t)}\right )^2 = H^2(t) = \frac{\kappa_{5}^4}{36} \, \rho_b^2  - \frac{k_\text{curv}}{a^2_0 (t)}  + \frac{{\cal C}(t)}{a^4_0 (t)}
\end{equation}
with $a_0 (t) = a(t,0)$. The function ${\cal C}(t)$ can be determined by solving in $y$ Eq.~\eqref{eq:Langloismethodfora2}, and then equating the result with Eq.~\eqref{eq:a2flatcase}. We get
\begin{equation}
    {\cal C}(t) = (\pi\, r_c)^2 \, \left [\nu^2(t) + k_{\rm curv}  \right]^2 \, 
    \left ( \frac{1}{\pi\, r_c \, \alpha_{\rm DD} \eta} - 1\right ) \, .
\end{equation}
It is clear from this that ${\cal C}$ can only vanish if $\alpha_{\rm DD} \, \eta$ is fine-tuned to $1/(\pi\, r_c)$. For this very specific case, ${\cal C}$ vanishes and we get a Friedmann equation quadratic in $\rho_0$. For ${\cal C} \neq 0$, though, by inserting it into Eq.~\eqref{eq:FriedmannFlat}, the quadratic term cancels exactly and we get a (4D-like) linear relation between the Hubble parameter and the energy density on the brane for the particular case $k_{\rm curv} = 0$: 
\begin{equation} \label{eq:FriedmannDD}
    H^2 = \frac{\alpha_{\rm DD}}{\pi\, r_c} \, \left (1 + \frac{\rho_0}{\sigma_0} \right ) = \frac{\sigma_0}{3 M_{\rm P}^2} + \frac{\rho_0}{3 M_{\rm P}^2} \, ,
\end{equation}
where we have used the LED relation $M_{\rm P}^2 = (2 \pi\, r_c) \, M_5^3$. 

If we assume that the brane tension cannot introduce a new fine tuning problem for the 4D cosmological constant induced by bulk effects (following the ``swampland conjectures'' inspired approach described in Section~\ref{sec:darkdimpheno}), we have: 
\begin{equation}
\frac{\sigma_0}{3 M_{\rm P}^2} \lesssim \frac{\Lambda_{4}}{3} = \frac{1}{3} \, \left (\frac{\lambda}{r_c} \right )^2 \ , 
\end{equation}
and it is determined  by the experimental value, Eq.~\eqref{eq:lambda4_DD}, from which:
\begin{equation} \label{eq:sigma0expDD}
    \sigma_0 \lesssim M_{\rm P}^2 \, \left ( \frac{\lambda}{r_c}\right)^2 = 10^{-61} \, M_{\rm P}^4 = 1.6 \times 10^{12} \;{\rm GeV}^4 \, ,
\end{equation}
where we used the relation between the fundamental scale $M_5$, the Planck mass and the compactification radius in Eq.~\eqref{eq:MplanckLED} for $n=1$. Therefore, even though $\sigma_0$ is a free parameter of the LED model, in the DD approach it is indeed quite constrained, as it should be at most equal to $\Lambda_4$. In fact, for any other choice of $\sigma_0$, $\Lambda_4$ would differ from its experimentally observed value.

\subsection{RS2 setup: anti-de Sitter with one brane} 
\label{sec:RS2cosmo}
The Einstein equations differ from the previous case in that there is a 5D cosmological constant term:
\begin{align}
    &\frac{\dot{a}^2}{a^2}-n^2\left(\frac{a^{\prime\prime}}{a}+\frac{a^{\prime2}}{a^2}\right)+k_\text{curv}\frac{n^2}{a^2}=\frac{\Lambda_{\rm RS}}{3}\,n^2-\frac{\kappa_{5}^2}{3}\,\rho_{b}\,
    \delta\left(y\right) \, , \label{G00 Einstein RS}\\
    &a^2\left(2\frac{a^{\prime\prime}}{a}+\frac{n^{\prime\prime}}{n}+\frac{a^{\prime2}}{a^2}+2\frac{a^{\prime}}{a}\frac{n^{\prime}}{n}\right)+\frac{a^2}{n^2}\left(2\frac{\dot{a}}{a}\frac{\dot{n}}{n}-2\frac{\ddot{a}}{a}-\frac{\dot{a}^2}{a^2}\right)-k_\text{curv} \nonumber\\
    &\qquad\qquad\qquad\qquad =- \frac{\Lambda_{\rm RS}}{3}\,a^2+\frac{\kappa_{5}^2}{3}\,p_{b}\,
    \delta(y),\\
    &\frac{n^{\prime}}{n}\frac{\dot{a}}{a}-\frac{\dot{a}^{\prime}}{a}=0\,, \\
    &\left(\frac{a^{\prime2}}{a^2}+\frac{a^{\prime}}{a}\frac{n^{\prime}}{n}\right)-\frac{1}{n^2}\left(\frac{\ddot{a}}{a}+\frac{\dot{a}^2}{a^2}-\frac{\dot{a}}{a}\frac{\dot{n}}{n}\right)-\frac{k_{\text{curv}}}{a^2} = - \frac{\Lambda_{\rm RS}}{3}. \label{Gyy Einstein RS}
\end{align}

The junction conditions are the same as in the DD case, but due to the orbifold condition $y \leftrightarrow - y$, we apply them at $y = 0$, only:
\begin{align} \label{eq:junction}
    \left . \left ( \frac{a^\prime}{a} \right ) \right |_{y \to\, 0} & = - \frac{\kappa^2_5}{6} \rho_b\,,\\
    \left . \left ( \frac{n^\prime}{n} \right ) \right |_{y \to\, 0} & =  \frac{\kappa^2_5}{6} \left ( 3 p_b + 2 \rho_b \right ) \, ,
\end{align}
where $\rho_b = \sigma_0 + \rho_0$. Notice that the Einstein equations above can be explicitly solved, getting:
\begin{equation}
    a^2(t,y) = C_0 (t) + C_1(t) \, \cosh 2 k y + C_2 (t) \, \sinh 2 k |y| \, ,
\end{equation}
where the coefficients can be computed using the boundary conditions at $y = 0$ (see Ref.~\cite{Langlois:2002bb}). From this equation, an expression for $n(t,y)$ can be derived using the relation $\dot a(t,y) = \nu(t) \, n(t,y)$ and the condition $n(t,0) = 1$.

Considering Eq.~\eqref{G00 Einstein RS} and computing it in the bulk, (that is, for $y \neq 0$), one gets 
\begin{equation}
    \frac{\nu^2}{a^2} - \left ( \frac{a^{\prime \prime}}{a} + \frac{a^{\prime 2}}{a^2} \right ) + \frac{k_\text{curv}}{a^2} - \frac{\Lambda_{\rm RS}}{3}= 0 \, .
\end{equation}
This can be trivially written as
\begin{equation}
    \frac{d}{dy} \left ( a \, a^\prime \right ) + \frac{\Lambda_{\rm RS}}{3} a^2 = \nu^2(t) + k_\text{curv} \, .
\end{equation}
Performing the same change of variables as before:
\begin{equation}
    x(t,y) = a^2 (t,y) \, ,
\end{equation}
we obtain
\begin{equation}
    \frac{d}{dy} = \frac{dx}{dy} \, \frac{d}{dx} = 2 a \, a^\prime \frac{d}{d a^2}
\end{equation}
for fixed $t$, meaning that
\begin{equation}
    \frac{d}{da^2} (a \, a^\prime)^2 = (\nu^2 + k_\text{curv}) - \frac{\Lambda_{\rm RS}}{3} a^2.
\end{equation}
Thus, integrating over $a^2$, we are left with:
\begin{equation}
    \left ( a \, a^\prime \right )^2 +  {\cal C} = \left ( \nu^2 + k_\text{curv} \right ) a^2 - \frac{\Lambda_{\rm RS}}{6} a^4 \, ,
\end{equation}
with ${\cal C}$ an integration constant. Finally, using the junction condition in Eq.~\eqref{eq:junction}, we obtain the following for the Hubble parameter:
\begin{equation} \label{eq:FriedmannRS1}
    \left . \frac{\nu^2 (t)}{a^2(t,y)} \right |_{y = 0} = \left ( \frac{\dot a_0 (t)}{a_0 (t)}\right )^2 = H^2(t) = \frac{\kappa_{5}^4}{36} \, \rho_b^2 +  \frac{\Lambda_{\rm RS}}{6} - \frac{k_\text{curv}}{a^2_0 (t)}  + \frac{\cal C}{a^4_0 (t)}
\end{equation}
with $a_0 (t) = a(t,0)$.

The first Friedmann equation can be obtained by recalling that $\rho_b = \sigma_0 + \rho_0$, which leads to:
\begin{align}
    H^2 & =\frac{\kappa_5^4}{36}\rho_0^2+\frac{\kappa_5^4}{18}\rho_0\, \sigma_{\rm IR}+\frac{\kappa_5^4}{36}\sigma_0^2+\frac{\Lambda_{\rm RS}}{6}-\frac{k_\text{curv}}{a_0^2(t)}+\frac{{\cal C}}{a_0^4(t)}\nonumber \\
    & =\frac{\kappa_5^4}{18} \sigma_0 \,\rho_0 \, \left(1+\frac{\rho_0}{2\sigma_0}\right) + \frac{\kappa_5^4}{36} \sigma_0^2 + \frac{\Lambda_{\rm RS}}{6}-\frac{k_\text{curv}}{a_0^2(t)}+\frac{{\cal C}}{a_0^4(t)} \, .
    \label{almost H}
\end{align}
The 4D cosmological constant $\Lambda_4$ is defined as a combination of the brane-tension term and the 5D cosmological constant as follows:
\begin{equation}
    \frac{\Lambda_{4}}{3} = \frac{\kappa_{5}^{4}}{36} \sigma_0^2 + \frac{\Lambda_{\rm RS}}{6} \, ,
\end{equation}
and vanishes when $\Lambda_{\rm RS}$ and the IR-brane tension are fine-tuned as in Eqs.~\eqref{k curvature RS2} and~\eqref{eq:branetensionfinetuningRS} (see Appendix~\ref{sec:EEqand5D} for details on the correct definition of $\kappa_5^2$, $\Lambda_{\rm RS}$ and $\sigma_0$ to obtain the desired cancelation of $\Lambda_4$).

The Hubble parameter is then:
\begin{equation} \label{Hubble 1}
    H^2=\frac{\rho_0}{3\,M_{\rm P}^2}\left(1+\frac{\rho_0}{2 \sigma_0}\right) + \frac{\Lambda_{4}}{3} -\frac{k_\text{curv}}{a_0^2(t)}+\frac{{\cal C}}{a_0^4(t)} \, ,
\end{equation}
where the last term depends on the initial conditions, which allows us to fix ${\cal C}$, and the second to last term is the standard one that depends on the spatial curvature $k_\text{curv}$. In the anti-de Sitter case, it is possible to set ${\cal C} = 0$ by using suitable initial conditions (see Refs.~\cite{Langlois:2000ns, Langlois:2002bb}), differently from the flat case (where periodicity and boundary conditions fix it to a very specific value). Assuming that both $\Lambda_4$ and $k_\text{curv}$ vanish, the Hubble parameter is simply:
\begin{equation} \label{Hubble final}
    H^2 = \frac{\rho_0}{3\,M_\text{P}^2}\left(1+\frac{\rho_0}{2\, \sigma_0}\right) \, .
\end{equation}
We will use this equation when studying the difference between standard 4D cosmology and the RS2 extra-dimensional case. 

In the simple case $k_{\rm curv} = 0$ and ${\cal C} = 0$ the explicit expressions for $a(t,y)$ and $n(t,y)$ are (see Ref.~\cite{Langlois:2002bb} for the general solution away from this limit):
\begin{equation} \label{nRS2}
    \left \{
    \begin{array}{l}
        a(t,y) = a_0(t) \left [ \cosh k y - \eta \, \sinh k |y| \right ] \, , \\
        \\
        n(t,y) = \left [ \cosh k y - \eta \, \sinh k |y| \right ] \, ,
    \end{array}
        \right .
\end{equation}
under the assumption that $\dot \rho_0 = 0$ (and thus $\dot \eta = 0$).

\subsection{RS1 setup: anti-de Sitter with two branes} 
\label{sec:RS1cosmo}
The Einstein equations are the same as in the one-brane case but for the localized terms:
\begin{align}
    &\frac{\dot{a}^2}{a^2}-n^2\left(\frac{a^{\prime\prime}}{a}+\frac{a^{\prime2}}{a^2}\right)+k_\text{curv}\frac{n^2}{a^2}=\frac{\Lambda_{\rm RS}}{3}\,n^2-\frac{\kappa_{5}^2}{3}\,\rho_{0} \, \delta (y) + \frac{\kappa_{5}^2}{3}\,\rho_{\pi}
    \,
    \delta (y - \pi\, r_c ) \, , \\
    &a^2\left(2\frac{a^{\prime\prime}}{a}+\frac{n^{\prime\prime}}{n}+\frac{a^{\prime2}}{a^2}+2\frac{a^{\prime}}{a}\frac{n^{\prime}}{n}\right)+\frac{a^2}{n^2}\left(2\frac{\dot{a}}{a}\frac{\dot{n}}{n}-2\frac{\ddot{a}}{a}-\frac{\dot{a}^2}{a^2}\right)-k_\text{curv} \nonumber\\
    &\qquad\qquad\qquad\qquad =- \frac{\Lambda_{\rm RS}}{3}\,a^2+\frac{\kappa_{5}^2}{3}\,p_{0}\,
    \delta (y) -\frac{\kappa_{5}^2}{3}\,p_{\pi}\,
    \delta (y - \pi\, r_c), \\
    &\frac{n^{\prime}}{n}\frac{\dot{a}}{a}-\frac{\dot{a}^{\prime}}{a}=0\,, \\
    &\left(\frac{a^{\prime2}}{a^2}+\frac{a^{\prime}}{a}\frac{n^{\prime}}{n}\right)-\frac{1}{n^2}\left(\frac{\ddot{a}}{a}+\frac{\dot{a}^2}{a^2}-\frac{\dot{a}}{a}\frac{\dot{n}}{n}\right)-\frac{k_{\text{curv}}}{a^2} = - \frac{\Lambda_{\rm RS}}{3}. \label{Gyy Einstein RS1}
\end{align}

The junction conditions must be given at both singular orbifold points, $y = 0$ and $y = \pi\, r_c$:
\begin{align} \label{eq:junctionRS1}
    \left . \left ( \frac{a^\prime}{a} \right ) \right |_{y \to\, 0^+} & = - \frac{\kappa^2_5}{6} (\sigma_0 +  \rho_0) \,,\\
    \left . \left ( \frac{n^\prime}{n} \right ) \right |_{y \to\, 0^+} & =  \frac{\kappa^2_5}{6} \left [ 3 p_0 + 2 (\sigma_0 + \rho_0) \right ] \, , \\
     \left . \left ( \frac{a^\prime}{a} \right ) \right |_{y \to\, ( \pi\, r_c)^-} & = + \frac{\kappa^2_5}{6} (\sigma_\pi +  \rho_\pi) \,,\\
    \left . \left ( \frac{n^\prime}{n} \right ) \right |_{y \to\, ( \pi\, r_c)^-} & = - \frac{\kappa^2_5}{6} \left [ 3 p_\pi + 2 (\sigma_\pi + \rho_\pi) \right ] \, .
\end{align}

Notice that non-standard cosmology in RS braneworlds was first studied in Ref.~\cite{Binetruy:1999ut}. We follow here Ref.~\cite{Lesgourgues:2000tj}, instead, { where the standard 4D behavior was also recovered.} It is easy to derive as a general solution for the $G_{00}$ equation:
\begin{equation}
    a^2(t,y) = \left \{ a_0^2 (t) \, \omega_0^2(y) + a_\pi^2(t) \, \omega_\pi^2(y) + \frac{\nu^2(t)}{2 k^2} \, \left [ \omega_0^2(y) + \omega_\pi^2(y) - 1 \right ] \right \} \, , 
\end{equation}
where:
\begin{equation}
    \left \{
    \begin{array}{l}
        \omega^2_0(y) = \cosh 2 k y - \coth 2 \pi k r_c \, \sin 2 k y \, , \\
        \\
        \omega^2_\pi(y) = \frac{1}{\sinh 2 \pi k r_c} \, \sinh 2 k y\, .
    \end{array}
    \right .
\end{equation}

The time-dependent coefficients $a_0(t)$ and $a_\pi(t)$ can be fixed using the boundary conditions at $y = 0$ and $y = \pi\, r_c$, to get:
\begin{equation} \label{eq:Hubbletwobranesrelations}
    \left \{
    \begin{array}{l}
        \frac{\nu^2(t)}{a_0^2(t)} = H^2_0 = \frac{\kappa_5^2 \, k}{3 \left ( 1 - \Omega_0^2 \right ) } \,  \frac{\rho_0 \, + \, \Omega_0^4 \, \rho_\pi \, - \, \frac{\kappa_5^2}{12 k} \, \left ( 1 - \Omega_0^4 \right ) \, \rho_0 \, \rho_\pi}{1 - \left (1 - \Omega_0^2 \right ) \, \frac{\kappa_5^2}{12 k} \, \rho_\pi } \, ,
        \\
        \\
        \frac{\nu^2(t)}{a_\pi^2(t)} = H^2_\pi = \frac{\kappa_5^2 \, k}{3 \left ( 1 - \Omega_0^2 \right ) } \, \frac{\rho_0 \, + \ \Omega_0^4 \, \rho_\pi \, - \, \frac{\kappa_5^2}{12 k} \,\left ( 1 - \Omega_0^4 \right ) \, \rho_0 \, \rho_\pi}{\Omega_0^2 - \left (1 - \Omega_0^2 \right ) \, \frac{\kappa_5^2}{12 k} \, \rho_0 } \, ,
    \end{array}
    \right .
\end{equation}
where $\Omega_0 = \exp ( - \pi k r_c)$ is the RS warping factor. Notice that, as it was the case for the one-brane warped model, Eq.~\eqref{eq:Hubbletwobranesrelations} is quadratic in the matter densities on the brane. The main difference is that the quadratic terms imply contributions from both the matter densities $\rho_0$ and $\rho_\pi$. This means that if the inflaton is localized only on one brane, the quadratic terms vanishes. A second difference is the dependence of the matter densities on a brane located at $y=\pi\, r_c$ (or 0) on the Hubble parameter computed at $y=0$ (or $\pi\, r_c$). 

Under the assumption that the inflaton is located at $y = \pi\, r_c$ (where the SM lies) and that there is no matter at $y = 0$, we get: 
\begin{equation} \label{Hubble nonfinal RS1}
   \left . \left ( \frac{\dot a}{n a} \right )^2 \right |_{y = \pi\, r_c} = \frac{\kappa_5^2 \, k}{3 \left ( 1 - \Omega_0^2 \right ) } \, \Omega_0^2 \, \rho_\pi = \frac{1}{3 M_{\rm P}^2} \Omega_0^2 \rho_\pi \, ,
\end{equation}
after using the relation between the Planck mass and the fundamental parameters of the 5D model, $M_5$ and $k$. Introducing the Hubble parameter,\footnote{Notice that our results differ by a factor $\Omega_0^2$ with respect to Eq.~(94) of Ref.~\cite{Giudice:2002vh}.} 
\begin{equation} \label{Hubble final RS1}
    H^2 = \left . \left ( \frac{\dot a}{a} \right )^2 \right |_\pi = \frac{1}{3 M_{\rm P}^2 }\, n^2_\pi \Omega_0^2 \, \rho_\pi = \frac{1}{3 M_{\rm P}^2 }\, \Omega_0^4 \, \rho_\pi = \frac{1}{3 M_{\rm P}^2 }\,  \bar \rho_\pi\, ,
\end{equation}
where we used the fact that if $\rho_0$ and $\rho_\pi$ are constant in time, then it can be shown that $n^2_\pi = \Omega_0^2$ (see Ref.~\cite{Lesgourgues:2000tj}).

A closed expression~\cite{Giudice:2002vh} for $n(t,y)$ in the case in which $\rho_0 = 0$ can be derived from Eqs.~\eqref{eq:Hubbletwobranesrelations} and~\eqref{Hubble final RS1}:
\begin{equation} \label{nRS1}
    n^2(y) = e^{- 2 k |y|} \, \left [ 1 + \frac{\left ( 2 \Omega_0^2 - 1\right ) \, H^2}{4 \, k^2 \, \Omega_0^4}\right ] + \left ( \frac{e^{2 k |y|}}{2} - 1 \right )\, \frac{H^2}{2 \, k^2}\,.
\end{equation}

We will use Eqs.~(\ref{Hubble final RS1}) and (\ref{nRS1}) when studying the difference between standard 4D cosmology and the extra-dimensional two-brane case. 

\section{Inflation in 5D models with the inflaton on one brane} 
\label{sec:inflationBrane}
The presence of extra-dimensions may modify the Hubble expansion rate with respect to standard cosmology in 4D, depending on the particular scenario considered. In Sections~\ref{sec:extraD} and~\ref{sec:Friedmann}, we have studied three possible setups: a flat 5D space-time with one brane at $y = 0$, inspired by the Dark Dimension literature to justify the 4D cosmological constant in the Universe (see Sections~\ref{sec:darkdimpheno} and~\ref{sec:DarkDimcosmo}); an anti-de Sitter 5D space-time with one brane at $y = 0$ (see Sections~\ref{sec:RS2pheno} and~\ref{sec:RS2cosmo}), widely used in 5D braneworld cosmology; and an anti-de Sitter 5D space-time with two branes at $y=0$ and $y= \pi\, r_c$ (see Sections~\ref{sec:RS1pheno} and~\ref{sec:RS1cosmo}), traditionally adopted when studying 5D braneworld phenomenology as a solution to the hierarchy problem. In the three configurations, we have considered the inflaton located in a single brane position: $y=0$ for the DD and RS2 cases (in which there is only one brane in the setup) and $y= \pi\, r_c$ for the RS1 case (where there is a choice between two possible branes). In the latter case, we decided to put the inflaton on the IR-brane to easily compare with the existing literature. The case in which the inflaton is located at the UV-brane in the two-branes scenario or in the bulk is left for future work.

We have derived the Friedmann equation for the three setups, finding: 
\begin{equation} \label{SetOfH}
    H^2 =
    \begin{dcases}
        \frac{\rho_0}{3 M_{\rm P}^2} &\text{ for 4D},\\
        \frac{\rho_0}{3 M_{\rm P}^2} + \frac{\Lambda_4}{3}&\text{ for DD},\\
        \frac{\Omega_0^4\,\rho_\pi}{3 M_{\rm P}^2} = \frac{\bar \rho_\pi}{3 M_{\rm P}^2} &\text{ for RS1},\\
        \frac{\rho_0}{3 M_{\rm P}^2} \, \left ( 1 + \frac{\rho_0}{2 \sigma_0} \right )  &\text{ for RS2},
    \end{dcases}
\end{equation}
where $\rho_0$ is the inflaton density in the brane located at $y=0$ and $\bar \rho_\pi$ the rescaled inflaton density at $y= \pi\, r_c$ (see below). Notice that these equations are obtained under the assumption that $\mathcal{C}$ and the spatial curvature $k_\text{curv}$ vanish, and, for the two RS models, a fine-tuning between the 5D cosmological constant and the brane tensions makes $\Lambda_4 = 0$. On the other hand, in the DD case, $\Lambda_4$ is equal to its experimental value and $\sigma_0$ is assumed to be of the same order, see Eq.~\eqref{eq:sigma0expDD}.

However, one comment is in order: although we have derived the complete expressions for the SR parameters in the case of the DD scenario, including terms proportional to the 4D cosmological constant $\Lambda_ 4$ (see the Appendix~\ref{sec:AppSRDD}), it is easy to see that these terms are numerically irrelevant. For $n=1$ extra-dimension, in fact, we know that if the compactification radius $r_c$ saturates the present bound on deviations from Newton's law ({\em i.e.} $r_c = \lambda \leq 40~\mu$m), then necessarily the fundamental scale of gravity must be $M_5 \geq 10^9$~GeV. Therefore, if the potential $V \sim M_5^4$, we have the parameter $M_{\rm P}^2\, \Lambda_4 / V \sim 10^{-64}$. We could lower the fundamental scale of gravity (and, thus, $V$) by adding more than one extra-dimension. However, in this case, the bounds on $r_c$ from supernovae should be applied (since they are stronger than those on deviations from Newton's law). For $n=2$ we typically have $r_c \leq 10^{-3}~\mu$m (see Section~\ref{sec:darkdimpheno}) and thus $M_5 \geq 10^5$~GeV. Also in this case, $M_{\rm P}^2\, \Lambda_4 / V \sim 10^{-48}$. 

In addition to this, it should be reminded that a constant term proportional to the 4D cosmological constant $\Lambda_4$ should be included in the Friedmann equation also in the case of 4D cosmology and of the RS1 model, as a consequence of the fit to Cosmic Microwave Background (CMB) and Big Bang Nucleosynthesis (BBN) data and to the rise of the $\Lambda$CDM model as the ``standard'' model of cosmology. However, this term is traditionally neglected when studying inflationary models. For this reason, in the rest of the paper we will neglect the corrections proportional to $\Lambda_4$ and, thus, the DD and the RS1 scenarios will be identical to the 4D case. 

The only difference between these three models is the expected upper bound on the scalar potential $V$. In both DD and RS2 scenarios, the inflaton is located at $y = 0$, so no rescaling of matter fields in the brane is needed. Therefore, $\rho_0$ should be a dimension 4 operator whose value could be as large as $M_{\rm P}^4$. However, as soon as field fluctuations become larger than the fundamental 5D scale $M_5$ (which could be much lower than $M_{\rm P}$), the effective theory we used to derive the Einstein equations would break down, and quantum gravity effects should enter into play. For this reason, when computing the SR conditions in these two setups, we should compare our matter density with powers of $M_5$. For the DD dimension case, therefore, we get the standard cosmology as long as $\rho_0/M_5^4 \gg \Lambda_4 \times (M_{\rm P}^2/M_5^4) \sim 10^{-40}$ for $M_5 = {\cal O}(10^9)$~TeV. In the RS2 case we see that we recover standard cosmology as long as $\rho_0 \ll \sigma_0$, whereas for $\rho_0 \ge \sigma_0$ $H_{\rm RS2}$ scales quadratically with $\rho$ and the cosmology of the 5D Universe may differ from the 4D scenario. 

On the other hand, in the RS1 scenario the inflaton is located at $y= \pi\, r_c$ and therefore its kinetic term should be rescaled accordingly to get a canonical kinetic term. The rescaled matter density at $\pi\, r_c$ is $\bar \rho_\pi = \Omega_0^4 \, \rho_\pi$. We expect $\bar \rho_\pi$ to be as large as $\Lambda^4$, being $\Lambda =\Omega_0 \, M_{\rm P}$ the effective scale of gravity as seen from the IR brane. In this case, the Friedmann equation is formally as in standard cosmology, the only difference being that the matter density in the numerator is expected to be at most $\bar \rho_\pi \leq \Lambda^4$ (whereas in 4D we have $\rho \leq M_{\rm P}^4$).

In the rest of this section, we analyze how inflation proceeds on the brane in the three cases, describing general conditions for SR, together with scalar and tensor perturbations, emphasizing the differences between the 4D- and the 5D-cosmologies.

\subsection[Slow-roll conditions and number of $e$-folds]{Slow-roll conditions and number of $\boldsymbol e$-folds} \label{sec:SR}
We will consider in the next Sections a few models of SR inflation driven by a scalar field $\phi$ with energy density $\rho = \frac{\dot{\phi}}{2} + V\left(\phi\right)$ and pressure $p = \frac{\dot{\phi}}{2} - V\left(\phi\right)$, where $V\left(\phi\right)$ is the potential of $\phi$. The field satisfies the Klein-Gordon equation
\begin{equation} \label{KG}
    \ddot{\phi} + 3\, H\, \dot{\phi} + V'(\phi) = 0\,,
\end{equation}
where dots ($\dot{}$) and primes ($'$) denote derivatives with respect to time and to the field $\phi$, respectively, and $H$ is the Hubble parameter, which can be any of the expressions in Eq.~\eqref{SetOfH}, depending on the extra-dimensional model. For SR inflation to occur, two conditions must be met: $i)$ the kinetic energy of the inflaton must be smaller than its potential energy, that is, $\frac12\, \dot{\phi}^2\ll V(\phi)$, and $ii)$ the acceleration of the field must be small, that is, $\ddot{\phi} \ll 3\, H\, \dot{\phi}$, otherwise $i)$ will not be satisfied during a sufficiently long period. These two conditions constitute the so-called SR conditions, which ensure that the scalar-field evolution is overdamped long enough to allow for a quasi-exponential expansion. When both $i)$ and $ii)$ are satisfied, we can approximate the energy density of the inflaton by its potential $\rho_0 \simeq V\left(\phi\right)$, and Eq.~\eqref{KG} becomes:
\begin{equation} \label{velocity inf}
    \dot{\phi}\simeq-\frac{V'}{3\,H}\,.
\end{equation}
Usually, SR conditions are expressed in terms of a set of so-called ``SR parameters''. The first SR parameter, $\epsilon_1$, as a function of the Hubble parameter, is defined as
\begin{equation} \label{SR H}
    \epsilon_H = \epsilon_1 \equiv - \frac{\dot H}{H^2}\, . 
\end{equation}
Other SR parameters are defined recursively for $n \geq 2$ as (see, e.g., Refs.~\cite{Lidsey:1995np, Terrero-Escalante:2001zfi})
\begin{equation} \label{eq:otherSRparam}
    \epsilon_n \equiv \frac{1}{H} \, \frac{\dot\epsilon_{n-1}}{ \epsilon_{n-1}} \, .
\end{equation}
Hence, the second and third SR parameters are
\begin{equation} \label{epsilon2}
    \epsilon_2 \equiv \frac{1}{H}\, \frac{\dot \epsilon_1}{\epsilon_1} = - 2 \frac{\dot H}{H^2} + \frac{\ddot H}{H\, \dot H} = 2\,\left(\epsilon_{H}-\eta_{H}\right),
\end{equation}
and
\begin{equation} \label{thirdSR}
    \epsilon_3 \equiv \frac{1}{H}\, \frac{\dot \epsilon_2}{\epsilon_2} = \frac{1}{\epsilon_2} \left[3\, \epsilon_1\, \epsilon_2 - 2\, \epsilon_1^2 + \xi_H^2\right],
\end{equation}
where
\begin{equation} \label{eta_H}
   \eta_{H} \equiv-\frac{1}{2}\frac{\ddot{H}}{H\,\dot{H}}, 
\end{equation}
and
\begin{equation} \label{xiH}
    \xi_H^2 \equiv \frac{\dddot{H}}{H^2\,\dot{H}}-\frac{\ddot{H}^2}{H^2\,\dot{H}^2}\,.
\end{equation}

Once the Hubble parameter is expressed in terms of the potential of the inflaton field, $V(\phi)$, the SR parameters are conventionally introduced as $\epsilon_V$, $\eta_V$, $\xi_V$, respectively. The relation between the $H$-dependent and the $V$-dependent SR parameters is: 
\begin{equation} \label{eq:SRparamV}
    \left \{
    \begin{array}{l}
        \epsilon_V \simeq \epsilon_H + \dots \, , \\
        \\
        \eta_V \simeq \eta_H + \epsilon_H + \dots  \, , \\
        \\
            \xi^2_V  \simeq \frac{1}{2}\, \xi^2_H + 3 \, \epsilon_H (\eta_H -\epsilon_H) + \dots \; ,
    \end{array}
    \right .
\end{equation}
where $\dots$ stands for higher-order terms in $\epsilon_H, \eta_H$ and $\xi_H^2$. In principle, $H$-dependent SR parameters are better suited to study SR inflation (see Ref.~\cite{Liddle:1994dx}): while the SR parameters of Eqs.~\eqref{SR H}, \eqref{eta_H} and~\eqref{xiH} are univocally defined in terms of the Hubble parameter, the corresponding parameters in terms of the inflaton potential $\epsilon_V, \eta_V$ and $\xi^2_V$ will depend on the particular expression that relates $H$ with $V(\phi)$, that is model-dependent. However, their expressions are usually much easier to compute numerically.

\subsection{Inflationary parameters}
We now compute the SR parameters, the scalar and tensor perturbations, and the tensor-to-scalar ratio using the expressions we have written in Section~\ref{sec:SR}. The time-derivatives of the Hubble parameter for the different scenarios are given in Appendix~\ref{sec:AppHdot}. Since DD and RS1 have the same Friedmann equation as in the 4D case, these three scenarios are presented together. In the case of the RS2 scenario, we present the full result and their approximations for $V \ll \sigma_0$ or $V \gg \sigma_0$. In Appendix~\ref{sec:AppSRDD}, we present the SR parameters when a constant is added to the Friedmann equation. These parameters could be useful, for example, in the case of a model in which the constant is non-negligible with respect to $V/M_{\rm P}^2$ during the inflationary phase but becomes compatible with the $\Lambda$CDM value afterwards. 

\subsubsection{Slow-roll conditions} \label{sec:slowroll}
In Appendix~\ref{sec:AppHdot}, the explicit expressions of $\dot H$, $\ddot H$ and $\dddot H$ as a function of the inflaton potential are given for the standard 4D cosmology and for the three scenarios studied here, the DD, the RS1, and the RS2 models. In all cases, the assumption is taken that the terms proportional to $\ddot \phi$ and $\dddot \phi$ are irrelevant. The only difference between the 4D, DD and RS1 models is that in standard cosmology $\phi$ can be as large as $M_{\rm P}$, in the case of DD $\phi$ can be as large as the fundamental scale of gravity $M_5$ and that in the latter case $\phi$ can be as large as the effective gravitational scale $\Lambda$. The RS2 scenario is the only one that can give an inflationary model that is not a slight correction to the standard 4D case. This is because, in the limit $V \gg \sigma_0$, a quadratic term in the inflaton potential dominates the Hubble parameter. The corresponding expressions for the SR parameters are generically more involved, so we have also derived useful limits in the case of $V \ll \sigma_0$ (which gives a small correction to the standard case) and of $V \gg \sigma_0$ (for which we derive results significantly different from the 4D case). 

Expressions that relate the $V$-dependent SR parameters, $\epsilon_V, \eta_V$ and $\xi^2_V$, as a function of the $H$-dependent ones, $\epsilon_H, \eta_H$ and $\xi^2_H$, are given in Eq.~\eqref{eq:SRparamV}. For the standard 4D cosmology (taken as a comparison), the DD scenario, and the RS scenarios, we get:
\begin{equation} \label{first SR}
    \epsilon_V \simeq \frac{M_{\rm P}^2}{2} \left(\frac{V'}{V}\right)^2 \times
    \begin{dcases}
        1 &\text{ for 4D, \, DD \; and \; RS1},\\
        \frac{\left ( 1 + \frac{V}{\sigma_0} \right ) }{\left ( 1 + \frac{V}{2 \sigma_0} \right )^2} &\text{ for RS2}\,,
    \end{dcases}
\end{equation}
with $\epsilon_H \simeq \epsilon_V$ and $V^\prime = dV/d \phi$. Notice that in order to derive these expressions, we have taken $\rho_0 = V(\phi)$ in the 4D and DD cases, and $\bar \rho_\pi = V(\bar \phi)$ in RS1, where $\bar \rho_\pi = \Omega_0^4 \, \rho_\pi$ is the rescaled energy density at $y = \pi\, r_c$ (and $\bar \phi = \Omega_0 \varphi$ is the rescaled inflaton field). In the case of the RS2 scenario, we get for $\epsilon_V$ (see Ref.~\cite{Maartens:1999hf}) in the two relevant limits: 
\begin{align} \label{first SR RS2}
    \epsilon_V & \longrightarrow \frac{M_{\rm P}^2}{2} \left(\frac{V'}{V}\right)^2 \, 
    \begin{dcases}
        \left [  1 - \frac{1}{4} \left ( \frac{V}{\sigma_0}\right )^2 + \dots \right ] &\text{ for } V \ll \sigma_0 \,,\\
        4 \, \frac{\sigma_0}{V} \left( 1 - 3\, \frac{\sigma_0}{V} + \dots \right) &\text{ for } V \gg \sigma_0 \,.
    \end{dcases}
\end{align}

The second SR parameter is:
\begin{equation} \label{second SR}
    \eta_V =
    \begin{dcases}
        M_{\rm P}^2\, \left ( \frac{V''}{V} \right ) &\text{ for \; 4D, \, DD \; and \; RS1}\,,\\
        \frac{M_{\rm P}^2}{\left (1 + \frac{V}{2 \sigma_0} \right )}\, \left [\left ( \frac{V''}{V} \right ) + \frac{1}{2} \, \left ( \frac{V^\prime}{V}\right )^2 \, \left ( \frac{V}{\sigma_0}\right ) \, \frac{1}{\left ( 1 + \frac{V}{\sigma_0} \right ) }\right ] &\text{ for \; RS2}\,.
        \end{dcases}
\end{equation}
The latter expression becomes, in the interesting limits $V \ll \sigma_0$ and $V \gg \sigma_0$:
\begin{equation}
    \left . \eta_V \right |_{\rm RS2} \longrightarrow 
    \begin{dcases}
        \left . \eta_V \right |_{\rm 4D} - \frac{M_{\rm P}^2}{2} \, \left ( \frac{V}{\sigma_0}\right ) \, \left [ \frac{V^{\prime\prime}}{V} - \left ( \frac{V^\prime}{V}\right )^2 \right ] + \dots  &{\rm for} \, V  \ll \sigma_0 \, , \\
        2 \, M_{\rm P}^2 \, \left ( \frac{\sigma_0}{V}\right )\, \left [ \frac{V^{\prime\prime}}{V} + \frac{1}{2} \, \left ( \frac{V^\prime}{V}\right )^2 \right ] + \dots  &{\rm for} \, V  \gg \sigma_0 \, .
    \end{dcases}
\end{equation}

For the third SR parameter, we get in the standard 4D case (and for the DD and RS1 scenarios, as well)
\begin{equation} \label{third SRV}
    \xi_V^2 \equiv M_{\rm P}^4\, \left ( \frac{V'\, V'''}{V^2} \right ) \qquad \text{for 4D, \, DD \; and \; RS1},
\end{equation}
and
\begin{align} \label{eq:xiV2RS2}
    \left . \xi^2_V \right |_{\rm RS2} &= \frac{M_{\rm P}^4}{\left ( 1 + \frac{V}{2 \sigma_0}\right )^2} \, \left \{ \left ( \frac{V'\, V'''}{V^2} \right ) + \frac{7}{4} \, \left ( \frac{V}{\sigma_0} \right ) \, \frac{\left [ 1 + \frac{8}{7} \,\left ( \frac{V}{\sigma_0}\right ) + \frac{3}{7} \, \left ( \frac{V}{\sigma_0}\right )^2\right ] }{\left ( 1 + \frac{V}{\sigma_0}\right ) \,\left ( 1 + \frac{V}{2 \sigma_0}\right ) } \,\left (\frac{V^\prime}{V} \right )^2 \, \left ( \frac{V^{\prime \prime}}{V}\right ) \right . \nonumber \\
    &\quad - \left . \frac{3}{4} \, \left ( \frac{V}{\sigma_0} \right ) \,\frac{\left [ 1 + \frac{8}{3} \, \left ( \frac{V}{\sigma_0}\right )+ \frac{4}{3} \, \left ( \frac{V}{\sigma_0}\right )^2\right ]}{\left ( 1 + \frac{V}{\sigma_0}\right )^2 \, \left ( 1 + \frac{V}{\sigma_0}\right )} \, \left ( \frac{V^\prime}{V}\right )^4\right \}
\end{align}
for the RS2 scenario. Taking into account the two limits $V \ll \sigma_0$ and $V \gg \sigma_0$, the last expression becomes: 
\begin{equation} \label{eq:xiV2RS2approx}
    \left . \xi^2_V \right |_{\rm RS2} =
    \begin{dcases}
        \left ( 1 - \frac{V}{\sigma_0} \right ) \, \left . \xi^2_V \right |_{\rm 4D} + \frac{7}{4} \, M_{\rm P}^4 \,  \left ( \frac{V}{\sigma_0}\right ) \, \left (\frac{V^\prime}{V} \right )^2 \,\left ( \frac{V^{\prime \prime}}{V}\right ) - \frac{3}{4} \, M_{\rm P}^4 \,  \left ( \frac{V}{\sigma_0}\right ) \,\left ( \frac{V^\prime}{V}\right )^4 + \dots\\
        \hspace{8.5cm} {\rm for} \, V \ll \sigma_0 \, , \\
        6 M_{\rm P}^4 \, \left ( \frac{\sigma_0}{V} \right ) \, \left (\frac{V^\prime}{V} \right )^2 \, \left ( \frac{V^{\prime \prime}}{V}\right ) + \dots \hspace{3cm} {\rm for} \, V \gg \sigma_0 \, .
    \end{dcases}
\end{equation}

As soon as the SR conditions are no longer satisfied, inflation ends. The end of the inflationary period is defined as the moment where $\epsilon_V(\phi_\text{end}) = 1$. The number of $e$-folds before inflation ends, $N_\star$, is given by
\begin{align}
    N_\star &= \int H\, dt =\int_{\phi_{\star}}^{\phi_\text{end}}\frac{H}{\dot{\phi}}\,d\phi \nonumber\\
    &\simeq \frac{1}{M_{\rm P}^2} \times
    \begin{dcases}
        \int^{\phi_\star}_{\phi_\text{end}} \frac{V}{V'}\, d\phi &\text{ for 4D, DD, RS1 and RS2} \, (V \ll \sigma_0) \,,\\
        \int^{\phi_\star}_{\phi_\text{end}}
        \frac{V}{V'}\, \frac{V}{2\, \sigma_0}\, d\phi &\text{ for RS2}
        \, (V \gg \sigma_0)\,,
    \end{dcases}
\end{align}
where $\phi_{\star}$ is the value of $\phi$ at the horizon crossing. Interestingly, in the RS2 scenario with $V \gg \sigma$, the rate expansion (the Hubble parameter) is enhanced by a factor $\frac{V}{2\, \sigma}$, which means that for a given $N_\star$, we can have a lower initial value of the inflaton, $\phi_\star$~\cite{Maartens:1999hf}.

\subsubsection{Cosmological perturbations}
\subsubsection*{Scalar perturbations}
In 5D, the amplitude of the scalar perturbations is similar to that in 4D, since the conservation of the curvature perturbation does not depend on the form of the gravitational equations, as long as there is local conservation of the energy-momentum tensor~\cite{Langlois:2000ns, Giudice:2002vh, Bassett:2005xm, Bento:2008yx}. In the SR approximation, the amplitude $\Delta_s^2$ of the scalar perturbations is given by:
\begin{equation} \label{scalar modes}
    \Delta_s^2 \equiv \frac{1}{4\, \pi^2}\, \frac{H_\star^4}{\dot\phi_\star^2} \simeq \frac{9}{4\, \pi^2}\, \frac{H_\star^6}{V_\star'^2} \, ,
\end{equation}
with $H_{\star} \equiv H(\phi_\star)$, $V_\star \equiv V(\phi_\star)$, and we have used Eq.~\eqref{velocity inf}. 

Using the definitions of the SR parameters found in the previous Section, the amplitude $\Delta_s^2$ of the scalar perturbations in 4D, DD and RS1 scenarios is given by:
\begin{equation} \label{scalar modes 4D}
    \Delta_s^2 \simeq \frac{1}{24 \pi^2\, \epsilon_V(\phi_\star)} \frac{V_\star}{M_{\rm P}^4} 
    \qquad {\rm for \, 4D, \, DD \, and \, RS1}\, ,
\end{equation}
whereas in the RS2 case we have:
\begin{equation} \label{scalar modes RS2}
    \left . \Delta_s^2 \right |_{\rm RS2} \simeq \frac{1}{24 \pi^2\, \epsilon_V(\phi_\star)} \frac{V_\star}{M_{\rm P}^4} \left (1 + \frac{V_\star}{\sigma_0}\right ) \, \left (1 + \frac{V_\star}{2 \sigma_0} \right ) \qquad \text{ for RS2}\,.
\end{equation}
In the relevant limits, it becomes: 
\begin{equation}
    \left . \Delta_s^2 \right |_{\rm RS2} \simeq \frac{1}{24 \pi^2\, \epsilon_V(\phi_\star)} \frac{V_\star}{M_{\rm P}^4} \times
    \begin{dcases}
        \left ( 1 + \frac{3}{2} \, \frac{V_\star}{\sigma_0} + \dots \right ) & {\rm for} \, V \ll \sigma_0 \, , \\
        \frac{1}{2} \left ( \frac{V_\star}{\sigma_0} \right )^2 \left (1 + 3 \, \frac{\sigma_0}{V_\star} + \dots \right ) & {\rm for} \, V \gg \sigma_0 \, .
    \end{dcases}
\end{equation}
The spectral index for scalar perturbations,  $n_s$, is defined as a function of the scale $k_\star$ at the end of inflation. We then have:
\begin{align}
    n_s - 1 &\equiv \frac{d\ln \Delta_s^2}{d\ln k_\star} \simeq - \frac{V'(\phi)}{3\, H^2}\, \frac{1}{\Delta_s^2}\, \frac{d\Delta_s^2}{d\phi} \nonumber \\
    &\simeq
    \begin{dcases}
        2\, \eta_V(\phi_\star) - 6\, \epsilon_V(\phi_\star) &\text{ for 4D, \, DD \, and \, RS1}\,,\\
        2\, \eta_V(\phi_\star)
        - 6\, \epsilon_V(\phi_\star) \, 
        \left [ 1 + \frac{1}{3} \, \left ( \frac{V_\star}{\sigma_0} \right ) \, \frac{\left ( 1 + \frac{V_\star}{2 \sigma_0}\right )}{\left ( 1 + 
        \frac{V_\star}{\sigma_0}\right )^2}
        \right ]
         &\text{ for RS2}\,,
    \end{dcases}
\end{align}
where we have used the relation $d/d \ln k = - M_{\rm P}^2 \, \left ( V^\prime/V \right ) \, d/d \phi$~\cite{Liddle:1992wi}. In the limits $V \ll \sigma_0$ and $V \gg \sigma_0$, we get:
\begin{equation}
    n_s - 1 \simeq 
    \begin{dcases}
        \left ( n_s - 1 \right )_{\rm 4D} - \frac{V_\star}{\sigma_0 } \left .\eta_V (\phi_\star) \right |_{\rm 4D} + \dots & {\rm for} \, V_\star \ll \sigma_0 \, , \\
        4 \left (\frac{\sigma_0}{V_\star} \right ) \left [ \eta_V (\phi_\star) - 6 \epsilon_V (\phi_\star)\right ]_{\rm 4D} + \dots & {\rm for} \, V_\star \gg \sigma_0 \, ,
    \end{dcases}
\end{equation}
where $\dots$ stands for quadratic terms in both $V_\star/\sigma_0$ and $\sigma_0/V_\star$. Notice that, at leading order in the expansion parameter, the 4D expressions for $\epsilon_V$, $\eta_V$ and $(n_s-1)$ must be used, as they differ from the RS2 expressions by higher order terms.

In turn, the running of the spectral index, $\alpha$, is given by:
\begin{align}
    \alpha &\equiv \frac{d n_s}{d\ln k_\star} \simeq
    - M_{\rm P}^2 
    \left ( \frac{V^\prime_\star}{V_\star}\right )
    \left . \frac{d n_s}{d \phi} \right |_{\phi_\star} 
    \nonumber \\
    & \simeq 
    -2\, \xi_V^2(\phi_\star) + 
        16\, \epsilon_V(\phi_\star)\, \eta_V(\phi_\star) 
        - 24\, \epsilon_V^2(\phi_\star) \qquad \text{ for 4D, 
        \, DD \, and \, RS1 }\, .
\end{align}
The running of the spectral index in the case of the RS2 scenario gives: 
\begin{align}
    \left . \alpha \right |_{\rm RS2} &=  -2 \frac{\left . \xi_V^2 \right |_{\rm 4D}}{\left (1 + \frac{V}{2 \sigma_0} \right )} + 16 \frac{H_1 \left ( \frac{V}{\sigma_0}\right)}{ \left ( 1+\frac{V}{\sigma_0} \right ) \, \left (1 + \frac{V}{2 \sigma_0} \right)^3} \,  \,  \left .\epsilon_V \right |_{\rm 4D} \,  \left .\eta_V \right |_{\rm 4D} \nonumber\\
    &\qquad - 24 \frac{H_2 \left ( \frac{V}{\sigma_0}\right)}{\left ( 1+\frac{V}{\sigma_0}\right )^2 \, \left ( 1 + \frac{V}{2 \sigma_0}\right)^4} \,  \left . \epsilon_V^2 \right |_{\rm 4D} \, ,
\end{align}
where:
\begin{align}
    H_1 \left ( \frac{V}{\sigma_0}\right) &\equiv 1 - \frac{96}{32} \, \left ( \frac{V}{\sigma_0}\right) + \frac{116}{32} \, \left ( \frac{V}{\sigma_0}\right)^2 - \frac{66}{32} \, \left ( \frac{V}{\sigma_0}\right)^3 - \frac{15}{32} \, \left ( \frac{V}{\sigma_0}\right)^4, \\
    H_2 \left ( \frac{V}{\sigma_0}\right) &\equiv 1 + \frac{432}{96} \, \left ( \frac{V}{\sigma_0}\right) + \frac{816}{96} \, \left ( \frac{V}{\sigma_0}\right)^2 + \frac{864}{96} \, \left ( \frac{V}{\sigma_0}\right)^3 + \frac{560}{96} \, \left ( \frac{V}{\sigma_0}\right)^4 \nonumber\\
    &\qquad- \frac{216}{96} \, \left ( \frac{V}{\sigma_0}\right)^5 - \frac{39}{96} \, \left ( \frac{V}{\sigma_0}\right)^6.
\end{align}
The two relevant limits for $\alpha$ in the RS2 case give: 
\begin{equation}
    \left . \alpha \right |_{\rm RS2} \to 
    \begin{dcases}
        - 2 \, M_{\rm P}^4 \, \left [\left (1 - \frac{V}{2 \sigma_0} \,  \right ) \, \frac{V^\prime \, V^{\prime \prime \prime}}{V^2}- 4 \,  \left ( \frac{V^\prime}{V}\right )^2 \left ( \frac{V^{\prime \prime}}{V}\right ) + 3 \, \left ( \frac{V^\prime}{V}\right )^4 \right ] \\
        \qquad\qquad + {\cal O} \left ( \frac{V^2}{\sigma_0^2}\right ) \qquad {\rm for} \; V \ll \sigma_0 \, , \\
        - 4 \, M_{\rm P}^4 \, \left ( \frac{\sigma_0}{V} \right ) \,\left [\frac{V^\prime \, V^{\prime \prime \prime}}{V^2}- 15 \,\left ( \frac{V^\prime}{V}\right )^2 \, \left ( \frac{V^{\prime \prime}}{V}\right ) + \frac{39}{2} \, \left ( \frac{V^\prime}{V}\right )^4 \right ] \\
        \qquad\qquad + {\cal O} \left ( \frac{\sigma_0^2}{V^2}\right ) \qquad {\rm for} \; V \gg \sigma_0 \, .
    \end{dcases}
\end{equation}
Notice that, for $V \gg \sigma_0$, the parameter $\alpha$ in the RS2 scheme is naturally small, as it is suppressed by a power of $\sigma_0/V$.

\subsubsection*{Tensor perturbations}
Although the amplitude of the scalar perturbations is the same regardless of the geometry we choose for the extradimensional setup (see, e.g., Ref.~\cite{Giudice:2002vh}), being also equal to the 4-dimensional scenario, the same is not true for the tensor amplitude perturbations. A full treatment of the tensor perturbations can be found in Ref.~\cite{Langlois:2000ns} for the RS2 setup, in Ref.~\cite{Im:2017eju} for the flat (DD) scenario, and in Ref.~\cite{Giudice:2002vh} for the RS1 framework.

The amplitude of tensor perturbations in 4D is defined as
\begin{equation}
    \left . \Delta_t^2 \right |_{\rm 4D} \equiv \frac{2}{\pi^2} \, \frac{H_\star^2}{M_{\rm P}^2} = \frac{2}{3 \pi^2} \, \frac{V_\star}{M_{\rm P}^4} \, .
\end{equation}
On the other hand, when we compute it in a 5D framework, our expression must be replaced as follows: 
\begin{equation}
    \left . \Delta_t^2 \right |_{\rm 5D} \equiv \frac{2}{\pi^2} \, \frac{H_\star^2}{M_{\rm eff}^2} \, ,
\end{equation}
where the ``effective Planck mass'', $M_{\rm eff}$, takes into account the different propagation of gravitons between a 4D and a 5D space-time~\cite{Bassett:2005xm}. It is defined as: 
\begin{equation} \label{Meff}
    M_{\rm eff}^2 = M_5^3 \int dy \, n(t, y)^2 \, .
\end{equation}
Since $n(t, y)$ is different in the DD, RS1, and RS2 scenarios, the amplitude of the tensor perturbations also differs. For the DD case, we have: 
\begin{equation} \label{eq:MeffDD}
    \left . M_{\rm eff}^2 \right |_{\rm DD} = M_5^3 \, \int_0^{2 \pi\, r_c} dy \, n^2(t,y) = M_{\rm P}^2 \, \left ( 1 - \frac{2}{3} \, \alpha_{\rm DD} \, \eta \, \pi \, r_c \right ),
\end{equation}
where we have used Eqs.~\eqref{eq:MplanckLED} and~\eqref{nDD}. In the RS1 case, on the other hand, we get: 
\begin{equation} \label{eq:MeffRS1}
    \left . M_{\rm eff}^2 \right |_{\rm RS1} = 2 M_5^3 \, \int_0^{\pi\, r_c} dy \, n^2(t,y) = M_{\rm P}^2 \, \left [ 1 - \frac{H^2_\star}{4 k^2 \Omega_0^4} \, \left (1 - 3 \, \Omega_0^2 + \frac{4 \, \pi \, k \, r_c \, \Omega_0^4}{(1- \Omega_0^2)}\right )\right ],
\end{equation}
where we have used Eqs.~\eqref{eq:MplanckRS} and~\eqref{nRS1}. In turn, in the RS2 case, using Eq.~(\ref{nRS2}) and the relation $M_{\rm P}^2 = M_5^3/k$, we obtain instead:
\begin{equation} \label{eq:MeffRS2}
    \left . M_{\rm eff}^2 \right |_{\rm RS2} = 2 M_5^3 \, \int_0^{\pi\, r_c} dy \, n^2(t,y) = M_{\rm P}^2 \, F(x),
\end{equation}
where $F(x)$ is given by~\cite{Langlois:2000ns, Bento:2008yx}:
\begin{equation}
    F(x) = \left[\sqrt{1 + x^2} - x^2\, \ln\left(\frac{1}{x} + \sqrt{\frac{1}{x^2} + 1}\right)\right],
\end{equation}
and the argument of the function is:
\begin{equation}
    x^2 = 6 \, \left ( \frac{M_{\rm P}^2 H^2}{\sigma} \right ) = \eta^2 - 1 \, .
\end{equation}

Once we have computed the effective Planck mass in the three 5D cases, we can write the amplitude of the tensor perturbations: 
\begin{equation}
    \left . \Delta_t^2 \right |_{\rm 5D} = 
    \begin{dcases}
        \frac{2}{\pi^2}\,\frac{H_\star^2}{M_{\rm P}^2 \, \left [ 1 -\frac{1}{6}\, \left ( \frac{M_{\rm P}^4}{M_5^6}\right ) \, H_\star^2\right ]} & \text{ for DD,} \\
        \\
        \frac{2}{\pi^2}\,\frac{H_\star^2}{M_{\rm P}^2 \, \left[1 - \frac{H_\star^2}{4\,k^2 \, \Omega_0^4} \left(1 - 3 \, \Omega_0^2 + \frac{4\,\pi\,k\,r_c \, \Omega_0^4}{\left ( 1 - \Omega_{0}^2 \right )} \right)\right]} & \text{ for RS1,} \\
        \\
        \frac{2}{\pi^2} \, \frac{H^2_\star}{M_{\rm P}^2} \,\left [1 - \frac{V_\star}{\sigma_0}\left ( 1 - \ln \frac{2 \sigma_0}{V_\star}\right ) + \dots \right ] & \text{ for RS2 with} \; V \ll \sigma_0, \\
        \\
        \frac{2}{\pi^2} \, \frac{H^2_\star}{M_{\rm P}^2} \,\left [ \frac{3}{2} \left ( 1 + \frac{V_\star}{\sigma_0}\right ) + \dots \right ] &\text{ for RS2 with} \; V \gg \sigma_0,
    \end{dcases}
\end{equation}
where in the RS2 case we have expanded $F(x)$ in the low-energy ($V_\star \ll \sigma_0$) and
 high-energy ($V_\star \gg \sigma_0$) limits~\cite{Langlois:2000ns, Bassett:2005xm, Bento:2008yx}. 

\subsubsection*{Tensor-to-scalar ratio}
Finally, the tensor-to-scalar ratio $r$ is defined as:
\begin{equation} \label{tensor-to-scalar}
    r \equiv \frac{\Delta_t^2}{\Delta_s^2} \simeq
    \begin{dcases}
        16\, \epsilon_V(\phi_\star) &\text{ for 4D}\,,\\
        \\
        16\,\epsilon_{V}(\phi_{\star})\,\frac{1}{\left [ 1
        -\frac{1}{6}\, \left ( \frac{M_{\rm P}^4}{M_5^6}\right ) \, H_\star^2\right ]} &\text{ for DD}\,,\\
        \\
        16\,\epsilon_{V}(\phi_{\star})\,\frac{1}{ \left[ 1 - \frac{H_\star^2}{4\,k^2 \, \Omega_0^4} \, \left(1 - 3 \, \Omega_0^2 + \frac{4 \,\pi \,k\,r_c \, \Omega_0^4}{
         \left (1 - \Omega_0^2\right) }\right )\right]} &\text{ for RS1}\,,\\
        \\
        16 \, \epsilon_V(\phi_\star) \, \frac{F^2(x)}{\left ( 1 + \frac{V_\star}{\sigma_0}
        \right ) } &\text{ for RS2}\, ,
    \end{dcases}\\
\end{equation}
where for RS2, in the low- and high-energy limits, we get: 
\begin{equation}
    r = 
    \begin{dcases}
        16 \epsilon_V(\phi_\star) \, \left [ 1 - \frac{V_\star}{\sigma_0} + \frac{V_\star}{\sigma_0} \, \ln \frac{2 \sigma_0}{V_\star} + \dots\right] \, &\text{for } V_\star \ll \sigma_0, \\
        24 \epsilon_V(\phi_\star) \, \left [ 1 + {\cal O} \left ( \frac{\sigma_0^2}{V^2_\star}\right ) \right ] &\text{for } V_\star \gg \sigma_0.
    \end{dcases}
\end{equation}

As a useful reminder, in Table~\ref{tab:datasets} we recall the measurements of these four parameters by the Planck, BICEP, and ACT collaborations~\cite{Planck:2018jri, BICEP:2021xfz, AtacamaCosmologyTelescope:2025blo, AtacamaCosmologyTelescope:2025nti}. We will see in the analysis of the inflationary models below that using one data set or the other makes a difference in the results. 
\begin{table}[h]
    \centering
    \begin{tabular}{|c||c|c|c|c|}
        \hline
        Data Set & $\boldsymbol{\Delta_s^2}$ & $\boldsymbol{n_s - 1}$ & $\boldsymbol{\alpha}$ & $\boldsymbol{r}$ \\
        \hline\hline
        & & & & \\
        \cite{Planck:2018jri, BICEP:2021xfz} 
        & $(2.1 \pm 0.1) \times 10^{-9}$
        & $- 0.035(4)$ 
        & $-0.0045(67)$
        & $< 0.036$ at 95\% CL \\
        & & & & \\
        \cite{Planck:2018jri, BICEP:2021xfz, AtacamaCosmologyTelescope:2025blo, AtacamaCosmologyTelescope:2025nti} 
        & $(2.1 \pm 0.1) \times 10^{-9}$
        & $- 0.026(3)$ 
        & $ 0.0062(52)$
        & $ < 0.038$ at 95\% CL \\
        & & & & \\
        \hline
    \end{tabular}
    \caption{Experimental results for $\Delta_s^2$, $n_s - 1$, $\alpha$ and $r$ using either Planck'18  and BICEP'21 data, only~\cite{Planck:2018jri, BICEP:2021xfz} or complementing them with the newest Atacama Cosmology Telescope data~\cite{Planck:2018jri, BICEP:2021xfz, AtacamaCosmologyTelescope:2025blo, AtacamaCosmologyTelescope:2025nti}.}
    \label{tab:datasets}
\end{table}

\section{Inflationary models in extra-dimensions in one brane} \label{sect:inflapot}
In this Section, we examine two examples of inflationary models in extra-dimensions: monomial inflation (Section~\ref{Monomial inf}) and $\alpha$-attractor inflation (Section~\ref{sec:alpha}), checking their viability by taking into account the data from observations.

We assume that inflation takes place after the 4D-branes are stabilized (see Refs.~\cite{Khoury:2001wf, Buchbinder:2007ad, Lehners:2008vx} for other possibilities). We also assume, in the case of RS1, that the inflaton is confined to the same brane as the SM particles (alternative options of an inflaton in the UV-brane or the bulk, will be considered elsewhere, see Ref.~\cite{Im:2017eju}). 

In order to simplify the notation, we introduce a common scale $\Lambda_{\rm I}$ (for the ``inflationary scale''). As was stressed in Section~\ref{sec:slowroll}, $\Lambda_{\rm I} = M_{\rm P}$ for 4D, $\Lambda_{\rm I} = M_5$ for DD and RS2, and $\Lambda_{\rm I} = \Lambda$ for RS1. 

\subsection{Monomial inflation}
\label{Monomial inf}
We start by considering the simplest option for an inflationary model, corresponding to a monomial potential with power $n$~\cite{Lin:2018kjm}:
\begin{equation} \label{Vmon}
    V(\phi) =  \lambda\, \frac{\phi^n}{\Lambda_I^{n-4}} \, ,
\end{equation}
where $\Lambda_{\rm I}$ is the energy scale to which fluctuations of the field $\phi$ are compared.

For this potential, the first $V$-dependent SR parameter in 4D and the three cases with extradimensions (DD, RS1, and RS2), is:
\begin{align}
    \epsilon_V(\phi) & = 
    \begin{dcases}
        \frac{n^2}{2} \left ( \frac{M_{\rm P}}{\phi} \right )^2
        + \dots &\text{ for 4D, DD, RS1 and RS2 } (V \ll \sigma_0) \, , \\
        \\
        \frac{12 \, n^2}{\lambda} \, \left ( \frac{M_5}{\phi}\right )^{n+2} &\text{ for RS2 } (V \gg \sigma_0)\, ,
    \end{dcases}
\end{align}
where for RS2 we took into account Eq.~\eqref{eq:branetensionfinetuningRS2}, the relation $M_{\rm P}^2 = M_5^3/k$ and $\Lambda_I = M_5$. In the first line, the dots stand for first order corrections to the 4D case, which depend on the particular 5D scenario considered. The expressions in first order in $V/\sigma_0$ for the RS2 scenario are given in Appendix~\ref{sec:FOcorrectionsRS2small}. We do not give analog expressions for DD and RS1, since the expansion parameter in both cases is usually much smaller (being $\Lambda_4/M_{\rm P}^2$ for the former and $(\Lambda/M_{\rm P})^2$ for the latter). 

The second SR parameter is:  
\begin{align}
    \eta_V(\phi) & =
    \begin{dcases}
        n \, (n-1) \left ( \frac{M_{\rm P}}{\phi}\right )^2 + \dots &\text{ for 4D, DD, RS1 and RS2 } (V \ll \sigma_0) \, , \\
        \\
        \frac{18}{\lambda} \, n \left ( n - \frac{2}{3} \right ) \, \left ( \frac{M_5}{\phi}\right )^{n+2} &\text{ for RS2 } (V \gg \sigma_0)\, ,
    \end{dcases}
\end{align}
and the third SR parameter is: 
    \begin{align}
    \xi_V^2(\phi) & =
    \begin{dcases}
        n^2 \, (n-1) \, (n-2) \, \left ( \frac{M_{\rm P}}{\phi}\right )^4 + \dots &\text{ for 4D, DD, RS1 and RS2 } (V \ll \sigma_0) \, , \\
        \\
        \frac{36}{\lambda} \, n^3 ( n - 1 ) \, \left ( \frac{M_{\rm P}}{M_5}\right )^2 \left ( \frac{M_5}{\phi}\right )^{n+4} &\text{ for RS2 } (V \gg \sigma_0)\, .
    \end{dcases}
\end{align}
It can be seen that, for all scenarios, the leading term for the SR parameters is identical to the 4D one. The only exception is RS2 for $V \gg \sigma_0$. In this case, the Friedmann equation is significantly modified (it is quadratic in $V$, contrary to the other cases for which it is linear in $V$) and the cosmological evolution differs significantly from the standard 4D one. For this reason, we show RS2 for $V \ll \sigma_0$ together with 4D, DD, and RS1 and present RS2 for $V \gg \sigma_0$ as a separate case.

The end of inflation occurs when $\epsilon_V(\phi_\text{end}) = 1$, which corresponds to:
\begin{equation} \label{eq:phiendphistarmonomial}
    \phi_\text{end} \simeq
    \begin{dcases}
        \frac{n}{\sqrt{2}}\, M_\text{P} + \dots &\text{ for 4D, DD, RS1 and RS2 } (V \ll \sigma_0) \, ,
        \\
        \left ( \frac{12 \, n^2}{\lambda} \right )^{1/(n+2)} \, M_5 &\text{ for RS2 } (V \gg \sigma_0) \, ,
    \end{dcases}
\end{equation}
where in the RS2 case we only considered the expression at ${\cal O}(\sigma_0/V)$, for simplicity. The inflaton field at the horizon-crossing, $\phi_\star$, is given in terms of the number of $e$-folds, $N_\star$, as:
\begin{equation}
    \phi_\star \simeq
    \begin{dcases}
        \sqrt{2\, n\, N_\star + \frac{n^2}{2}}\, M_\text{P} + \dots &\text{ for 4D, DD, RS1 and RS2 } (V \ll \sigma_0) \, ,\\
        \\
        \left [ \left ( \frac{12 \, n}{\lambda} \right ) \, \left [ (n+2) N_\star + n \right ] \right ]^{1/(n+2)} \, M_5 &\text{ for RS2 } (V \gg \sigma_0)\, ,
    \end{dcases}
\end{equation}
where in the expression for $\phi_\star$ in 4D, DD, RS1 and RS2 with $V \ll \sigma_0$ we have retained the term proportional to $\phi_{\rm end}^2$ which is usually neglected. 

Notice that for RS1 all the expressions above should be understood as functions of $\bar \phi = \Omega_0 \phi$, the rescaled field, since physics must be computed on the IR brane where all fields have canonical (rescaled) kinetic terms. Once we have computed $\phi_{\rm end}$ and $\phi_\star$, we can extract two additional constraints for the scenario RS2: 
\begin{equation}
    \begin{dcases}
        \frac{V(\phi_\star)}{\sigma_0} \ll 1 & \longrightarrow \lambda \ll \frac{6}{\left (2 \, n \, N_\star + \frac{n^2}{2}\right )^{n/2}} \, \left ( \frac{M_5}{M_{\rm P}}\right )^{n+2} \, , \\
        & \\
        \frac{V(\phi_{\rm end})}{\sigma_0} \gg 1 & \longrightarrow \lambda \gg  \, \frac{6}{(\sqrt{2} \, n)^n}\left ( \frac{M_5}{M_{\rm P}}\right )^{n+2} \, .
    \end{dcases}
\end{equation}

Armed with the expressions above for the SR parameters and the values of $\phi_\star$ and $\phi_{\rm end}$ for each scenario, we can now compute the physical observables: the amplitude of scalar perturbations $\Delta_s^2$, the spectral index $n_s - 1$, the running of the spectral index $\alpha$ and the tensor-to-scalar ratio $r$. First of all, we give the amplitude of the scalar perturbations for the four models: 
 \begin{equation}
    \Delta_s^2 \simeq
    \begin{dcases}
        \frac{\lambda}{3 \pi^2} \, (2\, n)^\frac{n-2}{2} \, \left (N_\star + \frac{n}{4} \right )^\frac{n+2}{2}  &\text{for 4D} \, , \\
        \frac{\lambda}{3 \pi^2} \, (2\, n)^\frac{n-2}{2} \, \left (N_\star + \frac{n}{4} \right )^\frac{n+2}{2}\, \left(\frac{M_{\rm P}}{M_5} \right)^{n-4} &\text{for DD} \, , \\
        \frac{\lambda}{3 \pi^2} \, (2\, n)^\frac{n-2}{2} \, \left (N_\star + \frac{n}{4} \right )^\frac{n+2}{2}\, \left(\frac{M_{\rm P}}{\Lambda} \right)^{n-4} &\text{for RS1} \, , \\
        \frac{\lambda}{3 \pi^2} \, (2\, n)^\frac{n-2}{2} \, \left (N_\star + \frac{n}{4} \right )^\frac{n+2}{2}\, \left(\frac{M_{\rm P}}{M_5} \right)^{n-4} &\text{for RS2 } (V \ll \sigma_0) \, , \\
        \frac{1}{\pi^2} \left(\frac{\lambda}{12}\right)^\frac{6}{n+2}
        \, n^\frac{2n -2}{n+2} \left[(n+2)\, N_\star + n\right]^\frac{4n+2}{n+2} &\text{for RS2 } (V \gg \sigma_0) \, , 
    \end{dcases}
\end{equation}
where we have split the RS2 scenario into two different regimes, $V \ll \sigma_0$ and $V \gg \sigma_0$. Notice that $\Delta_s^2$ differ in the five scenarios: not only the RS2 for $V \gg \sigma_0$ gives a different dependence on $n$ and $N_\star$, but also the other four scenarios (which otherwise have the same SR parameters, as we have seen in Section~\ref{sec:SR}) give three different results (since DD and RS2 depend on the same set of parameters), due to the different mass scales in the potential, $\Lambda_I=M_{\rm P}$ for 4D, $\Lambda_I=M_5$ for DD and RS2, and 
$\Lambda_I = \Lambda$ for RS1.

For the spectral index we get:
\begin{equation}
    n_s -1 \simeq 
    \begin{dcases}
        - \frac{(n+2)}{\left ( 2 \, N_\star + \frac{n}{2} \right )} &\text{ for 4D, DD, RS1 and RS2 } (V \ll \sigma_0)\, ,\\
        \\
        - \frac{2 \, (2 n  + 1)}{ [(n + 2)\, N_\star + n]} &\text{ for RS2 } (V \gg \sigma_0) \, ,
    \end{dcases}
\end{equation}
where we can see that, in this case, the results for the 4D, DD, RS1 and RS2 (for $V \ll \sigma_0$) scenarios are the same. Analogously,  for the running of the spectral index we get: 
\begin{equation}
    \alpha \simeq 
    \begin{dcases}
         - \frac{2 \, (n+2)}{\left ( 2 \, N_\star + \frac{n}{2} \right )^2}  = \frac{n_s - 1}{\left ( 2 \, N_\star + \frac{n}{2} \right )}
         &\text{ for 4D, DD,  RS1} \\
         & \text{ and RS2 } (V \ll \sigma_0)\, ,\\
         \\
        - 11 \, \left ( \frac{M_{\rm P}}{M_5} \right )^2 \, 
        \left ( \frac{\lambda}{12} \right )^{2/(n+2)} \, 
        \frac{n^{n/(n+2)} \, \left ( n + \frac{2}{11} \right ) \, (n+2)}{
        \left [ (n+2) N_\star  + n \right ]^{(n+4)/(n+2)}
        }
        &\text{ for RS2 } (V \gg \sigma_0) \, .
    \end{dcases}
\end{equation}
Comparing $n_s - 1$ and $\alpha$, we notice two things: 
\begin{enumerate}
    \item In 4D  and in the extra-dimensional scenarios that are perturbations of 4D, the two quantities are strongly correlated. If we vary $n$ and $N_\star$ to fit $n_s - 1$, the expected value of $\alpha$ should be of the same sign (since $N_\star$ is supposed to be larger than $n$ and positive). For this reason, since the present experimental results give a negative $(n_s -1)_{\rm exp}$ and a positive $\alpha_{\rm exp}$, we expect the fit to be poor. 
    \item Notice that, for RS2 in the limit $V \gg \sigma_0$, the running of the spectral index also depends on the coupling $\lambda$, differently from the other four scenarios.
\end{enumerate}
The tensor-to-scalar ratio for the 4D and DD model is: 
\begin{equation} \label{r mono}
    r \simeq 
    \begin{dcases}
        \frac{4\, n}{\left (N_\star + \frac{n}{4} \right )}  &\text{ for 4D} \, ,\\
        \frac{4\, n}{\left (N_\star + \frac{n}{4} \right )} \left [ 1 - \frac{\lambda}{18} \,  \left ( 2 \, n \, N_\star + \frac{n^2}{2}\right )^{n/2} \, \left ( \frac{M_{\rm P}}{M_5} \right )^2 \right ]^{-1} &\text{ for DD},
    \end{dcases}
\end{equation}
whereas for the RS1 model is: 
\begin{equation} \label{rmonoRS1}
    r \simeq \frac{4\, n}{\left (N_\star + \frac{n}{4} \right )} \left\{1 - \frac{x_1^2 \, \lambda}{12}\, \left ( 2\, n\, N_\star + \frac{n^2}{2} \right )^\frac{n}{2}\, \left (\frac{\Lambda}{m_1} \right )^2 \left(\frac{M_{\rm P}}{\Lambda}\right)^n\, G \left ( \frac{\Lambda}{M_{\rm P}}\right ) \right\}^{-1},
\end{equation}
where $x_1 \simeq 3.83$ is the first zero of the Bessel function $J_1(x)$, 
\begin{equation}
    G (\Omega_0) = \left[1 - 3 \Omega_0^2 - 4 \Omega_0^4 \frac{\ln \Omega_0 }{1 - \Omega_0^2} \right]
\end{equation}
and we have replaced $\Omega_0$ with the scale ratio $(\Lambda/M_{\rm P})$. Notice that we have used Eq.~\eqref{eq:kkgrav1} in order to trade the dependence on $k$ for a dependence on $m_1$. Using this parameterization, LHC bounds on $m_1$ and $\Lambda$ for RS1 can be applied immediately (see Fig.~\ref{fig:sigma}, left panel). On the other hand, we have no direct constraint on $k$, which is assumed to be ${\cal O}(M_{\rm P})$.

For the RS2 scenario with $V \ll \sigma_0$, we get:
\begin{align}
    r & \simeq \frac{4 \, n}{\left ( N_\star + \frac{n}{4} \right )} \, \left \{ 1 - \frac{\lambda}{6} \,  \left ( \frac{M_{\rm P}}{M_5}\right )^{n+2} \,  \left ( 2 n N_\star + \frac{n^2}{2} \right )^{n/2} \, \right . \nonumber\\
    &\qquad \times \left . \left [ 1 - \ln \left ( \frac{12}{\lambda} \left ( \frac{M_5}{M_{\rm P}}\right )^{n+2} \, \left ( 2 n N_\star + \frac{n^2}{2} \right )^{-n/2} \right ) \right ] \right \} \, ,
\end{align}
while for the RS2 scenario with $V \gg \sigma_0$ we instead get:
 \begin{equation} \label{rmonoRS2}
    r \simeq \frac{24\, n}{n + (2+n)\, N_\star} \, .
\end{equation}
In this case, as was the case for the 4D model, the tensor-to-scalar ratio does not depend on the coupling $\lambda$. 

These four experimental observables ($\Delta_s^2, n_s - 1, \alpha$ and $r$) should be fitted simultaneously as a function of the free parameters of the model: $\lambda, n, N_\star$ and the scale $\Lambda_I$ ({\em i.e.}, $M_5$ for DD and RS2 and $\Lambda$ for RS1) for all models, plus $m_1$ for RS1. However, we must also satisfy the theoretical constraint for which the largest value of the potential $V(\phi)$ is smaller than the fundamental scale of the model, $\Lambda_I^4$. Since, in the monomial inflationary model, the potential grows monotonically with $\phi$, the theoretical upper bound must be computed for $\phi = \phi_\star$. We then get: 
\begin{equation} \label{eq:thbounds}
    \begin{dcases}
        \lambda \, \left (2 n N_\star + \frac{n^2}{2}\right )^{n/2} \leq 1  & {\rm for \; 4D} \, , \\
        \\
        \lambda \, \left (2 n N_\star + \frac{n^2}{2} \right )^{n/2} \,  \left ( \frac{M_{\rm P}}{M_5}\right )^n \leq 1 & {\rm for \; DD} \, , \\
        \\
        \lambda \, \left (2 n N_\star + \frac{n^2}{2} \right )^{n/2} \, \left ( \frac{M_{\rm P}}{\Lambda}\right )^n \leq 1 & {\rm for \; RS1} \, , \\
        \\
        \lambda \, \left (2 n N_\star + \frac{n^2}{2} \right )^{n/2} \, \left ( \frac{M_{\rm P}}{M_5}\right )^n \leq 1 & {\rm for \; RS2 \; with } \; V \ll \sigma_0 \, , \\
        \\
        \lambda^2 \, \left \{ 12 \, n \, \left [ N_\star \, (n+2) + n \right ] \right \}^n \leq 1  & {\rm for \; RS2 \; with} \; V \gg \sigma_0  \, .
    \end{dcases}
\end{equation}

The results of a fit to the four observables in the four scenarios that are 4D or perturbations of 4D give very poor $\chi^2$. In principle, if we do not include $\alpha$ in the fit, $n_s - 1$ and $r$ demand a small $n$, approximately $n \leq 0.5$ (where the upper bound on $n$ comes from the upper bound on $r$) and $N_\star \sim 30$ using Planck and BICEP data~\cite{Planck:2018jri, BICEP:2021xfz} or $N_\star \sim 40$ adding ACT data~\cite{AtacamaCosmologyTelescope:2025blo, AtacamaCosmologyTelescope:2025nti}, in order to achieve $(n_s - 1)_{\rm exp}$. After that, $\Delta_s^2$ fixes the value of $\lambda$, so that in general these models would give a decent fit, with $\chi^2_{\rm min} \sim 0.5$ (with an {\em unpleasantly small} $n$ at the best fit point, usually $\bar n < 0.1$, though). However, when a measurement of $\alpha$ is added to the game, the fit worsens: since the present experimental result gives a non-vanishing {\em positive} $\alpha$, there is a tension between $n_s - 1$ and $\alpha$, with the result that all scenarios give a poorer $\chi^2_{\rm min}$. The same happens also for RS2 for $V \gg \sigma_0$. Disregarding $\alpha_{\rm exp}$, we could find an upper limit on $n$ from the upper limit on $r$, and then in turn fix $n_s - 1$ with $N_\star \sim 40$. Since $\alpha$ depends on both $\lambda$ and $M_5$, one could think that $\lambda$ is fixed by $(\Delta_s^2)_{\rm exp}$ and eventually $M_5$ can arrange for $\alpha_{\rm exp}$. However, this cannot happen because of the ``wrong'' sign of $\alpha_{\rm exp}$, which cannot be reproduced by moving $M_5$.

In summary, none of the considered extra-dimensional scenarios gives a very good fit to all the experimental results with a monomial inflationary potential. In addition, the preferred values of $N_\star$ are rather smallish, $N_\star \leq 40$, in conflict with post-inflationary cosmological evolution (we will come back to this later in the paper).

\subsection[$\alpha$-attractor]{\boldmath $\alpha$-attractor} \label{sec:alpha}
After studying the case of the monomial potential, which is severely constrained in 4D, we perform the same analysis in the simplest $\alpha$-attractor $T$-model of inflation~\cite{Kallosh:2013hoa, Kallosh:2013yoa}. In this case, we parametrize the inflaton potential as\footnote{In the literature, different parametrizations of this potential can be found. For example, in Ref.~\cite{Figueroa:2024yja}, the $\alpha$-attractor potential is written as: 
\begin{equation}
    V(\phi) = \frac{1}{p} \, \Lambda_{\alpha}^4 \, \tanh^p \left ( \frac{\phi}{\Lambda_I}\right ) ,
\end{equation}
with $\Lambda_{\alpha}$ an energy scale independent from $\Lambda_I$ and $p \geq 2$.}
\begin{equation}
    V(\phi) = \frac{1}{p} \, \Lambda_I^2 \, M^2\, \tanh^p\left[\frac{\phi}{\Lambda_I}\right] \qquad {\rm for} \; p = 2 \, , 
\end{equation}
with $M$ the inflaton mass and $\Lambda_I$ an energy scale. For the 4D model, we have $\Lambda_I = M_{\rm P}$. On the other hand, for DD and RS2 the scale is $\Lambda_I = M_5$ and for RS1 it is $\Lambda_I = \Lambda = \Omega_0 \, M_{\rm P}$. Notice that, for RS1, the mass of the inflaton $M$ is naturally of the scale $\Lambda$, due to the warping of all dimensionful quantities located on a brane at $y = \pi\, r_c$. This potential is the simplest of a class of inflationary potentials with $p \geq 2$ that share the feature of acting as a constant for $\phi/\Lambda_I \gg 1$ and as a monomial for $\phi/\Lambda_I \ll 1$. 

The first SR parameter, in this case, is: 
\begin{align}
    \epsilon_V(\phi) =  
    \begin{dcases}
        8 \, \left(\frac{M_{\mathrm{P}}}{\Lambda_I}\right)^2 \, \text{csch}^2 \left ( \frac{2 \, \phi}{\Lambda_I} \right ) & \text{for 4D, DD, RS1 and RS2 for} \; V \ll \sigma_0  \, , \nonumber \\
        \\
        96 \, \left(\frac{M_5}{M}\right)^2 \, \text{csch}^4 \left ( \frac{\phi}{M_5} \right ) & \text{for RS2 for} \; V \gg \sigma_0 \, . \nonumber 
    \end{dcases}
\end{align}
Notice the extremely different scale dependence between the first four models and RS2 with $V \gg \sigma_0$: $(M_{\rm P}/\Lambda_I)$ in the former case, while in the latter it is $(M_5/M)$. This dependence is due to the factor $\sigma_0/V$ in Eq.~\eqref{first SR RS2}. We recall that, in this expression and in the following, the field $\phi$ for the RS1 model is understood to be the rescaled field $\bar \phi = \Omega_0 \phi$.

The second SR parameter is: 
\begin{align}
    \eta_V(\phi) =
    \begin{dcases}
        -4 \left (\frac{M_{\rm P}}{\Lambda_I}\right)^2\, \text{sech}^2\left(\frac{\phi}{\Lambda_I}\right) \left [1 - \frac12 \text{csch}^2\left(\frac{\phi}{\Lambda_I}\right)\right] &\text{4D, DD, RS1, RS2 } (V \ll \sigma_0),\\
        \\
        - 96 \, \left ( \frac{M_5}{M} \right )^2 \, \text{csch}^2 \left ( \frac{\phi}{M_5} \right ) \,  \left [ 1 - \text{csch}^2 \left ( \frac{\phi}{M_5} \right ) \right ]& \text{for RS2 } (V \gg \sigma_0) \,,
    \end{dcases}
\end{align}
and, for the third SR parameter, we get:
\begin{align}
    \xi^2_V(\phi) =
    \begin{dcases}
        16 \, \left ( \frac{M_{\rm P}}{\Lambda_I} \right )^4\, \text{sech}^4 \left ( \frac{\phi}{\Lambda_I} \right )\, \left [ 1 - 2 \text{csch}^2 \left ( \frac{\phi}{\Lambda_I} \right )\right ] & \text{4D, DD, RS1}  \nonumber \\
        &  \text{RS2 } (V \ll \sigma_0) \\
        \\
        -1152 \left(\frac{M_{\rm P}}{M}\right)^2 \text{csch}^4\left(\frac{\phi}{M_5}\right) \text{sech}^2\left(\frac{\phi}{M_5}\right) \left[1 - \frac12 \text{csch}^2 \left(\frac{\phi}{M_5}\right)\right]& \text{RS2} (V \gg \sigma_0).
    \end{dcases}
\end{align}
Notice that the RS2 case for $V \gg \sigma_0$ differs significantly from the other four scenarios, as it is not a smooth perturbation over the 4D case. 

Once we have computed the SR parameters, we can determine the value of $\phi$ for which inflation ends: 
\begin{align} \label{eq:phiendalphaattractor}
    \phi_\text{end} \simeq 
    \begin{dcases}
        \frac{\Lambda_I}{2}\, \text{arcsinh} \left [ 2 \, \left ( \frac{\sqrt{2}\,M_{\mathrm{P}}}{\Lambda_I} \right ) \right ] &\text{ for 4D, DD, RS1 and RS2 for} \; V \ll \sigma_0,\\
        M_5 \, \text{arcsinh} \left [ 2 \, \left(\frac{ \sqrt{6} \, M_5}{M}\right)^{1/2} \right ] &\text{ for RS2 for} \; V \gg \sigma_0 \, .
    \end{dcases}
\end{align}
and at the horizon crossing, $\phi_\star$: 
\begin{align}
\label{eq:phistaralphaattractor}
    \phi_\star \simeq
    \begin{dcases}
        \frac{\Lambda_I}{2}\, \text{arccosh}\left[8 N_\star \left(\frac{M_{\mathrm{P}}}{\Lambda_I}\right)^2 + \sqrt{1+ 8 \left(\frac{M_{\mathrm{P}}}{\Lambda_I}\right)^2} \right] &\text{4D, DD, RS1, RS2 } (V \ll \sigma_0),\\
        \frac{M_5}{2} \, \ln \left \{ 384 \, \left ( \frac{M_5}{M} \right )^2 \,  \left [ N_\star + \frac{1}{4 \sqrt{6}} \, \left ( \frac{M}{M_5} + \dots \right ) \right ] \right \} &\text{RS2 } (V \gg \sigma_0),
    \end{dcases}
\end{align}
where, in order to obtain a closed expression for $\phi_\star$ for RS2 in the limit $V \gg \sigma_0$, we have assumed that $M_5 \gg M$ and $\phi_\star > M_5$. We have checked the latter assumption numerically, finding that the error in $\phi_\star$ computed using Eq.~\eqref{eq:phistaralphaattractor} with respect to the numerical value is less than 2\% for $N_\star \in [20,70]$ and $M_5/M \in [1, 200]$.

In addition, we can compute the theoretical expressions for the four observables to be compared to the experimental data, ($\Delta_s^2, n_s - 1, \alpha$ and $r$). The spectral index is: 
\begin{align} \label{eq:nsm1alphaattractor}
    n_s - 1 \simeq 
    \begin{dcases}
        - 16 \left(\frac{M_{\rm P}}{\Lambda_I}\right)^2 \left[8 N_\star \left(\frac{M_{\rm P}}{\Lambda_I}\right)^2 + \sqrt{1 + 8 \left(\frac{M_{\rm P}}{\Lambda_I}\right)^2} - 1\right ]^{-1} &\text{for 4D, DD, RS1} \, ,\\
        &  \text{and RS2 for} \; V \ll \sigma_0 \\
        \\
        -192 \left ( \frac{M_5}{M}\right )^2  \,  \text{csch}^2 \left ( \frac{\phi_\star}{M_5} \right ) \, \left [ 1 + \frac{5}{2} \, \text{csch}^2 \left (\frac{\phi_\star}{M_5}\right )  \right ] &\text{for RS2 for} \; V \gg \sigma_0 \, ,
    \end{dcases}
\end{align}
and the running of the spectral index is: 
\begin{align} \label{eq:alphaalphaattractor}
    \alpha \simeq 
    \begin{dcases}
        - 32 \left ( \frac{M_{\rm P}}{\Lambda_I}\right )^4 \,  \text{csch}^4 \left ( \frac{\phi_\star}{\Lambda_I}\right ) &\text{for 4D, DD, RS1},\\
        &  \text{and RS2 for} \; V \ll \sigma_0, \\
        \\
        -384\,\left(\frac{M_{\mathrm{P}}}{M}\right)^{2} \,  \text{csch}^{6} \, \left(\frac{\phi_\star}{M_5}\right) \,  \left[ 23 + \cosh  \left( \frac{2\,\phi_{\star}}{M_{5}} \right) \right] &\text{for RS2 for} \; V \gg \sigma_0 \, .
    \end{dcases}
\end{align}
On the other hand, the amplitude of the scalar perturbation is: 
\begin{align} \label{eq:Deltas2alphaattractor}
    \Delta_s^2 \simeq 
    \begin{dcases}
        \frac{1}{4 \, \pi^2}  \left [ \frac{M}{ \sqrt{24} \, \Lambda_I} \right ]^2 \, \left ( \frac{\Lambda_I}{M_{\rm P}}\right )^6  \,  \sinh^4 \left ( \frac{\phi_\star}{\Lambda_I}\right ) &\text{for 4D, DD, RS1} \, ,\\
        &  \text{and RS2 for} \; V \ll \sigma_0 \\
        \\
        \frac{1}{4 \, \pi^2} \left [ \frac{M}{\sqrt{24} \, M_5}\right ]^8 \, \sinh^4 \left ( \frac{\phi_\star}{M_5} \right ) \, \tanh^6 \left ( \frac{\phi_\star}{M_5} \right ) &\text{for RS2 for} \; V \gg \sigma_0 \, .
    \end{dcases}
\end{align}
In turn, the tensor-to-scalar ratio should be computed for each scenario. For 4D and DD, we have: 
\begin{align} \label{eq:ralphaattractor4DDD}
    r \simeq 
    \begin{dcases}
        \frac{2}{N_\star^2} \, \left ( 1 + \frac{3}{ 4 \, N_\star} + \frac{1}{8 \, N_\star^2}\right )^{-1} & \text{ for 4D} \, ,\\
        \\
        \frac{2}{N_\star^2} \, \left ( \frac{M_5}{M_{\rm P}}\right )^2 \, \left [  1 + \frac{1}{\sqrt{2} \, N_\star} \,  \left (\frac{M_5}{M_{\rm P}} \right) \, \sqrt{1 + \frac{1}{8} \left ( \frac{M_5}{M_{\rm P}} \right)^2} + \frac{1}{8 \, N_\star^2} \, \left ( \frac{M_5}{M_{\rm P}}\right )^2\right ]^{-1} \\
        \qquad \qquad \qquad \times  \left [ 1 - \frac{1}{36} \, \left ( \frac{M_{\rm P}}{M_5}\right )^2 \, \left ( \frac{M}{M_5} \right )^2 \,  \tanh^2 \left ( \frac{\phi_\star}{M_5 }\right ) \right ]^{-1} & \text{ for DD} \, ,
    \end{dcases}
\end{align}
whereas for RS1: 
\begin{align} \label{eq:ralphaattractorRS1}
    r \simeq  \frac{2}{N_\star^2} \, \left ( \frac{\Lambda}{M_{\rm P}}\right )^2 \, \left [  1 + \frac{1}{\sqrt{2} \, N_\star} \,  \left (\frac{\Lambda}{M_{\rm P}} \right) \,  \sqrt{1 + \frac{1}{8} \left ( \frac{\Lambda}{M_{\rm P}} \right)^2} + \frac{1}{8 \, N_\star^2} \, \left ( \frac{\Lambda}{M_{\rm P}}\right )^2\right ]^{-1} \\
    \qquad \qquad \qquad \times \left [ 1 - \frac{x_1^2}{24} \, \left ( \frac{M}{m_1} \right )^2 \, \tanh^2 \left ( \frac{\phi_\star}{\Lambda}\right ) G\left ( \frac{\Lambda}{M_{\rm P}}\right )\right ]^{-1} & \text{ for RS1} \, ,
\end{align}
and finally, for RS2 we have:
\begin{align}
    r \simeq
    \begin{dcases}
        \frac{2}{N_\star^2} \, \left ( \frac{M_5}{M_{\rm P}}\right )^2 \,\left [ 1 + \frac{1}{\sqrt{2} \, N_\star} \, \left (\frac{M_5}{M_{\rm P}} \right) \, \sqrt{1 + \frac{1}{8} \left ( \frac{M_5}{M_{\rm P}} \right)^2} + \frac{1}{8 \, N_\star^2} \, \left ( \frac{M_5}{M_{\rm P}}\right )^2\right ]^{-1} \\
        \qquad \qquad \times \left [ 1 - \frac{V(\phi_\star)}{\sigma_0} + \frac{V(\phi_\star)}{\sigma_0} \, \ln \left ( 2 \frac{\sigma_0}{V(\phi_\star)}\right ) \right ]^{-1} \quad \text{ for RS2 for} \; V \ll \sigma_0 \, ,\\
        \\
        \frac{1}{4 \, N_{\star}^{2}}\,\left(\frac{M}{M_{5}}\right)^{2}\, \left(1 + \frac{1}{4 \, \sqrt{6} \, N_\star} \, \frac{M}{M_5} \right)^{-2} \hspace{2.6cm} \text{ for RS2 for} \; V \gg \sigma_0 \, ,
    \end{dcases}
\end{align}
where, in the last line, we used the approximate expression in Eq.~\eqref{eq:phistaralphaattractor} under the assumptions that $\phi_\star > M_5$ and $M_5 \gg M$. For RS2 with $V \ll \sigma_0$, we have not explicitly written the terms in $V/\sigma_0$ as the corresponding expression for $r$ would become rather cumbersome.

The potential must be bounded by the fundamental scale of the model, $\left . V (\phi) \right|_{\rm max} \leq \Lambda_I^4$. Since the potential of the $\alpha$-attractor inflationary model in the window $\phi \in [\phi_{\rm end}, \phi_\star]$ increases monotonically with $\phi$ for all extra-dimensional scenarios considered, we must apply the bound at $\phi = \phi_\star$ (as was the case for monomial inflation). 

The general formula valid for the 4D, the DD, the RS1 and RS2 with $V \ll \sigma_0$ scenarios is: 
\begin{equation}
    \label{eq:constraintoverpotentials}
    \frac{1}{2} \, \left ( \frac{M}{\Lambda_I} \right )^2 \, \left \{ 1 - \frac{1}{4} \, \left ( \frac{\Lambda_I}{M_{\rm P}} \right )^2\, \frac{\left [  N_\star + \sqrt{1 + 8 \left ( \frac{M_{\rm P}}{\Lambda_I}\right )^2 } - 1\right ]}{\left [8 N_\star^2 \left ( \frac{M_{\rm P}}{\Lambda_I}\right )^2 + 2 N_\star \, \sqrt{1 + 8 \left ( \frac{M_{\rm P}}{\Lambda_I}\right )^2 } + 1\right ]}\right \} \leq 1 \, ,
\end{equation}
where $\Lambda_I = M_{\rm P}$, $M_5$, $\Lambda$ and $M_5$ for the 4D, DD, RS1 and RS2 scenarios, respectively.

For the RS2 scenario with $V \gg \sigma_0$, in the same approximation as we used to compute $\phi_\star$, the bound is simply
\begin{equation}
    \label{eq:constraintoverpotentialRS2}
    M^2 \leq 2 M_5^2 \, ,
\end{equation}
that is looser than the condition $M_5 \gg M$ for which we get Eq.~\eqref{eq:phistaralphaattractor}. Notice that for the two regime of RS2 we should {\it a posteriori} check that conditions $V/\sigma_0 \ll 1$ or $V/\sigma_0 \gg 1$ are fulfilled. We then have the following additional constraints: 
\begin{equation} \label{eq:Vsigmaconditions}
    \begin{dcases}
        \frac{V(\phi_{\star})}{\sigma_0} = \frac{1}{12} \, \frac{M_{\rm P}^2 \, M^2}{M_5^4} \, \tanh^2 \left ( \frac{\phi_\star}{M_5}\right) \ll 1 \, .\\
        \frac{V(\phi_{\rm end})}{\sigma_0} = \frac{1}{12} \, \frac{M_{\rm P}^2 \, M^2}{M_5^4} \, \tanh^2 \left ( \frac{\phi_{\rm end}}{M_5}\right) \gg 1 \, .
    \end{dcases}
\end{equation}

\begin{figure}[th!]
    \def\sepf{0.45}
    \centering
    \begin{tabular}{cc}
    \includegraphics[scale=\sepf]{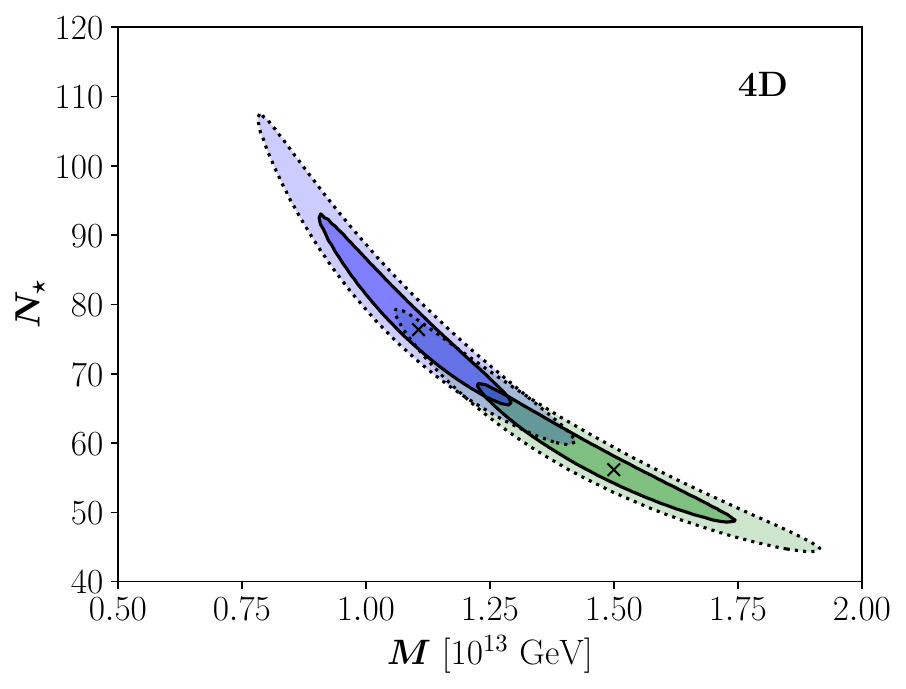} &
    \includegraphics[scale=\sepf]{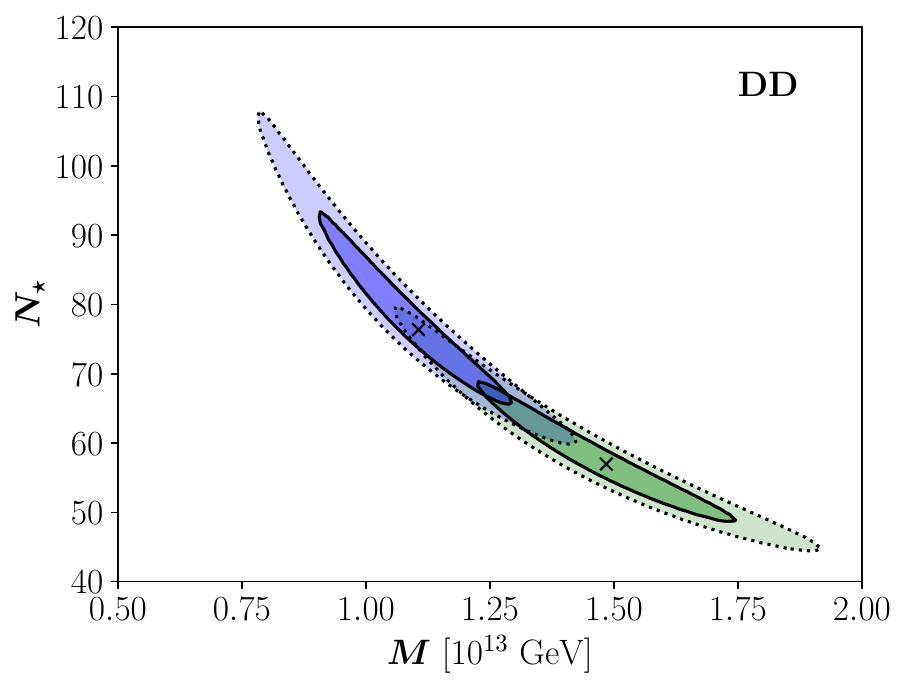} \\
    \includegraphics[scale=\sepf]{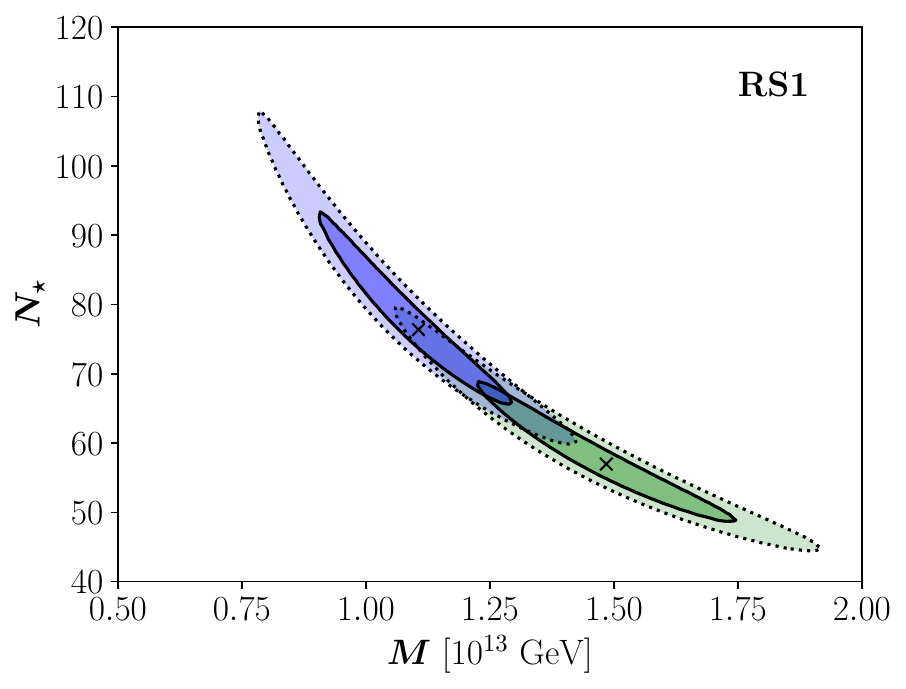} &
    \includegraphics[scale=\sepf]{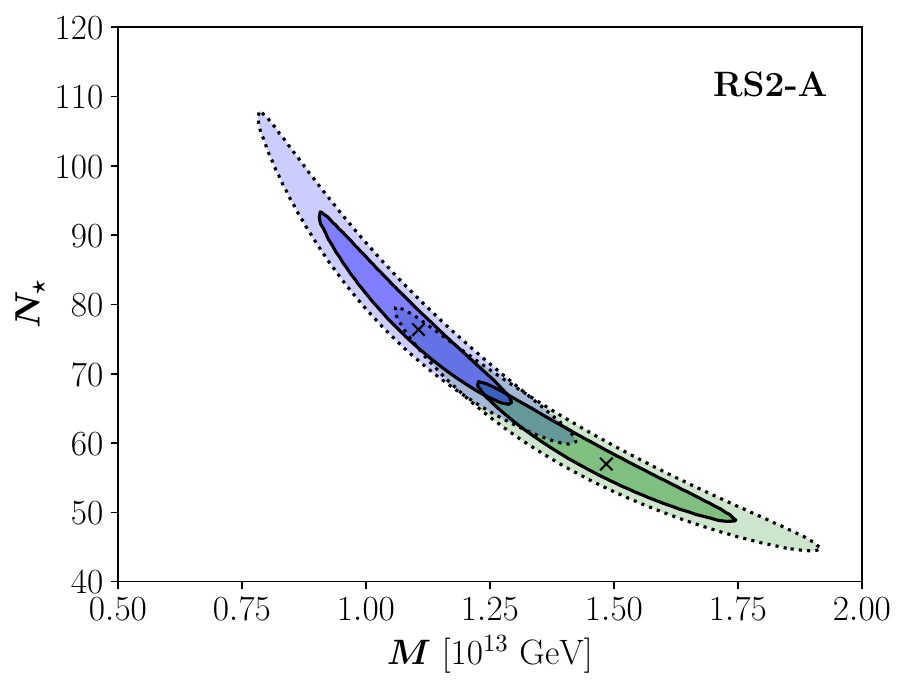}
    \end{tabular}
    \caption{Results for the simplest $\alpha$-attractor inflationary model for: $a)$ 4D (top left); $b)$ DD (top right); $c)$ RS1 (bottom left); $d)$ RS2 with $V \ll \sigma_0$, called RS2-A (bottom right). In all cases, the vertical axis represents $N_\star$ and the horizontal axis the inflaton mass $M$. The green contours have been obtained using Planck and BICEP data only, whereas the blue contours include the newest ACT dataset. The numerical results are given in Table~\ref{tab:alpharesults}.}
    \label{fig:alphaattractorfit}
\end{figure}
\begin{figure}[h]
    \def\sepf{0.50}
    \centering
    \includegraphics[scale=\sepf]{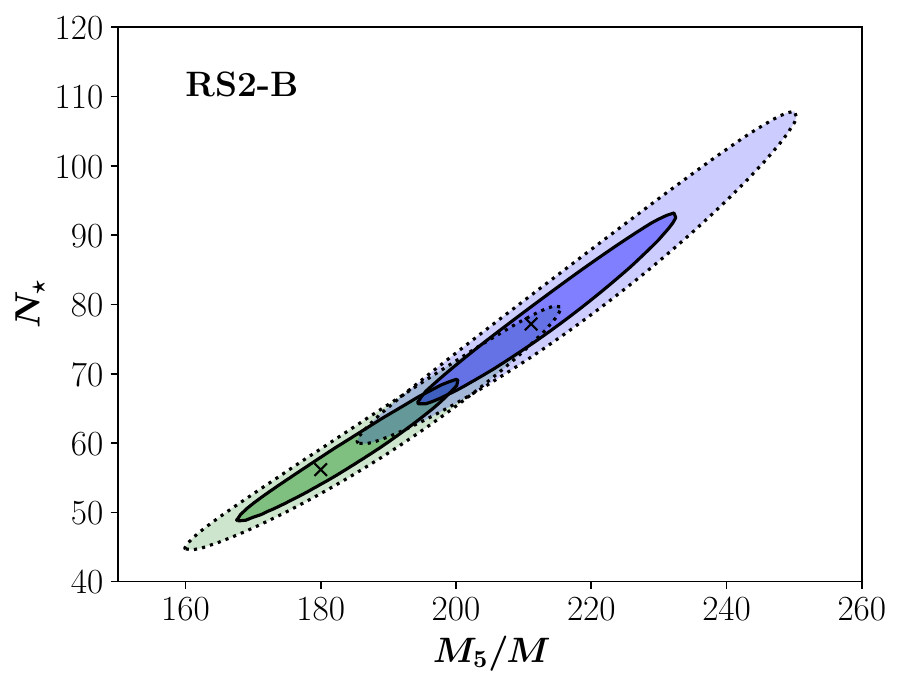}
    \caption{Results for the simplest $\alpha$-attractor inflationary model for RS2 with $V \gg \sigma_0$, called RS2-B. The vertical axis represents $N_\star$ and the horizontal axis the ratio of $M_5$ and the inflaton mass $M$. The green contours have been obtained using Planck and BICEP data, only, whereas the blue contours include the newest ACT dataset. The numerical results are given in Table~\ref{tab:alpharesults}.}
    \label{fig:RS2quadraticalphaattractorfit}
\end{figure}

Our results for the five scenarios considered in the simplest $\alpha$-attractor inflationary potential are are given in Figs.~\ref{fig:alphaattractorfit} for 4D, DD, RS1 and RS2 with $V \ll \sigma_0$, in Fig.~\ref{fig:RS2quadraticalphaattractorfit} for RS2 with $V \gg \sigma_0$ and in Table~\ref{tab:alpharesults}. In each panel of the figures, the blue contours in the plane ($M, N_\star$) refer to a $\chi^2$ analysis using the latest data set (Planck, BICEP, and ACT) while the green contours are obtained using the oldest data set (no ACT data). Notice that for RS2 with $V \gg \sigma_0$ the contours are given in the plane ($M_5/M, N_\star$), since the ratio between the inflaton mass and the fundamental scale of the theory is the relevant parameter in that case. The contours represent 1$\sigma$ and 2$\sigma$ regions with respect to the corresponding best fit, calculated for the number of free parameters in each scenario.
\begin{table}[h]
\centering
{\small
\begin{tabular}{|c||c|c|c|c|c|}
\hline
Scenario & $\boldsymbol{\frac{\chi^2_{\rm min}}{\rm dof}}$ & $\boldsymbol{N_\star}$ & $\boldsymbol{M}$~[$10^{13}$~GeV] & $\boldsymbol{\Lambda_I}$~[GeV] & {\bf Dataset} \\
\hline\hline
& & & & & \\
4D 
& 
$
\begin{array}{l}
0.17 \\
0.79
\end{array}
$
&
$
\begin{array}{l}
57 \, \pm \, 3\\
77 \, \pm \, 4
\end{array}
$
& 
$
\begin{array}{l}
1.50 \, \pm \, 0.08\\
1.11 \, \pm \, 0.06
\end{array}
$& 
& 
$
\begin{array}{c}
\text{\cite{Planck:2018jri, BICEP:2021xfz}} \\
\text{\cite{Planck:2018jri, BICEP:2021xfz, AtacamaCosmologyTelescope:2025blo, AtacamaCosmologyTelescope:2025nti}}
\end{array}
$
\\
& & & & & \\
DD
& 
$
\begin{array}{l}
0.11 \\
0.53
\end{array}
$
&
$
\begin{array}{l}
57 \, \pm \, 3 \\
77 \,  \pm \, 5
\end{array}
$
& 
$
\begin{array}{l}
1.50 \, \pm \, 0.09 \\
1.11 \, \pm \, 0.07
\end{array}
$ 
&
$
\begin{array}{l}
 M_5 \in [3 \times 10^{15}, M_{\rm P}]\\
 M_5 \in [2 \times 10^{15}, M_{\rm P}]
\end{array}
$ 
& 
$
\begin{array}{c}
\text{\cite{Planck:2018jri, BICEP:2021xfz}} \\
\text{\cite{Planck:2018jri, BICEP:2021xfz, AtacamaCosmologyTelescope:2025blo, AtacamaCosmologyTelescope:2025nti}}
\end{array}
$
\\
& & & & & \\
RS1
& 
$
\begin{array}{l}
0.08 \\
0.40
\end{array}
$
&
$
\begin{array}{l}
57 \, \pm \, 4 \\
77 \, \pm \, 5
\end{array}
$
& 
$
\begin{array}{l}
1.51 \, \pm \, 0.11\\
1.12 \, \pm \, 0.08
\end{array}
$
& 
$
\begin{array}{l}
 \Lambda \in [9 \times 10^{12}, M_{\rm P}] \\
 \Lambda \in [7 \times 10^{12}, M_{\rm P}]
\end{array}
$ 
& 
$
\begin{array}{c}
\text{\cite{Planck:2018jri, BICEP:2021xfz}} \\
\text{\cite{Planck:2018jri, BICEP:2021xfz, AtacamaCosmologyTelescope:2025blo, AtacamaCosmologyTelescope:2025nti}}
\end{array}
$
\\
& & & & & \\
RS2-A 
& 
$
\begin{array}{l}
0.11 \\
0.53
\end{array}
$
&
$
\begin{array}{l}
57\, \pm \, 3\\
77\, \pm \, 5
\end{array}
$
& 
$
\begin{array}{l}
1.50\, \pm \, 0.09 \\
1.11\, \pm \, 0.07 
\end{array}
$& 
$\begin{array}{l}
M_5 \in [2\times 10^{15}, M_{\rm P} ] \\
M_5 \in [3\times 10^{15}, M_{\rm P} ]
\end{array}$
& 
$
\begin{array}{c}
\text{\cite{Planck:2018jri, BICEP:2021xfz}} \\
\text{\cite{Planck:2018jri, BICEP:2021xfz, AtacamaCosmologyTelescope:2025blo, AtacamaCosmologyTelescope:2025nti}}
\end{array}
$
\\
& & & & & \\
RS2-B
& 
$
\begin{array}{l}
0.02 \\
0.50
\end{array}
$
&
$
\begin{array}{l}
57 \, \pm \, 4 \\
77\, \pm \, 5
\end{array}
$
& 
$
\begin{array}{l}
(5.5 \pm 1.7)\times 10^{-3} M_5 \\
(4.7 \pm 1.5) \times 10^{-3} M_5
\end{array}
$& 
$\begin{array}{l}

M_5 \in [7 \times 10^{12}, 0.02 \, M_{\rm P}]\\
M_5 \in [5 \times 10^{12}, 0.02 \, M_{\rm P}]
\end{array}$
& 
$
\begin{array}{c}
\text{\cite{Planck:2018jri, BICEP:2021xfz}} \\
\text{\cite{Planck:2018jri, BICEP:2021xfz, AtacamaCosmologyTelescope:2025blo, AtacamaCosmologyTelescope:2025nti}}
\end{array}
$
\\
& & & & & \\
\hline
\end{tabular}
}
\caption{Results of a $\chi^2$ fit to the physical observables $n_s - 1$, $\alpha$, $\Delta_s^2$ and $r$ using Planck and BICEP data, only~\cite{Planck:2018jri, BICEP:2021xfz}, or including the newest ACT data~\cite{Planck:2018jri, BICEP:2021xfz, AtacamaCosmologyTelescope:2025blo, AtacamaCosmologyTelescope:2025nti}. For all scenarios, $M_{\rm P}$ was considered an ultimate upper bound for the fundamental scale $\Lambda_I$, with the exception of RS2 for $V \gg \sigma_0$ (RS2-B, in the Table), for which Eq.~\eqref{eq:Vsigmaconditions} gives a tighter upper bound on $M_5$. Notice that, whereas all scenarios give a precise measurement of ($N_\star,M$), for RS2-B
we measure ($N_\star,M/M_5$), instead.}
\label{tab:alpharesults}
\end{table}

Consider first, as an illustrative case, the 4D scenario. Performing a $\chi^2$ fit to the two data sets (Planck and BICEP~\cite{Planck:2018jri, BICEP:2021xfz}, or Planck, BICEP and the newest data from the ACT collaboration~\cite{Planck:2018jri, BICEP:2021xfz, AtacamaCosmologyTelescope:2025blo, AtacamaCosmologyTelescope:2025nti}) we obtain the results in the first two lines of Table~\ref{tab:alpharesults}. We can see that in both cases the minimum value of the $\chi^2$ (normalized to the number of free parameters of the model, $N_{\rm par} = 2$) is rather good. The fit to the oldest data set is a bit better because of the measured value of $\alpha$: in that case, $\alpha_{\rm exp}$ is negative, as it should be comparing Eqs.~\eqref{eq:nsm1alphaattractor} and~\eqref{eq:alphaalphaattractor}; on the other hand, when the ACT data are included, $\alpha_{\rm exp}$ becomes positive, thus creating tension with the theoretical expectation. However, this tension is not so severe as to destroy the goodness of the fit. On the other hand, the best fit values for the two free parameters of the model ($N_\star$ and the mass of the inflaton $M$) change significantly. This is particularly striking for the number of e-folds: $N_\star$ at minimum goes from $\bar N_\star = 57 \, \pm \, 3$ for the old data set to $\bar N_\star = 77 \, \pm \, 4$ for the newest one. We can see in Fig.~\ref{fig:alphaattractorfit} (top left panel) that the best fit value for the two datasets is compatible only at 3$\sigma$. The black cross within the blue region represents the best fit for the set that includes ACT, whereas the black cross within the green region is the best fit of the oldest data set, with no ACT data. Such a high value for $N_\star$ is troubling, compared to the expectations of generic inflationary models. If we add a flat prior on $N_\star$, $N_\star \in [40,70]$, motivated by post-inflationary cosmological evolution independent of the specific inflationary model considered~\cite{Liddle:2003as,Martin:2013nzq}, we find that the minimum of the $\chi^2$ using the latest data set worsens to $\chi^2_{\rm min} = 1.12$ (since the fit would prefer a value of $\bar N_\star$ larger than the allowed value). 

If we now study the results for the different extra-dimensional scenarios considered in this paper, we get quite similar results. Consider first the case of the Dark Dimension scenario, with one brane and one flat extra-dimension (Fig.~\ref{fig:alphaattractorfit}, top right panel and corresponding lines in Table~\ref{tab:alpharesults}). In this case, we obtain for $\chi^2_{\rm min} = 0.11$ and $0.53$ using the oldest and the latest dataset, respectively (with three parameters to be fitted, $N_\star, M$ and $M_5$). As in 4D, for the DD scenario the best fit values for $N_\star$ and $M$ differ significantly using the two datasets, too: without the ACT data we get $\bar N_\star = 57 \pm 3$, while after adding the new data we get $\bar N_\star = 77 \pm 5$. The central value for $M$ also changes: $\bar M = (1.50 \pm 0.09) \times 10^{13}$~GeV for the oldest set and $\bar M = (1.11 \pm 0.07) \times 10^{13}$~GeV for the newest. When we introduce a flat prior on $N_\star$, though, the minimum of the $\chi^2$ becomes $\chi^2_{\rm min} = 0.78$. The third parameter of this scenario, $M_5$, makes it easier to adjust the value of $\bar N_\star$ a bit, with respect to the 4D case. A lower bound on the fundamental scale $M_5$ is also found, with $M_5 \geq 10^{15}$~GeV, coming from the tensor-to-scalar ratio being strictly positive, $r \geq 0$.

The other two extra-dimensional scenarios that are smooth deformation of 4D, {\em i.e.} RS1 and RS2 with $V \ll \sigma_0$ (RS2-A, in Table~\ref{tab:alpharesults} and in Fig.~\ref{fig:alphaattractorfit}), give quite similar results. In both cases ($\bar N_\star = 57$, $\bar M/10^{13} \sim 1.5$) and ($\bar N_\star = 77$, $\bar M/10^{13} \sim 1.1$) for the oldest (newest) dataset, respectively. The lower bound on the scale is $\Lambda \geq 10^{13}$~GeV for RS1 (due to Eq.~\eqref{eq:constraintoverpotentials}), and $M_5 \geq 10^{15}$~GeV for RS2-A (due to the upper bound on the tensor-to-scalar ratio). In the RS1 scenario, although the tensor to scalar ratio $r$ also depends on $m_1$, in practice the fit does not, so this parameter remains basically unconstrained by cosmological data (and it is only constrained by the LHC). 

The only scenario somewhat different is RS2 with $V \gg \sigma_0$ (RS2-B, in Table~\ref{tab:alpharesults} and in Fig.~\ref{fig:RS2quadraticalphaattractorfit}). In this case, the relevant parameters to be fit are not $M$ and $M_5$ but rather their ratio $R = M/M_5$ and one of the other two scales. For this reason, in the fourth column of Table~\ref{tab:alpharesults} we present the value of $M$ as $M = (\bar R \pm \Delta R) \times M_5$. Unlike the other cases, the data allow any value of $M$ as long as $R \sim 5 \times 10^{-3}$. A lower bound on $M_5$ is obtained at 2$\sigma$ due to $\alpha$, the only physical observable that does not depend only on the ratio $M/M_5$. On the other hand, the condition $V(\phi_{\rm end})/\sigma_0 \gg 1$ in Eq.~\eqref{eq:Vsigmaconditions} gives an upper bound on $M_5$ that, using the best fit value on the ratio $M/M_5$, $\bar R \sim 5 \times 10^{-3}$, translates into $M_5$ ``much lower than''  $0.02 \, M_{\rm P}$. 

We have checked that it is not possible to obtain $\bar N_\star \sim 60$ using the latest dataset when considering a modified $\alpha$-attractor potential, with $p > 2$. We have thus performed a fit with a flat prior $N_\star \in [40,70]$ in all scenarios, finding that for all extra-dimensional scenarios, we may lower $\bar N_\star$ to the upper limit of the flat prior, $\bar N_\star = 70$, with a moderate increase of the minimum value of the $\chi^2$: we get $\chi^2_{\rm min} \sim 0.5$ without an upper bound on $N_\star$ and $\chi^2_{\rm min} \sim 0.8$ adding the prior. This is not the case for 4D, for which adding the prior worsens the $\chi^2$, whose minimum value becomes $\chi^2_{\rm min} = 1.12$.  As we have found, extra-dimensional models seem to be better suited to lower $\bar N_\star$ down to 70 due to the additional freedom they have in fixing the scale of the model with respect to 4D. Between the extra-dimensional models, RS2-B is the only one that allows for a varying inflaton mass, as long as the ratio with $M_5$ is kept fixed.

\section{Conclusions} \label{sec:conclusions}
In this work, we have studied inflation in the context of extra-dimensional scenarios, in two different frameworks: the so-called Dark Dimension (DD), {\em i.e.}, one flat extra-dimension with length $L = 2 \pi\, r_c$, with $r_c$ the compactification radius, with one brane located at $y = 0$, with the ultimate goal of explaining the observed smallness of the 4D cosmological constant $\Lambda_4$; and the Randall-Sundrum (RS) model, with an anti-de Sitter 5D metric. In the last case, we considered two options, commonly known as RS1 (two branes located at $y = 0$ and $y = \pi\, r_c$, the latter known as the IR brane) and RS2 (one brane at $y = 0$ and another located at infinity). We assumed that the inflaton field $\phi$ is confined to one brane (the IR one, in the case of RS1), as well as all SM particles, and studied two different inflationary models, the {\em monomial inflation} and the {\em $\alpha$-attractor inflation} (in its simplest version, with $p = 2$). 

In addition to reviewing the main features of each framework (Section~\ref{sec:extraD}), we have computed the 5D Einstein's equations for each scenario and, from there, got the Friedmann equation and the Hubble parameter for each case (Section~\ref{sec:Friedmann}). Confirming previous results that can be found in the literature, we have obtained that 5D evolution 
leads to a correction of the Hubble parameter on the brane, $H^2$, which is proportional to $\rho^2/\sigma$ (being $\rho$ the matter density on the brane and $\sigma$ the tension of the brane, either $\sigma_0$ or $\sigma_\pi$ depending on the extra-dimensional scenario considered), differently from the 4D case in which $H^2 \propto \rho$. In all scenarios in which the inflaton potential, $V$, is much smaller than the tension of the brane $\sigma$, the standard evolution is recovered and $H^2 \propto V [1 + {\cal O}(V/\sigma)]$ (taking $\rho \sim V$ during the inflationary phase). For this reason, results in the DD, RS1 and RS2 with $V \ll \sigma_0$ extra-dimensional scenarios are quite similar to those obtained in 4D. The only case in which a genuine non-standard evolution is found is in the RS2 with $V \gg \sigma_0$ scenario that, therefore, is always treated separately from the rest.

We then computed the SR inflationary parameters in 4D and 5D for each scenario (Section~\ref{sec:inflationBrane}), and contrasted their predictions with the Planck collaboration, BICEP, and ACT data. The results for monomial inflation have been given in Section~\ref{Monomial inf}, whereas those for the $\alpha$-attractor model are given in Section~\ref{sec:alpha}. 

In the case of monomial inflation, we have found that no considered scenario gives a very good fit to the observational data (neither the old set~\cite{Planck:2018jri, BICEP:2021xfz} nor the newest one~\cite{Planck:2018jri, BICEP:2021xfz, AtacamaCosmologyTelescope:2025blo, AtacamaCosmologyTelescope:2025nti}). In all scenarios, the preferred value of $N_\star$ needed to fit $n_s - 1$ is usually below 30 (for the oldest dataset) or below 40 (for the newest dataset), {\em i.e.} too small for a healthy post-inflationary cosmological evolution. Eventually, the preferred value of $n$ is very small, typically $n \leq 0.1$. For this choice of $n$, it is not really meaningful to talk about a monomial potential: it is indeed a constant potential with a very slow growth proportional to $n \times  \ln (\phi/\Lambda_I)$. Comparing the fit with the newest dataset with that of the oldest one, the latter usually gives a slightly better value for $\chi^2_{\rm min}$ due to the experimental value of the running of the spectral index, $\alpha_{\rm exp}$. In the oldest dataset, indeed, $\alpha_{\rm exp}$ has the same sign as $(n_s - 1)_{\rm exp}$, as theoretically expected. On the other hand, for the latest dataset we have a positive $\alpha_{\rm exp}$ and a negative $(n_s - 1)_{\rm exp}$.

The case of the $\alpha$-attractor inflationary potential is different: the fit is quite good for all scenarios, as can be seen in the second column of Table~\ref{tab:alpharesults}. However, the results change significantly using the oldest dataset~\cite{Planck:2018jri, BICEP:2021xfz} or the newest~\cite{Planck:2018jri, BICEP:2021xfz, AtacamaCosmologyTelescope:2025blo, AtacamaCosmologyTelescope:2025nti}. In the first case, $N_\star \sim 57$ for all scenarios, whereas adding ACT data we get $N_\star \sim 77$ for all scenarios. The mass of the inflaton $M$ also changes: from $M \sim 1.5 \times 10^{13}$~GeV using Planck and BICEP, only, to $M \sim 1.1 \times 10^{13}$~GeV adding ACT. Regarding the inflaton mass, the case of the RS2 scenario with $V \gg \sigma_0$ differs: in this case, the data fix the ratio $M/M_5$, giving $M/M_5 \sim 0.005$. Therefore, any value of the inflaton mass is allowed, as long as the ratio with the fundamental scale $M_5$ is kept fixed. We have tried to add a flat prior to the fit using the newest dataset, including the constraint $N_\star \in [40,70]$, in order to comply with expectations for the post-inflationary cosmological evolution~\cite{Liddle:2003as, Martin:2013nzq}. In this case, we have found that the 4D scenario is not able to cope with the constraint and the fit worsens significantly. On the other hand, all extra-dimensional scenarios can adjust the fit in order to lower $N_\star$ to $N_\star \sim 70$, using the additional freedom guaranteed by changing the fundamental scale $\Lambda_I$. 

It is worth to point out in the case of RS1 that since cosmological data can only be fitted by an effective scale $\Lambda \geq 10^{13}$~GeV, far above the TeV scale, within this scenario the solution of the hierarchy problem seems incompatible with the inflaton located at the IR brane.

In summary, our analysis shows that extra-dimensional realizations of inflation provide a consistent and, in some cases, phenomenologically advantageous framework when compared to standard 4D cosmology. Although monomial inflation is strongly disfavored by current observations in all considered setups, as they give way too small values of $\bar N_\star$ and unpleasantly small values of $\bar n$
when fitting the data, $\alpha$-attractor models remain robust and compatible with data, with extra-dimensional scenarios offering additional flexibility in accommodating observational constraints, especially on the number of $e$-folds. In particular, the RS2-B (high-energy) regime highlights how genuinely non-standard cosmological evolution can leave observable imprints, while the DD and RS1 scenarios smoothly recover the 4D limit. These results reinforce the relevance of extra dimensions as a viable extension of inflationary cosmology and motivate further investigations into their interplay with precision cosmological data.

\acknowledgments
The authors acknowledge useful discussions with D. Figueroa, S. Pastor, J. G. Rosa, and V. Sanz.
NB received funding from the grants PID2023-151418NB-I00 funded by MCIU/AEI/10.13039/501100011033/ FEDER and PID2022-139841NB-I00 of MICIU/AEI/ 10.13039/501100011033 and FEDER, UE. CC is supported by the FCT - Funda\c{c}\~{a}o para a Ci\^{e}ncia e Tecnologia, I.P. project Grant No. IN1234CEECINST/00099/2021 and through the FCT projects UID/04564/2025, and 2024.00252.CERN, with DOI identifiers 10.54499 /UID/04564/2025 and 10.54499/2024.00252.CERN, respectively. CC was previously supported by the Generalitat Valenciana APOSTD2022 Grant No. CIAPOS/2021/170. This project has received funding/support from the European Union's Horizon 2020 research and innovation program under the Marie Skłodowska-Curie grant agreement No 860881-HIDDeN and by the Spanish Agencia Estatal de Investigaci\'on AEI projetcs PID2020-113644GB-I00, PID2022-137268NA-C55, PID2023-148162NB-C21 and the Severo Ochoa project MCIU/AEI CEX2023-001292-S. We also acknowledge the funding of the Generalitat Valenciana, through grants PROMETEO/19/083, first, and CIPROM/2022/069 after.

\appendix

\section{Reconciling the Einstein equations with the 5D-action}
\label{sec:EEqand5D}
It is not completely straightforward to derive the Einstein equations, and consequently, the Friedmann equation, starting from the 5D action. This is because in the literature several assumptions are made and implicitly understood by everyone except for the newcomers to the field. Let us start with the RS case, in which we have introduced the action in Eq.~\eqref{eq:RSaction}. Integrating the cosmological constant term over the extra-dimension, assuming the metric in Eq.~\eqref{eq:metricRS}, we get: 
\begin{equation}
    - 2 \left(M_5^{\rm RS}\right)^3 \, \Lambda_5^{\rm RS} \,\int_0^{\pi\, r_c} dy \sqrt{g^{(5)}(x,y)} = - \frac{ \left(M_5^{\rm RS}\right)^3 \, \Lambda_5^{\rm RS}}{2 k}\, \left ( 1 - e^{-4 \pi k r_c}\right ).
    \label{eq:cosmocondition1}
\end{equation}
These two terms must be canceled by the corresponding brane tensions for the background metric to be a solution of the Einstein equations on the whole interval $y \in [- \pi\, r_c, \pi\, r_c]$. This requirement fixes the brane tensions: 
\begin{equation}
    \sum_{i=1,2} \int_0^{\pi\, r_c} dy \, \delta (y - y_i) \sqrt{- g^{(4)}(x,y)} \left [ - \sigma_i + \dots \right ] = \frac{\sigma_{\rm IR}}{2} e^{-4 \pi k r_c} + \frac{\sigma_{\rm UV}}{2} + \dots \, ,
    \label{eq:cosmocondition2}
\end{equation}
where the factor $1/2$ in the two brane terms comes from the fact that both $\delta$ functions are computed at the boundaries of the integration region and the labels IR- and UV- refer to the phenomenology of fields localized on them. In the RS model, to solve the hierarchy problem, SM fields are localized on the brane at $y = \pi\, r_c$, which is consequently called the IR-brane. If not for the possible matter content, the role played by the two branes is identical. For the brane terms to cancel the cosmological constant terms, we have: 
\begin{equation}
    \sigma_{\rm IR} = - \sigma_{\rm UV} = - \frac{\left(M_5^{\rm RS}\right)^3 \Lambda_5^{\rm RS}}{k} \, .
\end{equation}
Once these terms cancel the bulk cosmological constant, we can solve the Einstein equations to get Eq.~\eqref{k curvature RS1} (or Eq.~\eqref{k curvature RS2}), from which in turn we get Eq.~\eqref{eq:branetensionfinetuningRS}. 

Until now, everything seems to be clear enough. Unfortunately, the Einstein equations for the RS model are described by the usual RS action:
\begin{equation}
    S_\text{RS}=  - \frac{1}{ \kappa^2_5} \, \int 
    d^{4}x\int_0^{\pi\, r_c} dy\, \sqrt{g^{(5)}} \, \left[R^{\left(5\right)} + 2 \, \Lambda_{5}^{\rm RS}\right] \, ,
    \label{eq:RSactionAppA1}
\end{equation}
where $\kappa^2_5 = 8 \pi G^{(5)}_{\rm N}/c^4 = 1/\left(M_5^{\rm RS}\right)^3$, are not those given in Eq.~(\ref{eq:Einstein}). Instead, let us start with the following action:
\begin{equation}
    S_\text{RS}=  - \frac{1}{ \tilde \kappa^2_5} \, \int d^{4}x\int_0^{\pi\, r_c} dy\, \sqrt{g^{(5)}} \, \left[R^{\left(5\right)} + 2 \, \beta \Lambda_{5}^{\rm RS}\right],
    \label{eq:RSactionAppA2}
\end{equation}
where $\tilde \kappa^2_5 = 8 \pi G^{(5)}_{\rm N}/c^4 = \alpha/\left(M_5^{\rm RS}\right)^3$. The equations of motion for the metric that we derive from Eq.~\eqref{eq:RSactionAppA2} are (see Ref.~\cite{Shiromizu:1999wj}):
\begin{equation}
    \left (\frac{D}{2} - 1 \right ) G_{MN} + \tilde \Lambda_5^{\rm RS} g^{(5)}_{MN} = \tilde \kappa_5^2 \, T_{MN} \, ,
\end{equation}
where $D$ stands for the number of dimensions and $\tilde \Lambda_5^{\rm RS} = \beta \Lambda_5^{\rm RS}$. In order to get the 4D Einstein equations of Eq.~(\ref{eq:Einstein}), widely adopted through the RS brane cosmology literature (see, {\em e.g.}, Refs.~\cite{Langlois:2000ns, Langlois:2002bb}), we must rescale the dimensionful parameters. Let us introduce the following undetermined factors: 
\begin{equation} \label{eq:newscalesRS}
    \begin{dcases}
        \Lambda^\prime_{\rm RS} = \frac{\beta}{\frac{D}{2}-1} \, \Lambda_5^{\rm RS} \, , \\
        \kappa_5^2 =  \frac{1}{\frac{D}{2}-1} \, \frac{\alpha}{M_{5,\rm RS}^3} \, , \\
        \tilde \sigma_{\rm IR} = \gamma \, \sigma_{\rm IR} \, ,
    \end{dcases}
\end{equation}
where $\Lambda_5^{\rm RS}$, $\left(M_5^{\rm RS}\right)^3$ and $\sigma_{\rm IR}$ are the parameters that appear in Eqs.~(\ref{eq:RSaction}) and~\eqref{eq:RSbraneterms RS1} (or Eq.~\eqref{eq:RSbraneterms RS2}). Once this rescaling is introduced, we have:
\begin{equation}
    G_{MN} + \Lambda^\prime_{\rm RS} \, g^{(5)}_{MN} = \kappa_5^2 \, T_{MN} \, ,
\end{equation}
that formally has the same aspect as the 4D Einstein equations. Recall that, using this equation, we can go through Sections~\ref{sec:RS2cosmo} and \ref{sec:RS1cosmo} and get the following definition for the 4D cosmological constant: 
\begin{equation}
    \Lambda_4 = \frac{1}{36} \kappa_5^4 \, \tilde \sigma_{\rm IR}^2 + \frac{1}{6} \, \Lambda^\prime_{\rm RS} = \frac{1}{\left ( \frac{D}{2} - 1 \right )^2}\frac{(\alpha \gamma )^2}{36 \, \left(M_5^{\rm RS}\right)^6} \sigma^2_{\rm IR} + \frac{1}{\left ( \frac{D}{2} - 1\right ) } \, \frac{\beta}{6} \, \Lambda_5^{\rm RS} \, .
\end{equation}
If we plug into this equation the values for $\sigma_{\rm IR}$ and $\Lambda_5^{\rm RS}$ from Section~\ref{sec:extraD}, that is, $\Lambda_5^{\rm RS} = - 6 k^2$ and $\sigma_{\rm IR} = - 6 k \, \left(M_5^{\rm RS}\right)^3$, we get: 
\begin{equation}
    (\alpha\, \gamma )^2 = \beta \left ( \frac{D}{2} - 1 \right ) .
\end{equation}
Deriving Eqs.~\eqref{eq:cosmocondition1} and~\eqref{eq:cosmocondition2} for the action in Eq.~\eqref{eq:RSactionAppA2}, the condition that the brane terms cancel the 5D cosmological constant contribution to the 5D action gives a second constraint between the three rescaling factors: 
\begin{equation}
    \alpha\, \gamma = \beta \, .
\end{equation}
Putting together the two constraints, we get: 
\begin{equation}   
    \beta = \frac{D}{2} - 1 = \alpha\, \gamma\, .
\end{equation}
As we can see, we can always fix one of the three rescaling factors to 1, as long as the other two are fixed properly. Eventually, the correct action to be used in Sections~\ref{sec:RS2cosmo} and~\ref{sec:RS1cosmo} is: 
\begin{equation}
    S_\text{RS}=  - \frac{M_{5,\rm RS}^3}{\frac{D}{2} - 1} \, \int 
    d^{4}x\int_0^{\pi\, r_c} dy\, \sqrt{g^{(5)}} \, \left[R^{\left(5\right)} + 2 \, \left ( \frac{D}{2} - 1 \right ) \,  \Lambda_{5}^{\rm RS}\right],
    \label{eq:RSaction3}
\end{equation}
where the brane term is as in Eq.~\eqref{eq:RSbraneterms RS1} (or Eq.~\eqref{eq:RSbraneterms RS2}), since we can choose $\gamma = 1$.

\section{Time-derivatives of the Hubble parameter}
\label{sec:AppHdot}
In this Appendix, we compute explicitly the time derivatives of the Hubble parameter for the different scenarios considered in the main text, taking into account the different dependence on the inflaton potential. 

We get for the single time derivative
\begin{equation} \label{eq:firstHtimederivative}
    \dot H = - \frac{1}{6} \, \frac{\left ( V^\prime \right )^2}{V} \times
    \begin{dcases}
        1  &{\rm for \; 4D, RS1} \, , \\
        \left ( \frac{V}{M_{\rm P}^2 \Lambda_4 + V} \right ) &{\rm for \; DD} \, , \\
        \frac{\left ( 1 + V/\sigma_0 \right ) }{\left (1 + V/ 2\sigma_0 \right ) } &{\rm for \; RS2} \, ,
    \end{dcases}
\end{equation}
where we used the SR conditions $\rho \sim V(\phi)$ and $V^\prime (\phi) \sim - 3 H \dot \phi$. In RS2, we have two interesting limits: 
\begin{equation}
    \dot H = - \frac{1}{6} \, \frac{\left ( V^\prime \right )^2}{V} \times 
    \begin{dcases}
        1 + \frac{1}{2} \frac{V}{\sigma_0} + \dots \qquad {\rm for} \; V \ll \sigma_0\,, \\
        2 \left ( 1 - \frac{\sigma_0}{V} + \dots \right ) {\rm for} \; V \gg \sigma_0\,.
    \end{dcases}
\end{equation}

The second time derivative of $H$ is:
\begin{equation} \label{eq:secondHtimederivative}
    \ddot H = \frac13 \sqrt{ \frac{M_{\rm P}^2}{3}}\, \frac{{V'}^2}{V^\frac32} \times
    \begin{dcases}
        \left [ V^{\prime \prime} - \frac{1}{2}\, \frac{{V'}^2}{V}\right]  \hspace{6.3cm} {\rm for \; 4D, RS1} \, , \\
        \left ( \frac{V}{M_{\rm P}^2 \Lambda_4 + V} \right )^{3/2} \left[V^{\prime \prime} - \frac{1}{2} \frac{{V'}^2}{V} \left ( \frac{V}{M_{\rm P}^2 \Lambda_4 + V} \right) \right] \qquad 
        {\rm for \; DD} \, ,\\
        \frac{1}{\left ( 1 + \frac{V}{2 \sigma_0} \right)^{3/2}} \, \left \{
        \left [ V^{\prime \prime} - \frac{1}{2}\, \frac{ \left ( V' \right )^2}{V}\right] \left(1 + \frac{V}{\sigma_0}\right) 
        + \frac{V}{4 \sigma_0} \, \frac{ \left (V' \right)^2}{V}\, \frac{1}{\left(1 + \frac{V}{2 \sigma_0}\right)} 
        \right \} \\
        \hspace{8.7cm}{\rm for \; RS2} \, .
    \end{dcases}
\end{equation}
The third time-derivative of $H$ is more complicated in all scenarios. For the standard 4D case and for the two-brane RS scenario, we have:
\begin{equation} \label{eq:thirdHtimederivative4DRS1}
    \dddot H_{\rm 4D} = - \frac{M_{\rm P}^2}{9} \left( \frac{ V^\prime }{V} \right )^2 \left [ V^\prime \, V^{\prime \prime \prime} + 2 \left ( V^{\prime \prime} \right )^2 - \frac{7}{2} \frac{\left ( V^\prime \right )^2}{V} \, V^{\prime\prime} + \frac{5}{4} \frac{ \left ( V^\prime \right )^4}{V^2} \right] \qquad {\rm for \; 4D, RS1,} 
\end{equation}
where we call the expression $\ddot H_{\rm 4D}$ for simplicity. On the other hand, the constant term proportional to the 4D cosmological constant $\Lambda_4$ makes the expression a bit more involved:
\begin{align} \label{eq:thirdHtimederivativeDD}
    \dddot H_{\rm DD} &= - \frac{M_{\rm P}^2}{9} \left ( \frac{ V^\prime }{V} \right )^2 \, \left (\frac{V}{ M_{\rm P}^2 \, \Lambda_4 + V }\right )^2 \Bigg\{ V^\prime V^{\prime \prime \prime} + 2 \left ( V^{\prime \prime }\right )^2 \nonumber\\
    &\qquad - \frac{7}{2} \, \left (\frac{V}{ M_{\rm P}^2 \, \Lambda_4 + V }\right ) \, \frac{\left ( V^\prime\right )^2}{V} \, V^{\prime \prime} + \frac{5}{4} \, \left (\frac{V}{ M_{\rm P}^2 \, \Lambda_4 + V }\right )^2 \, \frac{\left ( V^\prime \right )^4}{V^2}\Bigg\} \qquad {\rm for \; DD} \, .
\end{align}
Notice that, when $V$ is as large as its maximum ($V \sim M_5^4$ for this particular scenario), the constant term can be safely neglected and the expression reduces to that of the 4D and the RS1 scenarios.  For lower values of $V$, whilst up to $V \gg M_{\rm P}^2 \, \Lambda_4$, we can expand:
\begin{align} \label{eq:thirdHtimederivativeDDapprox}
    \dddot H_{\rm DD} &= \dddot H_{\rm 4D}+ \frac{2 M_{\rm P}^2}{9} \left(\frac{V^\prime }{V}\right)^2 \frac{M_{\rm P}^2 \, \Lambda_4}{V} \left[V' \, V''' + 2 {V'''}^2 - \frac{21}{4} \frac{{V'}^2}{V} V'' + \frac52 \frac{{V'}^4}{V^2}\right] + {\cal O}\left(\frac{M_{\rm P}^4 \Lambda_4^2}{V^2}\right).
\end{align}
In the case of the single brane RS scenario, the expression is rather cumbersome:
\begin{align} \label{eq:thirdHtimederivativeRS2}
    \dddot H_{\rm RS2} &= - \frac{M_{\rm P}^2}{9} \, \left ( \frac{ V^\prime }{V} \right )^2 \, \frac{1}{\left ( 1 + \frac{V}{2 \sigma_0} \right )^2} \Bigg\{ \left [ V^\prime V^{\prime \prime \prime} + 2  \left ( V^{\prime \prime }\right )^2 \right ] \, \left ( 1 + \frac{V}{\sigma_0} \right ) \nonumber\\
    &\quad+ \left [ - \frac{7}{2} \, \frac{1}{\left ( 1 + \frac{V}{2 \sigma_0} \right )} - 4 \frac{V}{\sigma_0}\right ] \frac{\left ( V^\prime\right )^2}{V} \, V^{\prime \prime} + \frac{\left ( 1 + \frac{V}{\sigma_0}\right )}{\left ( 1 + \frac{V}{2 \sigma_0}\right )} \, \left [ \frac{5}{4} \, \frac{1}{\left ( 1 + \frac{V}{2 \sigma_0} \right )}+ \frac{3}{4} \, \frac{V}{\sigma_0}\right ] \, \frac{\left ( V^\prime \right )^4}{V^2} \Bigg\}
 \end{align}
for RS2. In order to derive these expressions, we have consistently neglected terms proportional to $\ddot \phi$ and $\dddot \phi$, assuming that within the SR paradigm they should be irrelevant.

It can be easily seen that Eq.~\eqref{eq:thirdHtimederivativeRS2} reduces to Eq.~\eqref{eq:thirdHtimederivative4DRS1} in the limit $V \ll \sigma_0$. Leading corrections to the standard scenario are:
\begin{equation} \label{eq:thirdHtimederivativeRS2smallV}
    \left . \dddot H_{\rm RS2} \right |_{V \ll \sigma_0} \simeq \dddot H_{\rm 4D} - \frac{M_{\rm P}^2}{9} \,  \left ( \frac{ V^\prime }{V} \right )^2 \,  \left ( \frac{V}{\sigma_0} \right ) \,  \left \{   \frac{5}{4} \, \frac{\left ( V^\prime \right )^2}{V} \,  V^{\prime \prime } - \frac{1}{2} \frac{\left ( V^\prime \right )^4}{V^2}\right \} + {\cal O} \left ( \frac{V^2}{\sigma_0^2}\right ).
 \end{equation}
On the other hand, for $V \gg \sigma_0$ we have: 
\begin{equation} \label{eq:thirdHtimederivativeRS2largeV}
    \left. \dddot H_{\rm RS2}\right|_{V \gg \sigma_0} \simeq -\frac49 M_{\rm P}^2 \left(\frac{V'}{V}\right)^2 \left(\frac{\sigma_0}{V}\right) \left[V'\, V''' + 2{V''}^2 - 4 \frac{{V'}^2}{V}\, V'' + \frac32 \frac{{V'}^4}{V^2} + {\cal O}\left(\frac{\sigma_0}{V}\right)\right].
\end{equation}

Once we have derived expressions as functions of the inflaton potential for the first three time derivatives of the Hubble parameter, it is easy to derive the SR parameters in Section~\ref{sec:SR}.

\section{First order corrections to the Slow-Roll parameters for DD}
\label{sec:AppSRDD}
In this appendix, we present our results for the SR parameters when a constant proportional to $\Lambda_4$ is added to the Friedmann equation (as in the case of the DD scenario). In principle, this constant should be neglected (as is commonly done in the 4D standard cosmology). However, explicit expressions in the presence of a constant for the SR parameters may be useful in the case of a time-varying cosmological constant. 

The first SR parameter is: 
\begin{equation} \label{firstSR DD}
    \epsilon_V \simeq \frac{M_{\rm P}^2}{2} \left(\frac{V'}{V}\right)^2 \, 
        \left(\frac{V}{M_\mathrm{P}^2\, \Lambda_4 + V}\right)^2 
        \longrightarrow 
        \frac{M_{\rm P}^2}{2} \left(\frac{V'}{V}\right)^2 \,
        \left ( 1 - 2 \frac{M_{\rm P}^2 \, \Lambda_4}{V} + \dots \right ) \,,
\end{equation}
the second SR parameter is:
\begin{equation} \label{second SR DD}
    \eta_V = M_{\rm P}^2\, \left ( \frac{V''}{V} \right )
    \, 
        \left ( \frac{V}{M_{\rm P}^2\, \Lambda_4 + V} \right ) 
        \longrightarrow 
        M_{\rm P}^2\, \left ( \frac{V''}{V} \right )
    \, \left (
        1 - \frac{M_{\rm P}^2 \, \Lambda_4}{V} + \dots \right )
        \,,
\end{equation}
and finally, for the third SR parameter, we get:
\begin{equation} \label{third SRV DD}
    \xi_V^2 \equiv M_{\rm P}^4\, \left ( \frac{V'\, V'''}{V^2} \right ) \, 
        \left ( \frac{V}{M_{\rm P}^2 \, \Lambda_4 + V}\right )^2
        \longrightarrow 
        M_{\rm P}^4\, \left ( \frac{V'\, V'''}{V^2} \right ) \,
        \left ( 
        1 - 2 \, \frac{M_{\rm P}^2 \, \Lambda_4}{V} + \dots 
        \right )  \, . 
\end{equation}

The amplitude of the scalar perturbation is, in this case: 
\begin{equation}
    \Delta_s^2 \simeq \frac{1}{24 \pi^2\, \epsilon_V(\phi_\star)} \frac{V_\star}{M_{\rm P}^4} \, \left ( \frac{M_{\rm P}^2 \, \Lambda_4 + V_\star}{V_\star}\right ) \, , 
\end{equation}
where the spectral index has the standard expression: 
\begin{equation}
    n_s - 1 = 2 \eta_V (\phi_\star) - 6 \epsilon_V(\phi_\star) \, , 
\end{equation}
although in terms of the modified SR parameters, Eqs.~(\ref{firstSR DD}) and~\eqref{second SR DD}. The running of the spectral index gives, in the presence of a constant term:
\begin{equation}
    \alpha = \frac{d n_s}{d \ln k} = \left (\frac{M_{\rm P}^2 \Lambda_4 + V_\star}{V_\star} \right ) \, \left [ - 2 \xi_V^2 (\phi_\star) + 16 \epsilon_V (\phi_\star) \, \eta_V (\phi_\star)- 24 \epsilon_V^2 (\phi_\star) \right ] \ .
\end{equation}

\section{\boldmath First order corrections to physical observables for RS2 with $V \ll \sigma_0$}
\label{sec:FOcorrectionsRS2small}
We give here the expressions for the SR parameters and the observables in the scenario RS2 in the limit $V \ll \sigma_0$, up to first order in $V/\sigma_0$, for the two inflationary models.  

\subsection{Monomial inflation}
\label{sec:FOcorrectionsMonomial}
We have: 
\begin{equation}
    \left \{
    \begin{array}{l}
        \epsilon_V(\phi) = \frac{n^{2}}{2}\,\left(\frac{M_{\mathrm{P}}}{\phi}\right)^{2} \left \{ 1 + {\cal O} \left [\left ( \frac{V(\phi)}{\sigma_0} \right )^2\right ]\right \} \, , \\
        \\
        \eta_{V}\left(\phi\right)=n\,(n-1)\left(\frac{M_{{\rm P}}}{\phi}\right)^{2}\left [1+\frac{\lambda}{12\,(n-1)}\,\left(\frac{M_{\mathrm{P}}}{M_{5}}\right)^{2}\left(\frac{\phi}{M_{5}}\right)^{n}\right] \, , \\
        \\
        \xi_V^2(\phi) = n^2\,(n-1)\,(n-2)\, \left(\frac{M_{{\rm P}}}{\phi}\right)^{4} \left[1 + \frac{\lambda}{24} \, \frac{5 n - 8}{(n-1) \, (n-2)} \, \left(\frac{M_{\mathrm{P}}}{M_5}\right)^2 \, \left(\frac{\phi}{M_{5}}\right)^n \right] \, .
    \end{array}
    \right .
\end{equation}

Since $\epsilon_V(\phi)$ is modified only at the second order in $V/\sigma_0$, $\phi_{\rm end}$ and $\phi_\star$ are the same as in Eq.~(\ref{eq:phiendphistarmonomial}):
\begin{equation}
    \left \{
    \begin{array}{l}
        \phi_{\rm end} = \frac{n}{\sqrt{2}} \, M_{\rm P}, \\
        \\
        \phi_\star = \left [ 2 \, n \, N_\star + \frac{n^2}{2}\right ]^{1/2} \, M_{\rm P}.
    \end{array}
    \right .
\end{equation}
In turn, the physical observables become: 
\begin{equation}
    \left \{
    \begin{array}{l}
        n_s - 1 = -\frac{(n+2)}{2\,N_{\star}}\left [1 + \frac{\lambda}{6} \, \left(\frac{M_{\rm P}}{M_5}\right)^{n+2}\, \frac{(n-2)\,(n-1)\,\left(2 \, n\,N_\star\right)^{n/2}}{(n+2)}\right ] \, , \\
        \\
        \alpha = -\frac{(n+2)}{2\,N_\star^2}\left [1- \frac{\lambda}{12}\left (\frac{M_{\rm P}}{M_5}\right)^{n+2} \, \frac{(n-2) \, (n-1) \,\left ( 2 \, n \, N_\star \right)^{n/2}}{(n+2)}\right ] \, , \\
        \\
        \Delta_s^2 = \frac{\lambda}{3\pi^{2}}\,(2\,n)^{\frac{n-2}{2}}\,N_{\star}^{\frac{n+2}{2}}\,\left(\frac{M_{{\rm P}}}{M_{5}}\right)^{n-4}\,\left [1 + \lambda \, \left(\frac{M_{{\rm P}}}{M_{5}}\right)^{n+2}(2 \, n\,N_{\star})^{\frac{n-2}{2}}\right ] \, , \\
        \\
        r = \frac{4\,n}{N_{\star}}\left [ 1- \frac{\lambda}{6} \, \left(\frac{M_{\rm P}}{M_5}\right)^{n+2}\,\left(2 \, n \,N_\star\right)^{n/2}\right ] \, , 
    \end{array}
    \right .
\end{equation}
with the additional constraint: 
\begin{equation}
    \frac{V(\phi_\star)}{\sigma_0} = \lambda \, \left ( \frac{M_{\rm P}}{M_5}\right )^{n+2} \, \left ( 2 n N_\star + \frac{n^2}{2}\right )^{n/2} \ll 1 \, .
\end{equation}

\subsection[$\alpha$-attractor model]{\boldmath $\alpha$-attractor model}
\label{sec:FOcorrectionsalpha}
We can compute the same quantities in the $\alpha$-attractor inflationary model. We get for the RS2 extra-dimensional scenario at first order in $V/\sigma_0$, for $V/\sigma_0 \ll 1$, the following SR parameters: 
\begin{equation}
    \left \{
    \begin{array}{l}
        \epsilon_V = 8\,\left(\frac{M_{\mathrm{P}}}{\Lambda_{I}}\right)^{2}\,\text{csch}^{2}\left(\frac{2\,\phi}{\Lambda_{I}}\right)\left[1-\frac{M^{4}\,M_{\mathrm{P}}^{4}\tanh\left(\frac{\phi}{M_{5}}\right)}{36\,M_{5}^{8}}\right],\\
        \\
        \eta_V = -4\,\left(\frac{M_{{\rm P}}}{M_{5}}\right)^{2}\,\text{sech}^{2}\left(\frac{\phi}{M_{5}}\right)\,\left[1-\frac{1}{2}\text{csch}^{2}\left(\frac{\phi}{M_{5}}\right)\right]\,\left[1+\frac{M^{2}\,M_{\mathrm{P}}^{2}\tanh\left(\frac{\phi}{M_{5}}\right)}{24\,M_{5}^{4}}\right],\\
        \\
        \xi_V^2=16 \left(\frac{M_{{\rm P}}}{M_{5}}\right)^4 \text{sech}^4 \left(\frac{\phi}{M_5}\right) \left[1 - 2\text{csch}^2 \left(\frac{\phi}{M_5}\right)\right] \left[1 + \frac{M^2 M_{\mathrm{P}}^2 \left[\text{csch}^2 \left(\frac{\phi}{M_5}\right) + 9 \text{sech}^2 \left(\frac{\phi}{M_5}\right) - 8\right]}{96 M_5^4 - 192 M_5^4 \text{csch}^2 \left(\frac{\phi}{M_5}\right)}\right]. 
    \end{array}
    \right .
\end{equation}
The values of $\phi_{\rm end}$ and $\phi_\star$ are the same as in Eqs.~(\ref{eq:phiendalphaattractor}) and (\ref{eq:phistaralphaattractor}). The physical observables are:
\begin{equation}
    \left \{
    \begin{array}{l}
        n_{s}-1=-8\left(\frac{M_{{\rm P}}}{M_{5}}\right)^{2}\,\text{csch}^{2}\left(\frac{\phi_{\star}}{M_{5}}\right)\left\{ 1-\frac{M^{2}M_{\mathrm{P}}^{2}\left[\cosh\left(\frac{2\phi_{\star}}{M_{5}}\right)-2\right]\,\text{sech}^{4}\left(\frac{\phi_{\star}}{M_{5}}\right)}{48\,M_{5}^{4}\,\text{csch}^{2}\left(\frac{\phi_{\star}}{M_{5}}\right)}\right\}, \\
        \\
        \alpha=-32\left(\frac{M_{{\rm P}}}{M_{5}}\right)^{4}\,\text{csch}^{4}\left(\frac{\phi_{\star}}{M_{5}}\right)\left\{ 1-\frac{M^{2}\,M_{\mathrm{P}}^{2}\left[\cosh\left(\frac{2\phi_{\star}}{M_{5}}\right)-5\right]\,\text{sech}^{6}\left(\frac{\phi_{\star}}{M_{5}}\right)}{48\,M_{5}^{4}\,\text{csch}^{4}\left(\frac{\phi_{\star}}{M_{5}}\right)}\right\}, \\
        \\
        \Delta_s^2  =  \frac{1}{4\,\pi^{2}}\left[\frac{M}{\sqrt{24}\,M_{5}}\right]^{2}\,\left(\frac{M_{5}}{M_{{\rm P}}}\right)^{6}\,\sinh^{4}\left(\frac{\phi_{\star}}{M_{5}}\right)\left[1+\frac{\left(\frac{M}{M_{\mathrm{5}}}\right)^{2}\left(\frac{M_{\mathrm{P}}}{M_{\mathrm{5}}}\right)^{2}\,\tanh^{2}\left(\frac{\phi_{\star}}{M_{5}}\right)}{8}\right],\\
        \\
        r = \frac{2}{N_\star^2} \, \left ( \frac{M_5}{M_{\rm P}}\right )^2 \, \left [ 1 + \frac{1}{\sqrt{2} \, N_\star} \, \left (\frac{M_5}{M_{\rm P}} \right) \, \sqrt{1 + \frac{1}{8} \left ( \frac{M_5}{M_{\rm P}} \right)^2} + \frac{1}{8 \, N_\star^2} \, \left ( \frac{M_5}{M_{\rm P}}\right )^2\right ]^{-1} \\
        \qquad \qquad \qquad \times \left [ 1 - \frac{V(\phi_\star)}{\sigma_0} + \frac{V(\phi_\star)}{\sigma_0} \, \ln \left ( 2 \frac{\sigma_0}{V(\phi_\star)}\right )\right ]^{-1} \, .
    \end{array}
    \right .
\end{equation}

\bibliography{biblio}

@article{Arcadi:2024ukq,
    author = "Arcadi, Giorgio and Cabo-Almeida, David and Dutra, Ma{\'\i}ra and Ghosh, Pradipta and Lindner, Manfred and Mambrini, Yann and Neto, Jacinto P. and Pierre, Mathias and Profumo, Stefano and Queiroz, Farinaldo S.",
    title = "{The Waning of the WIMP: Endgame?}",
    eprint = "2403.15860",
    archivePrefix = "arXiv",
    primaryClass = "hep-ph",
    doi = "10.1140/epjc/s10052-024-13672-y",
    journal = "Eur. Phys. J. C",
    volume = "85",
    number = "2",
    pages = "152",
    year = "2025"
}

@article{Nordstrom:1914ejq,
    author = "Nordström, Gunnar",
    title = "{On the possibility of unifying the electromagnetic and the gravitational fields}",
    eprint = "physics/0702221",
    archivePrefix = "arXiv",
    journal = "Phys. Z.",
    volume = "15",
    pages = "504--506",
    year = "1914"
}

@article{Kaluza:1921tu,
    author = "Kaluza, Th.",
    title = {{Zum Unit\"atsproblem der Physik}},
    eprint = "1803.08616",
    archivePrefix = "arXiv",
    primaryClass = "physics.hist-ph",
    reportNumber = "HUPD-8401",
    doi = "10.1142/S0218271818700017",
    journal = "Sitzungsber. Preuss. Akad. Wiss. Berlin (Math. Phys. )",
    volume = "1921",
    pages = "966--972",
    year = "1921"
}

@article{Klein:1926tv,
    author = "Klein, Oskar",
    editor = "Taylor, J. C.",
    title = "{Quantum Theory and Five-Dimensional Theory of Relativity. (In German and English)}",
    doi = "10.1007/BF01397481",
    journal = "Z. Phys.",
    volume = "37",
    pages = "895--906",
    year = "1926"
}

@article{Mohapatra:2000cm,
    author = "Mohapatra, R. N. and Perez-Lorenzana, Abdel and de Sousa Pires, Carlos Antonio",
    title = "{Inflation in models with large extra dimensions driven by a bulk scalar field}",
    eprint = "hep-ph/0003089",
    archivePrefix = "arXiv",
    reportNumber = "UMD-PP-00-062",
    doi = "10.1103/PhysRevD.62.105030",
    journal = "Phys. Rev. D",
    volume = "62",
    pages = "105030",
    year = "2000"
}

@article{Folgado:2019gie,
    author = "Folgado, Miguel G. and Donini, Andrea and Rius, Nuria",
    title = "{Gravity-mediated Dark Matter in Clockwork/Linear Dilaton Extra-Dimensions}",
    eprint = "1912.02689",
    archivePrefix = "arXiv",
    primaryClass = "hep-ph",
    reportNumber = "FTUV-19-1128.3781, IFIC/19-56",
    doi = "10.1007/JHEP04(2020)036",
    journal = "JHEP",
    volume = "04",
    pages = "036",
    year = "2020"
}

@article{Donini:2025cpl,
    author = "Donini, Andrea and Folgado, Miguel G. and Herrero-Garc{\'\i}a, Juan and Landini, Giacomo and Mu{\~n}oz-Ovalle, Alejandro and Rius, Nuria",
    title = "{Dark Matter in an evanescent three-brane Randall-Sundrum scenario}",
    eprint = "2505.13601",
    archivePrefix = "arXiv",
    primaryClass = "hep-ph",
    doi = "10.1007/JHEP11(2025)037",
    journal = "JHEP",
    volume = "11",
    pages = "037",
    year = "2025"
}

@article{Donini:2025qrf,
    author = "Donini, Andrea and Folgado, Miguel G. and Mu{\~n}oz-Ovalle, Alejandro",
    title = "{Dark Matter in a Three-Brane Randall-Sundrum Scenario out of the Evanescent Limit}",
    eprint = "2509.04580",
    archivePrefix = "arXiv",
    primaryClass = "hep-ph",
    month = "9",
    year = "2025"
}

@article{Bernal:2020yqg,
    author = "Bernal, Nicol\'as and Donini, Andrea and Folgado, Miguel G. and Rius, Nuria",
    title = "{FIMP Dark Matter in Clockwork/Linear Dilaton Extra-Dimensions}",
    eprint = "2012.10453",
    archivePrefix = "arXiv",
    primaryClass = "hep-ph",
    reportNumber = "PI/UAN-2020-***FT, FTUV-20-0908.8746, IFIC/20-43",
    doi = "10.1007/JHEP04(2021)061",
    journal = "JHEP",
    volume = "04",
    pages = "061",
    year = "2021"
}

@article{Shiromizu:1999wj,
    author = "Shiromizu, Tetsuya and Maeda, Kei-ichi and Sasaki, Misao",
    title = "{The Einstein equation on the 3-brane world}",
    eprint = "gr-qc/9910076",
    archivePrefix = "arXiv",
    reportNumber = "DAMTP-1999-150, OUTAP-103, UTAP-349, RESCEU-40-99",
    doi = "10.1103/PhysRevD.62.024012",
    journal = "Phys. Rev. D",
    volume = "62",
    pages = "024012",
    year = "2000"
}

@article{Binetruy:1999ut,
    author = "Binetruy, Pierre and Deffayet, Cedric and Langlois, David",
    title = "{Nonconventional cosmology from a brane universe}",
    eprint = "hep-th/9905012",
    archivePrefix = "arXiv",
    reportNumber = "LPT-ORSAY-99-25",
    doi = "10.1016/S0550-3213(99)00696-3",
    journal = "Nucl. Phys. B",
    volume = "565",
    pages = "269--287",
    year = "2000"
}

@article{Khoury:2001wf,
    author = "Khoury, Justin and Ovrut, Burt A. and Steinhardt, Paul J. and Turok, Neil",
    title = "{The Ekpyrotic universe: Colliding branes and the origin of the hot big bang}",
    eprint = "hep-th/0103239",
    archivePrefix = "arXiv",
    doi = "10.1103/PhysRevD.64.123522",
    journal = "Phys. Rev. D",
    volume = "64",
    pages = "123522",
    year = "2001"
}

@article{Buchbinder:2007ad,
    author = "Buchbinder, Evgeny I. and Khoury, Justin and Ovrut, Burt A.",
    title = "{New Ekpyrotic cosmology}",
    eprint = "hep-th/0702154",
    archivePrefix = "arXiv",
    doi = "10.1103/PhysRevD.76.123503",
    journal = "Phys. Rev. D",
    volume = "76",
    pages = "123503",
    year = "2007"
}

@article{Lehners:2008vx,
    author = "Lehners, Jean-Luc",
    title = "{Ekpyrotic and Cyclic Cosmology}",
    eprint = "0806.1245",
    archivePrefix = "arXiv",
    primaryClass = "astro-ph",
    doi = "10.1016/j.physrep.2008.06.001",
    journal = "Phys. Rept.",
    volume = "465",
    pages = "223--263",
    year = "2008"
}

@article{Maartens:1999hf,
    author = "Maartens, Roy and Wands, David and Bassett, Bruce A. and Heard, Imogen",
    title = "{Chaotic inflation on the brane}",
    eprint = "hep-ph/9912464",
    archivePrefix = "arXiv",
    reportNumber = "PU-RCG-99-24",
    doi = "10.1103/PhysRevD.62.041301",
    journal = "Phys. Rev. D",
    volume = "62",
    pages = "041301",
    year = "2000"
}

@article{Langlois:2002bb,
    author = "Langlois, David",
    editor = "Maeda, K. and Sasaki, M.",
    title = "{Brane cosmology: An Introduction}",
    eprint = "hep-th/0209261",
    archivePrefix = "arXiv",
    doi = "10.1143/PTPS.148.181",
    journal = "Prog. Theor. Phys. Suppl.",
    volume = "148",
    pages = "181--212",
    year = "2003"
}

@article{Lin:2018kjm,
    author = "Lin, Chia-Min and Ng, Kin-Wang and Cheung, Kingman",
    title = "{Chaotic inflation on the brane and the Swampland Criteria}",
    eprint = "1810.01644",
    archivePrefix = "arXiv",
    primaryClass = "hep-ph",
    doi = "10.1103/PhysRevD.100.023545",
    journal = "Phys. Rev. D",
    volume = "100",
    number = "2",
    pages = "023545",
    year = "2019"
}

@article{Bento:2008yx,
    author = "Bento, M. C. and Felipe, R. Gonzalez and Santos, N. M. C.",
    title = "{Brane assisted quintessential inflation with transient acceleration}",
    eprint = "0801.3450",
    archivePrefix = "arXiv",
    primaryClass = "astro-ph",
    doi = "10.1103/PhysRevD.77.123512",
    journal = "Phys. Rev. D",
    volume = "77",
    pages = "123512",
    year = "2008"
}

@article{Bassett:2005xm,
    author = "Bassett, Bruce A. and Tsujikawa, Shinji and Wands, David",
    title = "{Inflation dynamics and reheating}",
    eprint = "astro-ph/0507632",
    archivePrefix = "arXiv",
    doi = "10.1103/RevModPhys.78.537",
    journal = "Rev. Mod. Phys.",
    volume = "78",
    pages = "537--589",
    year = "2006"
}

@article{Liddle:1992wi,
    author = "Liddle, Andrew R. and Lyth, David H.",
    title = "{COBE, gravitational waves, inflation and extended inflation}",
    eprint = "astro-ph/9208007",
    archivePrefix = "arXiv",
    reportNumber = "SUSSEX-AST-92-6-1, LANC-TH-5-92",
    doi = "10.1016/0370-2693(92)91393-N",
    journal = "Phys. Lett. B",
    volume = "291",
    pages = "391--398",
    year = "1992"
}

@article{Liddle:1994dx,
    author = "Liddle, Andrew R. and Parsons, Paul and Barrow, John D.",
    title = "{Formalizing the slow roll approximation in inflation}",
    eprint = "astro-ph/9408015",
    archivePrefix = "arXiv",
    reportNumber = "SUSSEX-AST-94-8-1",
    doi = "10.1103/PhysRevD.50.7222",
    journal = "Phys. Rev. D",
    volume = "50",
    pages = "7222--7232",
    year = "1994"
}

@article{Langlois:2000ns,
    author = "Langlois, David and Maartens, Roy and Wands, David",
    title = "{Gravitational waves from inflation on the brane}",
    eprint = "hep-th/0006007",
    archivePrefix = "arXiv",
    doi = "10.1016/S0370-2693(00)00957-6",
    journal = "Phys. Lett. B",
    volume = "489",
    pages = "259--267",
    year = "2000"
}

@article{BICEP:2021xfz,
    author = "Ade, P. A. R. and others",
    collaboration = "BICEP, Keck",
    title = "{Improved Constraints on Primordial Gravitational Waves using Planck, WMAP, and BICEP/Keck Observations through the 2018 Observing Season}",
    eprint = "2110.00483",
    archivePrefix = "arXiv",
    primaryClass = "astro-ph.CO",
    doi = "10.1103/PhysRevLett.127.151301",
    journal = "Phys. Rev. Lett.",
    volume = "127",
    number = "15",
    pages = "151301",
    year = "2021"
}

@article{Kallosh:2013hoa,
    author = "Kallosh, Renata and Linde, Andrei",
    title = "{Universality Class in Conformal Inflation}",
    eprint = "1306.5220",
    archivePrefix = "arXiv",
    primaryClass = "hep-th",
    doi = "10.1088/1475-7516/2013/07/002",
    journal = "JCAP",
    volume = "07",
    pages = "002",
    year = "2013"
}

@article{Kallosh:2013yoa,
    author = "Kallosh, Renata and Linde, Andrei and Roest, Diederik",
    title = "{Superconformal Inflationary $\alpha$-Attractors}",
    eprint = "1311.0472",
    archivePrefix = "arXiv",
    primaryClass = "hep-th",
    doi = "10.1007/JHEP11(2013)198",
    journal = "JHEP",
    volume = "11",
    pages = "198",
    year = "2013"
}

@article{Planck:2018jri,
    author = "Akrami, Y. and others",
    collaboration = "Planck",
    title = "{Planck 2018 results. X. Constraints on inflation}",
    eprint = "1807.06211",
    archivePrefix = "arXiv",
    primaryClass = "astro-ph.CO",
    doi = "10.1051/0004-6361/201833887",
    journal = "Astron. Astrophys.",
    volume = "641",
    pages = "A10",
    year = "2020"
}

@article{Randall:1999ee,
    author = "Randall, Lisa and Sundrum, Raman",
    title = "{A Large mass hierarchy from a small extra dimension}",
    eprint = "hep-ph/9905221",
    archivePrefix = "arXiv",
    reportNumber = "MIT-CTP-2860, PUPT-1860, BUHEP-99-9",
    doi = "10.1103/PhysRevLett.83.3370",
    journal = "Phys. Rev. Lett.",
    volume = "83",
    pages = "3370--3373",
    year = "1999"
}

@article{Randall:1999vf,
    author = "Randall, Lisa and Sundrum, Raman",
    title = "{An Alternative to compactification}",
    eprint = "hep-th/9906064",
    archivePrefix = "arXiv",
    reportNumber = "MIT-CTP-2874, PUPT-1867, BUHEP-99-13",
    doi = "10.1103/PhysRevLett.83.4690",
    journal = "Phys. Rev. Lett.",
    volume = "83",
    pages = "4690--4693",
    year = "1999"
}

@inproceedings{Csaki:2004ay,
    author = "Csaki, Csaba",
    title = "{TASI lectures on extra dimensions and branes}",
    booktitle = "{Theoretical Advanced Study Institute in Elementary Particle Physics (TASI 2002): Particle Physics and Cosmology: The Quest for Physics Beyond the Standard Model(s)}",
    eprint = "hep-ph/0404096",
    archivePrefix = "arXiv",
    pages = "605--698",
    month = "4",
    year = "2004"
}

@article{Giudice:2016yja,
    author = "Giudice, Gian F. and McCullough, Matthew",
    title = "{A Clockwork Theory}",
    eprint = "1610.07962",
    archivePrefix = "arXiv",
    primaryClass = "hep-ph",
    reportNumber = "CERN-TH-2016-223",
    doi = "10.1007/JHEP02(2017)036",
    journal = "JHEP",
    volume = "02",
    pages = "036",
    year = "2017"
}

@article{Giudice:2017fmj,
    author = "Giudice, Gian F. and Kats, Yevgeny and McCullough, Matthew and Torre, Riccardo and Urbano, Alfredo",
    title = "{Clockwork/linear dilaton: structure and phenomenology}",
    eprint = "1711.08437",
    archivePrefix = "arXiv",
    primaryClass = "hep-ph",
    reportNumber = "CERN-TH-2017-219",
    doi = "10.1007/JHEP06(2018)009",
    journal = "JHEP",
    volume = "06",
    pages = "009",
    year = "2018"
}

@article{Bernal:2020fvw,
    author = "Bernal, Nicol\'as and Donini, Andrea and Folgado, Miguel G. and Rius, Nuria",
    title = "{Kaluza-Klein FIMP Dark Matter in Warped Extra-Dimensions}",
    eprint = "2004.14403",
    archivePrefix = "arXiv",
    primaryClass = "hep-ph",
    reportNumber = "PI/UAN-2020-669FT, FTUV-20-0415.2168, IFIC/20-15",
    doi = "10.1007/JHEP09(2020)142",
    journal = "JHEP",
    volume = "09",
    pages = "142",
    year = "2020"
}

@article{ATLAS:2017ayi,
    author = "Aaboud, Morad and others",
    collaboration = "ATLAS",
    title = "{Search for new phenomena in high-mass diphoton final states using 37 fb$^{-1}$ of proton--proton collisions collected at $\sqrt{s}=13$ TeV with the ATLAS detector}",
    eprint = "1707.04147",
    archivePrefix = "arXiv",
    primaryClass = "hep-ex",
    reportNumber = "CERN-EP-2017-132",
    doi = "10.1016/j.physletb.2017.10.039",
    journal = "Phys. Lett. B",
    volume = "775",
    pages = "105--125",
    year = "2017"
}

@article{ATLAS:2017fih,
    author = "Aaboud, Morad and others",
    collaboration = "ATLAS",
    title = "{Search for new high-mass phenomena in the dilepton final state using 36 fb$^{-1}$ of proton-proton collision data at $\sqrt{s}=13$ TeV with the ATLAS detector}",
    eprint = "1707.02424",
    archivePrefix = "arXiv",
    primaryClass = "hep-ex",
    reportNumber = "CERN-EP-2017-119",
    doi = "10.1007/JHEP10(2017)182",
    journal = "JHEP",
    volume = "10",
    pages = "182",
    year = "2017"
}

@article{CMS:2018dqv,
    author = "Sirunyan, A. M. and others",
    collaboration = "CMS",
    title = "{Search for physics beyond the standard model in high-mass diphoton events from proton-proton collisions at $\sqrt{s} =$ 13 TeV}",
    eprint = "1809.00327",
    archivePrefix = "arXiv",
    primaryClass = "hep-ex",
    reportNumber = "CMS-EXO-17-017, CERN-EP-2018-219",
    doi = "10.1103/PhysRevD.98.092001",
    journal = "Phys. Rev. D",
    volume = "98",
    number = "9",
    pages = "092001",
    year = "2018"
}

@article{ATLAS:2019erb,
    author = "Aad, Georges and others",
    collaboration = "ATLAS",
    title = "{Search for high-mass dilepton resonances using 139 fb$^{-1}$ of $pp$ collision data collected at $\sqrt{s}=$13 TeV with the ATLAS detector}",
    eprint = "1903.06248",
    archivePrefix = "arXiv",
    primaryClass = "hep-ex",
    reportNumber = "CERN-EP-2019-030",
    doi = "10.1016/j.physletb.2019.07.016",
    journal = "Phys. Lett. B",
    volume = "796",
    pages = "68--87",
    year = "2019"
}

@article{Antoniadis:1997zg,
    author = "Antoniadis, Ignatios and Dimopoulos, S. and Dvali, G. R.",
    title = "{Millimeter range forces in superstring theories with weak scale compactification}",
    eprint = "hep-ph/9710204",
    archivePrefix = "arXiv",
    reportNumber = "CERN-TH-97-260, CPTH-S562-0997",
    doi = "10.1016/S0550-3213(97)00808-0",
    journal = "Nucl. Phys. B",
    volume = "516",
    pages = "70--82",
    year = "1998"
}

@article{Arkani-Hamed:1998jmv,
    author = "Arkani-Hamed, Nima and Dimopoulos, Savas and Dvali, G. R.",
    title = "{The Hierarchy problem and new dimensions at a millimeter}",
    eprint = "hep-ph/9803315",
    archivePrefix = "arXiv",
    reportNumber = "SLAC-PUB-7769, SU-ITP-98-13",
    doi = "10.1016/S0370-2693(98)00466-3",
    journal = "Phys. Lett. B",
    volume = "429",
    pages = "263--272",
    year = "1998"
}

@article{Antoniadis:1990ew,
    author = "Antoniadis, Ignatios",
    title = "{A Possible new dimension at a few TeV}",
    reportNumber = "EP-CPTH-A978-0690",
    doi = "10.1016/0370-2693(90)90617-F",
    journal = "Phys. Lett. B",
    volume = "246",
    pages = "377--384",
    year = "1990"
}

@article{Arkani-Hamed:1998sfv,
    author = "Arkani-Hamed, Nima and Dimopoulos, Savas and Dvali, G. R.",
    title = "{Phenomenology, astrophysics and cosmology of theories with submillimeter dimensions and TeV scale quantum gravity}",
    eprint = "hep-ph/9807344",
    archivePrefix = "arXiv",
    reportNumber = "SLAC-PUB-7864, SU-ITP-98-142, IC-98-44",
    doi = "10.1103/PhysRevD.59.086004",
    journal = "Phys. Rev. D",
    volume = "59",
    pages = "086004",
    year = "1999"
}

@article{Antoniadis:1998ig,
    author = "Antoniadis, Ignatios and Arkani-Hamed, Nima and Dimopoulos, Savas and Dvali, G. R.",
    title = "{New dimensions at a millimeter to a Fermi and superstrings at a TeV}",
    eprint = "hep-ph/9804398",
    archivePrefix = "arXiv",
    reportNumber = "SLAC-PUB-7801, SU-ITP-98-28, CPTH-S608-0498, IC-98-39",
    doi = "10.1016/S0370-2693(98)00860-0",
    journal = "Phys. Lett. B",
    volume = "436",
    pages = "257--263",
    year = "1998"
}

@article{Lee:2013bua,
    author = "Lee, Hyun Min and Park, Myeonghun and Sanz, Veronica",
    title = "{Gravity-mediated (or Composite) Dark Matter}",
    eprint = "1306.4107",
    archivePrefix = "arXiv",
    primaryClass = "hep-ph",
    reportNumber = "CERN-PH-TH-2013-143, KIAS-P13032",
    doi = "10.1140/epjc/s10052-014-2715-8",
    journal = "Eur. Phys. J. C",
    volume = "74",
    pages = "2715",
    year = "2014"
}

@article{Lee:2014caa,
    author = "Lee, Hyun Min and Park, Myeonghun and Sanz, Veronica",
    title = "{Gravity-mediated (or Composite) Dark Matter Confronts Astrophysical Data}",
    eprint = "1401.5301",
    archivePrefix = "arXiv",
    primaryClass = "hep-ph",
    reportNumber = "CERN-PH-TH-2013-175",
    doi = "10.1007/JHEP05(2014)063",
    journal = "JHEP",
    volume = "05",
    pages = "063",
    year = "2014"
}

@article{Han:2015cty,
    author = "Han, Chengcheng and Lee, Hyun Min and Park, Myeonghun and Sanz, Veronica",
    title = "{The diphoton resonance as a gravity mediator of dark matter}",
    eprint = "1512.06376",
    archivePrefix = "arXiv",
    primaryClass = "hep-ph",
    doi = "10.1016/j.physletb.2016.02.040",
    journal = "Phys. Lett. B",
    volume = "755",
    pages = "371--379",
    year = "2016"
}

@article{Rueter:2017nbk,
    author = "Rueter, Thomas D. and Rizzo, Thomas G. and Hewett, JoAnne L.",
    title = "{Gravity-Mediated Dark Matter Annihilation in the Randall-Sundrum Model}",
    eprint = "1706.07540",
    archivePrefix = "arXiv",
    primaryClass = "hep-ph",
    doi = "10.1007/JHEP10(2017)094",
    journal = "JHEP",
    volume = "10",
    pages = "094",
    year = "2017"
}

@inproceedings{Rizzo:2018obe,
    author = "Rizzo, Thomas G.",
    title = "{Dark Photons, Kinetic Mixing and Light Dark Matter From 5-D}",
    booktitle = "{53$^{rd}$ Rencontres de Moriond on Electroweak Interactions and Unified Theories}",
    eprint = "1804.03560",
    archivePrefix = "arXiv",
    primaryClass = "hep-ph",
    reportNumber = "SLAC-PUB-17245",
    pages = "227--232",
    year = "2018"
}

@article{Rizzo:2018joy,
    author = "Rizzo, Thomas G.",
    title = "{Kinetic mixing, dark photons and extra dimensions. Part II: fermionic dark matter}",
    eprint = "1805.08150",
    archivePrefix = "arXiv",
    primaryClass = "hep-ph",
    reportNumber = "SLAC-PUB-17178",
    doi = "10.1007/JHEP10(2018)069",
    journal = "JHEP",
    volume = "10",
    pages = "069",
    year = "2018"
}

@article{Carrillo-Monteverde:2018phy,
    author = "Carrillo-Monteverde, A. and Kang, Yoo-Jin and Lee, Hyun Min and Park, Myeonghun and Sanz, Veronica",
    title = "{Dark Matter Direct Detection from new interactions in models with spin-two mediators}",
    eprint = "1803.02144",
    archivePrefix = "arXiv",
    primaryClass = "hep-ph",
    doi = "10.1007/JHEP06(2018)037",
    journal = "JHEP",
    volume = "06",
    pages = "037",
    year = "2018"
}

@article{Brax:2019koq,
    author = "Brax, Philippe and Fichet, Sylvain and Tanedo, Philip",
    title = "{The Warped Dark Sector}",
    eprint = "1906.02199",
    archivePrefix = "arXiv",
    primaryClass = "hep-ph",
    reportNumber = "UCR-TR-2019-FLIP-NCC-1804",
    doi = "10.1016/j.physletb.2019.135012",
    journal = "Phys. Lett. B",
    volume = "798",
    pages = "135012",
    year = "2019"
}

@article{deGiorgi:2021xvm,
    author = "de Giorgi, Arturo and Vogl, Stefan",
    title = "{Dark matter interacting via a massive spin-2 mediator in warped extra-dimensions}",
    eprint = "2105.06794",
    archivePrefix = "arXiv",
    primaryClass = "hep-ph",
    doi = "10.1007/JHEP11(2021)036",
    journal = "JHEP",
    volume = "11",
    pages = "036",
    year = "2021"
}

@article{deGiorgi:2022yha,
    author = "de Giorgi, Arturo and Vogl, Stefan",
    title = "{Warm dark matter from a gravitational freeze-in in extra dimensions}",
    eprint = "2208.03153",
    archivePrefix = "arXiv",
    primaryClass = "hep-ph",
    doi = "10.1007/JHEP04(2023)032",
    journal = "JHEP",
    volume = "04",
    pages = "032",
    year = "2023"
}

@article{Chivukula:2024nzt,
    author = "Chivukula, R. Sekhar and Gill, Joshua A. and Mohan, Kirtimaan A. and Sanamyan, George and Sengupta, Dipan and Simmons, Elizabeth H. and Wang, Xing",
    title = "{Limits on Kaluza-Klein portal dark matter models}",
    eprint = "2411.02509",
    archivePrefix = "arXiv",
    primaryClass = "hep-ph",
    doi = "10.1103/PhysRevD.111.075030",
    journal = "Phys. Rev. D",
    volume = "111",
    number = "7",
    pages = "075030",
    year = "2025"
}

@article{Aharony:1999ti,
    author = "Aharony, Ofer and Gubser, Steven S. and Maldacena, Juan Martin and Ooguri, Hirosi and Oz, Yaron",
    title = "{Large N field theories, string theory and gravity}",
    eprint = "hep-th/9905111",
    archivePrefix = "arXiv",
    reportNumber = "CERN-TH-99-122, HUTP-99-A027, LBNL-43113, RU-99-18, UCB-PTH-99-16, LBL-43113",
    doi = "10.1016/S0370-1573(99)00083-6",
    journal = "Phys. Rept.",
    volume = "323",
    pages = "183--386",
    year = "2000"
}

@article{Folgado:2019sgz,
    author = "Folgado, Miguel G. and Donini, Andrea and Rius, Nuria",
    title = "{Gravity-mediated Scalar Dark Matter in Warped Extra-Dimensions}",
    eprint = "1907.04340",
    archivePrefix = "arXiv",
    primaryClass = "hep-ph",
    reportNumber = "FTUV-19-0523.767, IFIC/19-26",
    doi = "10.1007/JHEP01(2020)161",
    journal = "JHEP",
    volume = "01",
    month = "1",
    year = "2020",
    pages = "161",
    note = "[Erratum: JHEP 02, 129 (2022)]"
}

@article{McDonald:2001vt,
    author = "McDonald, John",
    title = "{Thermally generated gauge singlet scalars as selfinteracting dark matter}",
    eprint = "hep-ph/0106249",
    archivePrefix = "arXiv",
    doi = "10.1103/PhysRevLett.88.091304",
    journal = "Phys. Rev. Lett.",
    volume = "88",
    pages = "091304",
    year = "2002"
}

@article{Choi:2005vq,
    author = "Choi, Ki-Young and Roszkowski, Leszek",
    editor = "Choi, Kiwoon and Kim, Jihn E. and Son, Dongchul",
    title = "{E-WIMPs}",
    eprint = "hep-ph/0511003",
    archivePrefix = "arXiv",
    doi = "10.1063/1.2149672",
    journal = "AIP Conf. Proc.",
    volume = "805",
    number = "1",
    pages = "30--36",
    year = "2005"
}

@article{Kusenko:2006rh,
    author = "Kusenko, Alexander",
    title = "{Sterile neutrinos, dark matter, and the pulsar velocities in models with a Higgs singlet}",
    eprint = "hep-ph/0609081",
    archivePrefix = "arXiv",
    reportNumber = "UCLA-06-TEP-23",
    doi = "10.1103/PhysRevLett.97.241301",
    journal = "Phys. Rev. Lett.",
    volume = "97",
    pages = "241301",
    year = "2006"
}

@article{Petraki:2007gq,
    author = "Petraki, Kalliopi and Kusenko, Alexander",
    title = "{Dark-matter sterile neutrinos in models with a gauge singlet in the Higgs sector}",
    eprint = "0711.4646",
    archivePrefix = "arXiv",
    primaryClass = "hep-ph",
    reportNumber = "UCLA-07-TEP-27",
    doi = "10.1103/PhysRevD.77.065014",
    journal = "Phys. Rev. D",
    volume = "77",
    pages = "065014",
    year = "2008"
}

@article{Hall:2009bx,
    author = "Hall, Lawrence J. and Jedamzik, Karsten and March-Russell, John and West, Stephen M.",
    title = "{Freeze-In Production of FIMP Dark Matter}",
    eprint = "0911.1120",
    archivePrefix = "arXiv",
    primaryClass = "hep-ph",
    reportNumber = "OUTP-09-18-P, UCB-PTH-09-32",
    doi = "10.1007/JHEP03(2010)080",
    journal = "JHEP",
    volume = "03",
    pages = "080",
    year = "2010"
}

@article{Bernal:2017kxu,
    author = "Bernal, Nicol\'as and Heikinheimo, Matti and Tenkanen, Tommi and Tuominen, Kimmo and Vaskonen, Ville",
    title = "{The Dawn of FIMP Dark Matter: A Review of Models and Constraints}",
    eprint = "1706.07442",
    archivePrefix = "arXiv",
    primaryClass = "hep-ph",
    reportNumber = "PI-UAN-2017-602FT, HIP-2017-08-TH, PI-UAN--2017--602FT, HIP--2017--08-TH",
    doi = "10.1142/S0217751X1730023X",
    journal = "Int. J. Mod. Phys. A",
    volume = "32",
    number = "27",
    pages = "1730023",
    year = "2017"
}

@article{Lesgourgues:2000tj,
    author = "Lesgourgues, Julien and Pastor, Sergio and Peloso, Marco and Sorbo, Lorenzo",
    title = "{Cosmology of the Randall-Sundrum model after dilaton stabilization}",
    eprint = "hep-ph/0004086",
    archivePrefix = "arXiv",
    reportNumber = "SISSA-39-2000-EP",
    doi = "10.1016/S0370-2693(00)00943-6",
    journal = "Phys. Lett. B",
    volume = "489",
    pages = "411",
    year = "2000"
}

@article{Giudice:2002vh,
    author = "Giudice, Gian F. and Kolb, Edward W. and Lesgourgues, Julien and Riotto, Antonio",
    title = "{Transdimensional Physics and Inflation}",
    eprint = "hep-ph/0207145",
    archivePrefix = "arXiv",
    reportNumber = "CERN-TH-2002-149, FNAL-PUB-02-137-A, FERMILAB-PUB-02-137-A, LAPTH-921-02",
    doi = "10.1103/PhysRevD.66.083512",
    journal = "Phys. Rev. D",
    volume = "66",
    pages = "083512",
    year = "2002"
}

@article{Im:2017eju,
    author = "Im, Sang Hui and Nilles, Hans Peter and Trautner, Andreas",
    title = "{Exploring extra dimensions through inflationary tensor modes}",
    eprint = "1707.03830",
    archivePrefix = "arXiv",
    primaryClass = "hep-ph",
    doi = "10.1007/JHEP03(2018)004",
    journal = "JHEP",
    volume = "03",
    pages = "004",
    year = "2018"
}

@article{Figueroa:2024yja,
    author = "Figueroa, Daniel G. and Loayza, Nicolas",
    title = "{Geometric reheating of the Universe}",
    eprint = "2406.02689",
    archivePrefix = "arXiv",
    primaryClass = "astro-ph.CO",
    doi = "10.1088/1475-7516/2025/03/073",
    journal = "JCAP",
    volume = "03",
    pages = "073",
    year = "2025"
}

@article{Liddle:2003as,
    author = "Liddle, Andrew R and Leach, Samuel M",
    title = "{How long before the end of inflation were observable perturbations produced?}",
    eprint = "astro-ph/0305263",
    archivePrefix = "arXiv",
    doi = "10.1103/PhysRevD.68.103503",
    journal = "Phys. Rev. D",
    volume = "68",
    pages = "103503",
    year = "2003"
}

@article{Martin:2013nzq,
    author = "Martin, J{\'e}r{\^o}me and Ringeval, Christophe and Trotta, Roberto and Vennin, Vincent",
    title = "{The Best Inflationary Models After Planck}",
    eprint = "1312.3529",
    archivePrefix = "arXiv",
    primaryClass = "astro-ph.CO",
    doi = "10.1088/1475-7516/2014/03/039",
    journal = "JCAP",
    volume = "03",
    pages = "039",
    year = "2014"
}

@article{Adelberger:2009zz,
      author         = "Adelberger, E.G. and Gundlach, J.H. and Heckel, B.R. and
                        Hoedl, S. and Schlamminger, S.",
      title          = "{Torsion balance experiments: A low-energy frontier of
                        particle physics}",
      journal        = "Prog.Part.Nucl.Phys.",
      volume         = "62",
      pages          = "102-134",
      doi            = "10.1016/j.ppnp.2008.08.002",
      year           = "2009",
      SLACcitation   = "%%CITATION = PPNPD,62,102;%%",
}

@article{Kapner:2006si,
      author         = "Kapner, D.J. and Cook, T.S. and Adelberger, E.G. and
                        Gundlach, J.H. and Heckel, Blayne R. and others",
      title          = "{Tests of the gravitational inverse-square law below the
                        dark-energy length scale}",
      journal        = "Phys.Rev.Lett.",
      volume         = "98",
      pages          = "021101",
      doi            = "10.1103/PhysRevLett.98.021101",
      year           = "2007",
      eprint         = "hep-ph/0611184",
      archivePrefix  = "arXiv",
      primaryClass   = "hep-ph",
      SLACcitation   = "%%CITATION = HEP-PH/0611184;%%",
}

@article{Lee:2020zjt,
      author         = "Lee, J. G. and Adelberger, E. G. and Cook, T. S. and
                        Fleischer, S. M. and Heckel, B. R.",
      title          = "{New Test of the Gravitational $1/r^2$ Law at Separations
                        down to 52 $\mu$m}",
      journal        = "Phys. Rev. Lett.",
      volume         = "124",
      year           = "2020",
      number         = "10",
      pages          = "101101",
      doi            = "10.1103/PhysRevLett.124.101101",
      eprint         = "2002.11761",
      archivePrefix  = "arXiv",
      primaryClass   = "hep-ex",
      SLACcitation   = "%%CITATION = ARXIV:2002.11761;%%"
}

@article{Donini:2016kgu,
      author         = "Donini, A. and Marim\'on, S. G.",
      title          = "{Micro-orbits in a many-brane model and deviations from
                        Newton’s $1/r^2$ law}",
      journal        = "Eur. Phys. J.",
      volume         = "C76",
      year           = "2016",
      number         = "12",
      pages          = "696",
      doi            = "10.1140/epjc/s10052-016-4537-3",
      eprint         = "1609.05654",
      archivePrefix  = "arXiv",
      primaryClass   = "hep-ph",
      reportNumber   = "IFIC-16-30",
      SLACcitation   = "%%CITATION = ARXIV:1609.05654;%%"
}

@article{Baeza-Ballesteros:2021tha,
    author = "Baeza-Ballesteros, J. and Donini, A. and Nadal-Gisbert, S.",
    title = "{Dynamical measurements of deviations from Newton\textquoteright{}s $1/r^2$ law}",
    eprint = "2106.08611",
    archivePrefix = "arXiv",
    primaryClass = "hep-ph",
    doi = "10.1140/epjc/s10052-022-10086-6",
    journal = "Eur. Phys. J. C",
    volume = "82",
    number = "2",
    pages = "154",
    year = "2022"
}

@article{Baeza-Ballesteros:2023par,
    author = "Baeza-Ballesteros, Jorge and Donini, Andrea and Molina-Terriza, Gabriel and Monrabal, Francesc and Sim{\'o}n, Ander",
    title = "{Towards a realistic setup for a dynamical measurement of deviations from Newton{\textquoteright}s $1/r^2$ law: the impact of air viscosity}",
    eprint = "2312.13736",
    archivePrefix = "arXiv",
    primaryClass = "hep-ph",
    doi = "10.1140/epjc/s10052-024-12943-y",
    journal = "Eur. Phys. J. C",
    volume = "84",
    number = "6",
    pages = "596",
    year = "2024"
}

@article{Hannestad:2003yd,
    author = "Hannestad, Steen and Raffelt, Georg G.",
    title = "{Supernova and neutron star limits on large extra dimensions reexamined}",
    eprint = "hep-ph/0304029",
    archivePrefix = "arXiv",
    reportNumber = "MPP-2003-19",
    doi = "10.1103/PhysRevD.69.029901",
    journal = "Phys. Rev. D",
    volume = "67",
    pages = "125008",
    year = "2003",
    note = "[Erratum: Phys.Rev.D 69, 029901 (2004)]"
}

@article{Fermi-LAT:2012zxd,
    author = "Ajello, M. and others",
    collaboration = "Fermi-LAT",
    title = "{Limits on Large Extra Dimensions Based on Observations of Neutron Stars with the Fermi-LAT}",
    eprint = "1201.2460",
    archivePrefix = "arXiv",
    primaryClass = "astro-ph.HE",
    reportNumber = "SLAC-PUB-14976",
    doi = "10.1088/1475-7516/2012/02/012",
    journal = "JCAP",
    volume = "02",
    pages = "012",
    year = "2012"
}

@article{Montero:2022prj,
    author = "Montero, Miguel and Vafa, Cumrun and Valenzuela, Irene",
    title = "{The dark dimension and the Swampland}",
    eprint = "2205.12293",
    archivePrefix = "arXiv",
    primaryClass = "hep-th",
    doi = "10.1007/JHEP02(2023)022",
    journal = "JHEP",
    volume = "02",
    pages = "022",
    year = "2023"
}

@article{Appelquist:1982zs,
    author = "Appelquist, Thomas and Chodos, Alan",
    title = "{Quantum Effects in Kaluza-Klein Theories}",
    reportNumber = "YTP-82-22",
    doi = "10.1103/PhysRevLett.50.141",
    journal = "Phys. Rev. Lett.",
    volume = "50",
    pages = "141",
    year = "1983"
}

@article{Appelquist:1983vs,
    author = "Appelquist, Thomas and Chodos, Alan",
    title = "{The Quantum Dynamics of Kaluza-Klein Theories}",
    reportNumber = "YTP 83-05",
    doi = "10.1103/PhysRevD.28.772",
    journal = "Phys. Rev. D",
    volume = "28",
    pages = "772",
    year = "1983"
}

@article{Branchina:2023ogv,
    author = "Branchina, Carlo and Branchina, Vincenzo and Contino, Filippo and Pernace, Arcangelo",
    title = "{Does the cosmological constant really indicate the existence of a dark dimension?}",
    eprint = "2308.16548",
    archivePrefix = "arXiv",
    primaryClass = "hep-th",
    doi = "10.1142/S0219887824503055",
    journal = "Int. J. Geom. Meth. Mod. Phys.",
    volume = "22",
    number = "04",
    pages = "2450305",
    year = "2025"
}

@article{Branchina:2024ljd,
    author = "Branchina, Carlo and Branchina, Vincenzo and Contino, Filippo and Pernace, Arcangelo",
    title = "{Dark dimension and the effective field theory limit}",
    eprint = "2404.10068",
    archivePrefix = "arXiv",
    primaryClass = "hep-th",
    doi = "10.1142/S0219887824503031",
    journal = "Int. J. Geom. Meth. Mod. Phys.",
    volume = "22",
    number = "04",
    pages = "2450303",
    year = "2025"
}

@article{Anchordoqui:2022svl,
    author = "Anchordoqui, Luis A. and Antoniadis, Ignatios and Lust, Dieter",
    title = "{Aspects of the dark dimension in cosmology}",
    eprint = "2212.08527",
    archivePrefix = "arXiv",
    primaryClass = "hep-ph",
    reportNumber = "MPP-2022-285, LMU-ASC 55/22",
    doi = "10.1103/PhysRevD.107.083530",
    journal = "Phys. Rev. D",
    volume = "107",
    number = "8",
    pages = "083530",
    year = "2023"
}

@article{Antoniadis:2023sya,
    author = "Antoniadis, Ignatios and Cunat, Jules and Guillen, Anthony",
    title = "{Cosmological perturbations from five-dimensional inflation}",
    eprint = "2311.17680",
    archivePrefix = "arXiv",
    primaryClass = "hep-ph",
    doi = "10.1007/JHEP05(2024)290",
    journal = "JHEP",
    volume = "05",
    pages = "290",
    year = "2024"
}

@article{Lidsey:1995np,
    author = "Lidsey, James E. and Liddle, Andrew R. and Kolb, Edward W. and Copeland, Edmund J. and Barreiro, Tiago and Abney, Mark",
    title = "{Reconstructing the inflation potential : An overview}",
    eprint = "astro-ph/9508078",
    archivePrefix = "arXiv",
    reportNumber = "SUSSEX-AST-95-8-3, FERMILAB-PUB-95-280-A",
    doi = "10.1103/RevModPhys.69.373",
    journal = "Rev. Mod. Phys.",
    volume = "69",
    pages = "373--410",
    year = "1997"
}

@inproceedings{Terrero-Escalante:2001zfi,
    author = "Terrero-Escalante, Cesar A. and Schwarz, Dominik J. and García, Alberto A.",
    title = "{Revisiting the calculations of inflationary perturbations}",
    booktitle = "{International School on Violation of CP Symmetry and Related Processes}",
    eprint = "astro-ph/0102174",
    archivePrefix = "arXiv",
    doi = "10.1007/0-306-47115-9_22",
    pages = "235--259",
    month = "2",
    year = "2001"
}

@article{AtacamaCosmologyTelescope:2025nti,
    author = "Calabrese, Erminia and others",
    collaboration = "Atacama Cosmology Telescope",
    title = "{The Atacama Cosmology Telescope: DR6 constraints on extended cosmological models}",
    eprint = "2503.14454",
    archivePrefix = "arXiv",
    primaryClass = "astro-ph.CO",
    reportNumber = "FERMILAB-PUB-25-0157-PPD",
    doi = "10.1088/1475-7516/2025/11/063",
    journal = "JCAP",
    volume = "11",
    pages = "063",
    year = "2025"
}

@article{AtacamaCosmologyTelescope:2025blo,
    author = "Louis, Thibaut and others",
    collaboration = "Atacama Cosmology Telescope",
    title = "{The Atacama Cosmology Telescope: DR6 power spectra, likelihoods and {\ensuremath{\Lambda}}CDM parameters}",
    eprint = "2503.14452",
    archivePrefix = "arXiv",
    primaryClass = "astro-ph.CO",
    reportNumber = "FERMILAB-PUB-25-0071-PPD",
    doi = "10.1088/1475-7516/2025/11/062",
    journal = "JCAP",
    volume = "11",
    pages = "062",
    year = "2025"
}
\end{document}